\RequirePackage{etoolbox}
\providebool{arxivversion}
\global\setbool{arxivversion}{true}

\ifbool{arxivversion}{
\documentclass{article}
}
{
\documentclass{llncs}
}

\newcommand{\arxiv}[1]{\ifbool{arxivversion}{#1}{}}
\newcommand{\sstd}[1]{\ifbool{arxivversion}{}{#1}}

\usepackage{amsmath, amssymb, amsfonts}
\usepackage{graphicx}
\graphicspath{{figs/}{figs/datasets/}{figs/pair_alignments/}{figs/histograms/}{figs/heatmaps/}}
\usepackage{dblfloatfix}
\usepackage{wrapfig}
\usepackage{tabularx}
\usepackage{cite}
\usepackage{url}
\usepackage{color}
\usepackage{latexsym}
\usepackage{euscript}
\usepackage{caption}
\usepackage{times}
\usepackage{xspace}
\arxiv{
\usepackage{fullpage}
}

\arxiv{
\newtheorem{theorem}{Theorem}[section]
\newtheorem{definition}{Definition}
}

\newcommand{\reals}{\mathbb{R}}

\def\etal{\textsl{et~al.}}

\makeatletter
\newcommand\mparagraph[1]{\vspace{1.0ex \@plus1ex \@minus.2ex}\noindent\textbf{#1}\hspace{0.5em}}

\long\def\@makecaption#1#2{
   \vskip 10pt
   \setbox\@tempboxa\hbox{{\footnotesize \textbf{#1.} #2}}
   \ifdim \wd\@tempboxa >\hsize         
       {\footnotesize \textbf{#1.} #2\par}
     \else                              
       \hbox to\hsize{\hfil\box\@tempboxa\hfil}
   \fi}
\makeatother

\DeclareMathOperator*{\argmax}{arg\,max}

\DeclareCaptionType{copyrightbox}

\def\frechet{Fr\'{e}chet }
\def\fr{\mathop{\mathrm{Fr}}}

\def\a{a}
\def\b{\Delta}
\def\c{c}
\def\l{l}
\newcommand{\T}{\Gamma} 
\def\score{\sigma} 
\def\ass{\Sigma}
\def\rms{\mathsf{rms}}

\def\gsymbol{{\scriptsize \bot}}

\def\snonany{\score_{\phi, *}}
\def\snonany{\score_{\phi, *}}
\def\sanynon{\score_{*,\phi}}
\def\sgapany{\score_{\gsymbol, *}}
\def\sanygap{\score_{*, \gsymbol}}
\def\snongap{\score_{\phi,\gsymbol}}
\def\sgapnon{\score_{\gsymbol, \phi}}
\def\sgapgap{\score_{\gsymbol, \gsymbol}}

\def\anonany{\ass_{\phi, *}}
\def\anonany{\ass_{\phi, *}}
\def\aanynon{\ass_{*,\phi}}
\def\agapany{\ass_{\gsymbol, *}}
\def\aanygap{\ass_{*, \gsymbol}}
\def\anongap{\ass_{\phi,\gsymbol}}
\def\agapnon{\ass_{\gsymbol, \phi}}
\def\agapgap{\ass_{\gsymbol, \gsymbol}}

\def\swami#1{}
\def\thomas#1{}
\def\pankaj#1{}

\begin{document}

\title{Computing Similarity between a Pair of Trajectories}

\newcommand*\samethanks[1][\value{footnote}]{\footnotemark[#1]}

\author{Swaminathan Sankararaman
\arxiv{
\thanks{Department of Computer Science, Duke University. email:\texttt{\{swami, pankaj, thomasm\}@cs.duke.edu}}
}
\sstd{\inst{1}} 
\and Pankaj K. Agarwal
\arxiv{\samethanks}
\sstd{\inst{1}} 
\and Thomas  M\o lhave
\arxiv{\samethanks}
\sstd{\inst{1}} 
\and Arnold P. Boedihardjo
\arxiv{\thanks{U.S. Army Corps of Engineers. email:\texttt{arnold.p.boedihardjo@usace.army.mil}}
}
\sstd{\inst{2}}
}

\sstd{
\institute{
Department of Computer Science, Duke University\\ 
\email{\{swami, pankaj, thomasm\}@cs.duke.edu}
\and U.S. Army Corps of Engineers\\
\email{arnold.p.boedihardjo@usace.army.mil}
}
}

\arxiv{\date{}}

\maketitle

\begin{abstract}
  With recent advances in sensing and tracking technology, trajectory
  data is becoming increasingly pervasive and analysis of trajectory
  data is becoming exceedingly important. A fundamental problem in
  analyzing trajectory data is that of identifying common patterns
  between pairs or among groups of trajectories. In this paper, we
  consider the problem of identifying similar portions between a pair
  of trajectories, each observed as a sequence of points sampled from
  it. 

  We present new measures of trajectory similarity --- both local and
  global --- between a pair of trajectories to distinguish between
  similar and dissimilar portions. Our model is robust under noise and
  outliers, it does not make any assumptions on the sampling rates on
  either trajectory, and it works even if they are partially
  observed. Additionally, the model also yields a scalar similarity
  score which can be used to rank multiple pairs of trajectories
  according to similarity, e.g. in clustering applications. We also
  present efficient algorithms for computing the similarity under our
  measures; the worst-case running time is quadratic in the number of
  sample points.

  Finally, we present an extensive experimental study evaluating the
  effectiveness of our approach on real datasets, comparing with it
  with earlier approaches, and illustrating many issues that arise in
  trajectory data. Our experiments show that our approach is highly
  accurate in distinguishing similar and dissimilar portions as
  compared to earlier methods even with sparse sampling.
\end{abstract}

\def\comm#1{\textit{#1}}

\section{Introduction}
\label{sec:intro}

\emph{Trajectories} are functions from a time domain---an interval on
the real line---to $\reals^{d}$ with $d>1$, and observed as sequences of 
points sampled from them. They arise in the
description of any system that evolves over time and are being
recorded or inferred from literally hundreds of millions of sensors
nowadays, from traffic monitoring systems~\cite{WZHYL} and recordings
of GPS sensors on cell phones~\cite{wsj} to cameras in surveillance
systems and those embedded in smart phones, helmets of soldiers in the
field~\cite{helmet}, medical devices, and scientific experiments and
simulations such as molecular dynamic simulations~\cite{model}.

Taking full advantage of these datasets for creating knowledge and
improving our decision making requires availability of effective algorithms and
analysis tools for processing, organizing, and querying these
trajectory datasets. This step has been lagging behind partly because
of huge scale of data, which is constantly growing, but because of
other challenges as well. Trajectory datasets are marred by
sensing uncertainty, and are heterogeneous in their quality, format,
and temporal support. For instance, GPS measurements are only
approximately accurate, and the inaccuracy can grow quite large in
conditions of poor satellite reception such as urban canyons; many
sensors operate on batteries, preferring to send aggregate or
approximate measurements to conserve energy (leading to nonuniform and
sparse sampling); trajectories retrieved from video typically arrive
in bits and pieces.  At the same time, individual trajectories can
have complex shapes, and even small nuances lead to big differences
in their understanding.

A central problem in analyzing trajectory data is to measure
similarity between a pair of trajectories and to identify portions
that are common between the two trajectories. Besides being
interesting in its own right, this problem often lies at the core of
classifying, clustering, and computing mean trajectories.  In some
applications (e.g. handwriting analysis, speech recognition), one of
the trajectories is the ground truth and the other is an observed
trajectory, while in some other applications (e.g. traffic analysis,
video surveillance), both trajectories are observed data.

\mparagraph{Problem statement.}  Let $P=\langle p_1,\ldots,p_m\rangle$
and $Q=\langle q_1,\ldots,q_n\rangle$ be two sequences of points in
$\reals^d$, sampled from two trajectories $\gamma_1$ and $\gamma_2$
defined over the time interval $[0,1]$ with added noise.  We assume
that the trajectories are defined over the same time interval only for
the simplicity of presentation; our models and algorithms work even
when they are defined over different time intervals. We also assume
that $P$ and $Q$ are points sampled from the images of the
trajectories and, for simplicity, we ignore the temporal
component;\footnote{Strictly speaking, a trajectory is the graph of
  the underlying function, and what we have are the curves traced by
  the two trajectories, but we will not distinguish between the two.}
again, our models and algorithms work even if we take the temporal
component into account.


Ideally, if we knew the two underlying trajectories $\gamma_1,
\gamma_2$, then their similar portions can be defined by
reparameterizing them using two functions $\alpha: [0,1] \rightarrow
\reals^d$ and $\beta: [0,1] \rightarrow \reals^d$ and identifying the
sub-intervals of $[0,1]$ over which the two of them are
similar.  
However, we only have finite sample points from each trajectory with
additional noise, so we identify similar portions between the two by
computing correspondences between sample points in one trajectory to
points in the other, and vice-versa. Such a set of correspondences
must satisfy the following criteria:
\begin{enumerate}
\item[(i)] Similar portions of the trajectories must be identified
  even if the sampling rates are different. For example, common routes
  taken by moving objects should be identified even if GPS coordinates
  are obtained at highly different times.
\item[(ii)] It must take into account noise/outliers, i.e., it must
  properly discriminate between noise and actual deviations of the two
  trajectories.
\item[(iii)] It should work even if the trajectories are partially
  observed, i.e., some portion of a trajectory is missing.
\end{enumerate}
For any given set of correspondences between two trajectories, it is
also advantageous to provide a score or measure indicating their
degree of similarity. Such a score is not only useful in identifying a
good set of correspondences but also as a way to rank multiple pairs
of trajectories according to similarity, e.g. in clustering
applications.

The objective is to define an appropriate model for correspondences
together with an appropriate scoring function to capture the above
requirements and given two trajectories, compute correspondences with
the optimal score efficient. In addition, we are also interested in
identifying most similar contiguous sub-trajectories of two given
trajectories. Thus, another objective is to compute {\bf\em local}
similarity between trajectories.

\begin{wrapfigure}{r}{0.25\textwidth}
\centering
\begin{picture}(0,0)%
\includegraphics{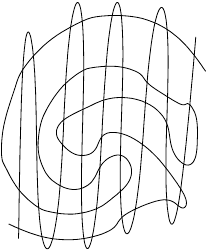}%
\end{picture}%
\setlength{\unitlength}{2072sp}%
\begingroup\makeatletter\ifx\SetFigFont\undefined%
\gdef\SetFigFont#1#2#3#4#5{%
  \reset@font\fontsize{#1}{#2pt}%
  \fontfamily{#3}\fontseries{#4}\fontshape{#5}%
  \selectfont}%
\fi\endgroup%
\begin{picture}(1891,2276)(8967,-8400)
\end{picture}%
\caption{A pair of trajectories with small Hausdorff distance but large Fr\'{e}chet distance.}
\label{fig:haus}
\end{wrapfigure}
\mparagraph{Related work.}  
Motivated by a wide range of applications, there is extensive work on
computing similarity between two objects (or rather points sampled
from the objects) in many fields, including computational geometry,
computer vision, computer graphics, and statistical learning. Two of
the commonly used methods for measuring similarity are Hausdorff
distance~\cite{agsurvey} and earth-movers
distance~\cite{Rubner:2000:EMD:365875.365881}, 
and their
variants. These methods are, however, not suitable for measuring
similarity between two trajectories because they only focus on the
location of points and not their ordering along the trajectories. A
better choice is the so-called Fr\'echet distance~\cite{ALT1995} (see
Fig.~\ref{fig:haus}). 
 Informally, consider a person and a dog
connected by a leash, each walking along a curve from its starting
point to its end point. Both are allowed to control their speed, but
they cannot backtrack. The Fr\'echet distance between the two curves
is the minimal length of a leash that is sufficient for traversing
both curves in this manner.  A {\em reparameterization} is a
continuous non-decreasing surjection $\alpha: [0,1] \rightarrow [0,1]$,
such that $\alpha(0)=0$ and $\alpha(1) = 1$.  The Fr\'echet distance
$\fr(\gamma_1,\gamma_2)$ between two trajectories $\gamma_1$ and
$\gamma_2$ is then defined as follows:
$$
\fr(\gamma_1, \gamma_2) = \inf_{\alpha, \beta}\max_{t \in [0,1]} \| \gamma_1(\alpha(t)) -  \gamma_2(\beta(t)) \|,
$$
where $\|\cdot\|$ is the underlying norm (typically, as above, the
Euclidean norm), and $\alpha$ and $\beta$ are reparameterizations of
$[0,1]$. The Fr\'echet distance between two polygonal curves with $m$
and $n$ vertices respectively can be computed in $O(mn\log (m+n))$
time~\cite{ALT1995}.

If we only have samples of points on each trajectory, then a simpler
variant, called \emph{discrete Fr\'echet distance}, is used, where the
dog and its owner are replaced by a pair of frogs that can only reside
on $n$ and $m$ specific stones $P$ and $Q$ on $\gamma_1$ and
$\gamma_2$, respectively. These frogs hop from one stone to the next
without backtracking, and the discrete Fr\'echet distance is the
minimum length of a ``leash'' that connect the frogs and allows them
to execute such a sequence of hops. The discrete Fr\'echet distance
can be computed in $O(mn)$ time by a straightforward quadratic
programming algorithm. Recently Agarwal \etal~\cite{pankaj_discretefr}
have presented a sub-quadratic algorithm for this problem.

\begin{figure*}[t]
\centering
\begin{tabularx}{\textwidth}{cccc}
\includegraphics[width=0.23\textwidth]{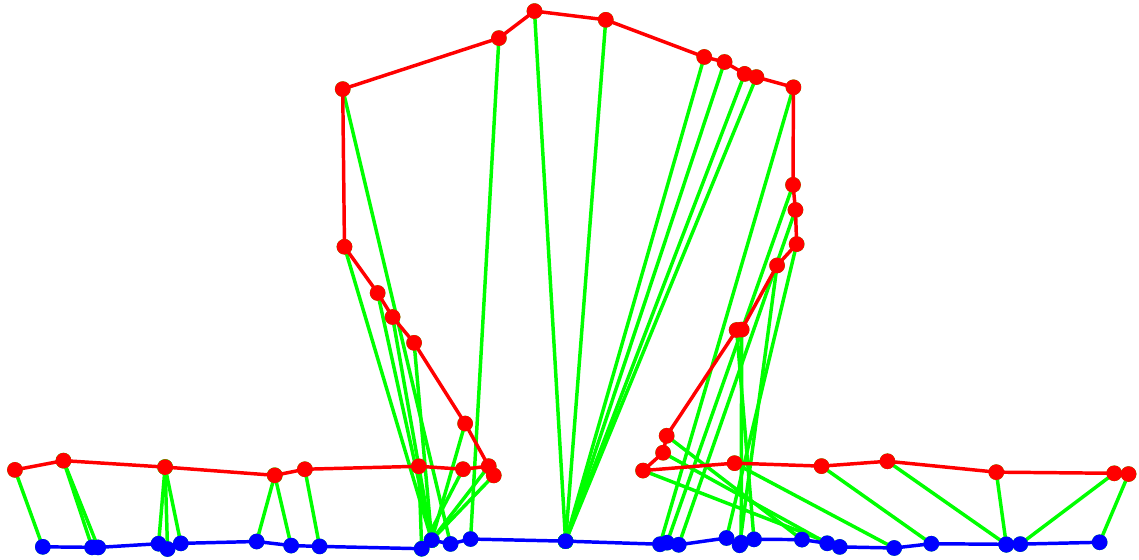} &
\includegraphics[width=0.23\textwidth]{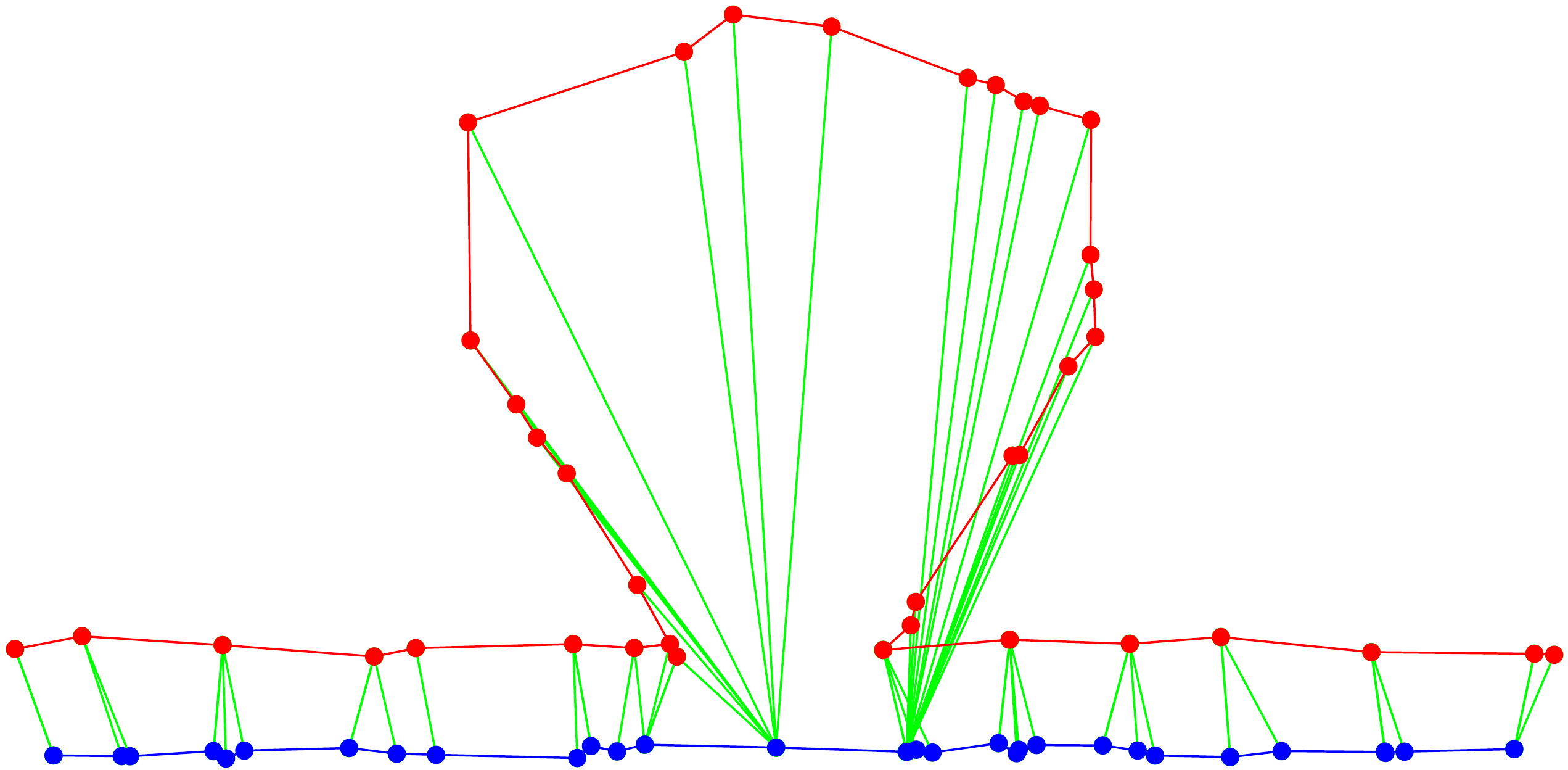} &
\includegraphics[width=0.23\textwidth]{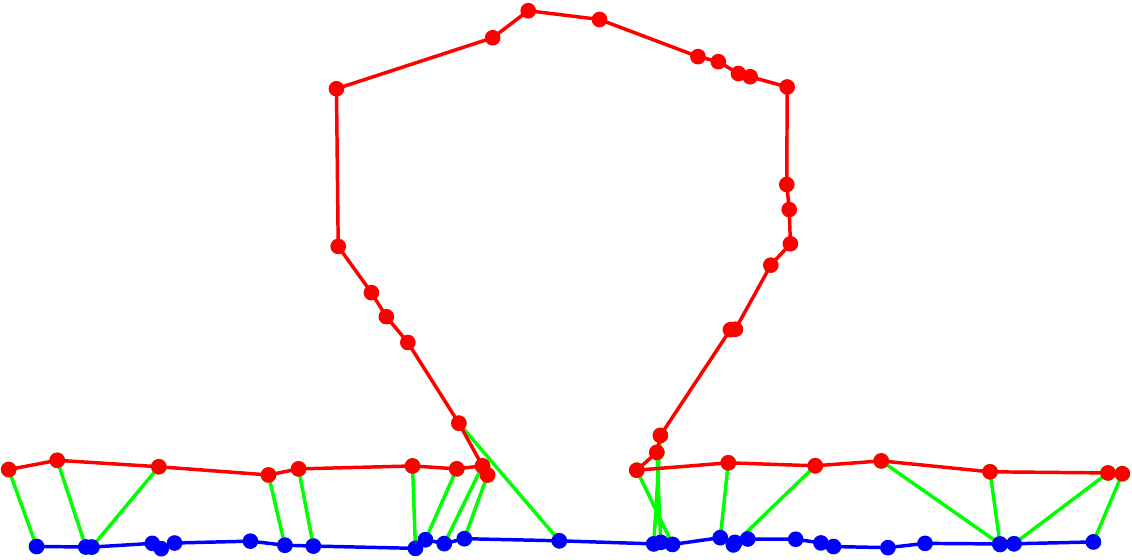} &
\includegraphics[width=0.23\textwidth]{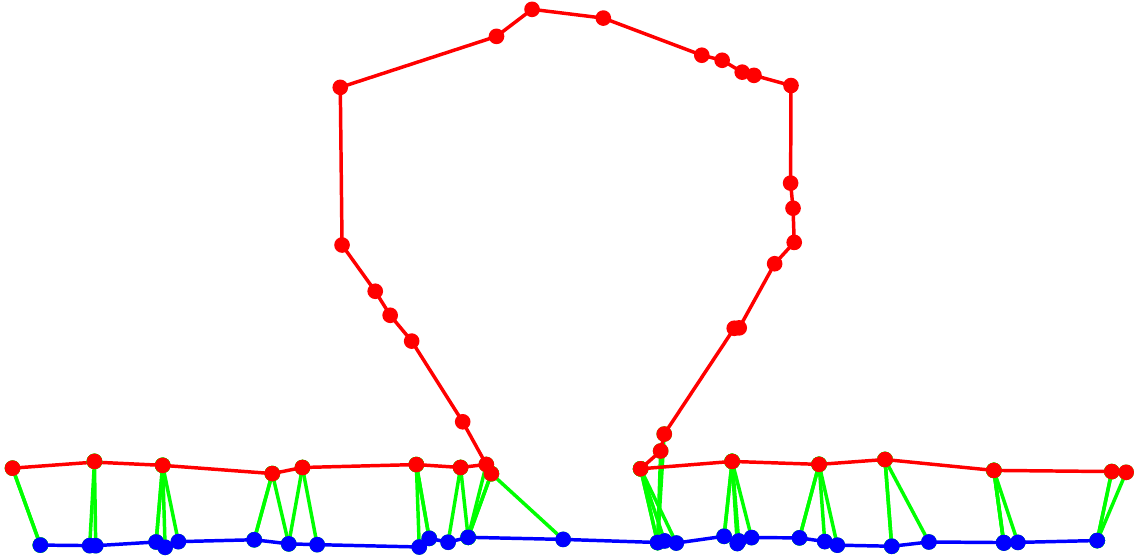} \\
(a) & (b) & (c) & (d)
\end{tabularx}
\caption{Computing similarity between two trajectories using different
  distance measures: (a) \frechet distance, (b) dynamic time warping
  or average \frechet distance, (c) sequence alignment based method,
  (d) our model.}
\label{fig:toy}
\end{figure*}

Although the \frechet distance is often used as a measure of curve
matching, in its traditional definition, an assignment yielding the
\frechet distance is not necessarily a good indicator of the
correspondences between the trajectories. The reason for this is
because there could possibly exist a large number of correspondences
which yield the optimal \frechet distance since the optimization
criteria is the distance between the two farthest points in the
coupling; see Fig.~\ref{fig:toy}(a).  Thus, in order to identify a
good correspondence between the two trajectories, we must resort to
the \emph{average \frechet distance}, more commonly known as {\em
  dynamic time warping distance}, where we use the analogy of a pair
of frogs hopping along $P$ and $Q$, but the goal is to minimize the
``average length'' of the leash instead of its maximum
length
.  \emph{Dynamic time warping (DTW)} was originally developed for
matching speech signals in speech
recognition~\cite{rabiner1993fundamentals} and computes good
correspondences between the two trajectories if they are similar for
the most part, as is the case for trajectories corresponding to pairs
of speech signals for a word or handwriting trajectories for a
particular words or letters.


However, if the trajectories contain significant dissimilar portions,
possibly due to actual deviations (such as different routes for GPS
trajectories), the results are not as meaningful. Consider
Fig.~\ref{fig:toy}(b) which shows two trajectories starting with
similar paths but where, one trajectory deviates from the other for a
significant portion and then rejoins the original path. DTW tries to
find a correspondence for all points and thus, gives correspondences
for points in the deviation for which no meaningful one
exists. Moreover, the actual measure is skewed by these portions and
it is difficult to distinguish them from the similar ones using the
obtained correspondences. It is even more difficult to distinguish
actual deviations from outliers caused by measurement errors.

There is a rich body of literature on pairwise sequence alignment in
computational biology, where the goal is to identify similar portions
between two DNA or protein sequences. Given two sequences $A$ and $B$,
their alignment is expressed by writing them in two rows respectively
such that at each position in the first (resp. second) row, there is
either a character in $A$ (resp. $B$) or a blank character (termed as
a \emph{gap character}). Characters in the same column are deemed
either identical or similar and characters in a sequence aligned to
gap characters do not have a correspondence in the other sequence.  A
maximal contiguous sequence of gap characters is termed a \emph{gap},
analogous to our notion of gaps; see
Def.~\ref{def:assignment}. The goal is to compute such an
alignment which optimizes a similarity scoring function which assigns
a score for aligning two characters and a penalty for gaps. The score
for an alignment of two characters is an incentive if they are deemed
similar and is a penalty if not. See \cite[cf. Chapter
2]{durbin1998biological} for further details. 

In computational biology, the algorithms for sequence alignments have
been extended to aligning two polygonal curves such as protein
backbones.  We may easily extend this model to the alignment of
trajectories with the choice of an appropriate scoring function. We
describe such a scoring function in Sec.~\ref{sec:global}. However,
as Fig.~\ref{fig:toy}(c) shows, non-uniform sampling rates cause
similar portions to be designated as gaps since correspondences are
restricted to be one-to-one.

Various approaches of defining distance in order to distinguish
similar and dissimilar portions exist~\cite{Ding2008} which aim to
combine the advantages of DTW as well as sequence alignment. These
include the longest common sub-sequence~\cite{Vlachos2002}, an
adaptation of the edit-distance measure~\cite{Chen2005} for sequences
where trajectory points are designated a match if they are closer than
a threshold distance and different otherwise, and the Edit-Distance
with real penalties measure~\cite{Chen2004}. The latter tries to
incorporate the distances by evaluating dissimilarity of points by
their distance to a fixed reference point. However, it is not clear
how this point should be chosen and why it is a good indicator of
dissimilarity. A drawback of all these measures is that they do not
distinguish between noise and actual dissimilarity. In fact, in the
application scenarios under consideration in these works, noise is the
primary reason for the dissimilarity since the underlying curves are
identical unlike the general setting.


Regarding the problem of computing partial or local similarity of
curves, Buchin \etal~\cite{Buchin2011} consider the problem of
sub-trajectory similarity under a continuous version of the average
\frechet distance where a user-defined parameter specifies the minimum
sub-trajectory duration. Further, in \cite{bbw-09}, the
authors show how to compute a ``partial'' \frechet distance, i.e., a
maximal portion of the trajectory curves which are close
enough. However, their algorithm does not work under the Euclidean
norm.

Apart from spatial similarity of trajectories, various other patterns
have also been considered. In \cite{Gudmundsson2007,Laube2005}, the
authors consider various formalizations of interesting patterns such
as \emph{flock}, where a set of trajectories follow each other for a
long duration and \emph{convergence} where a set of trajectories
converge at a particular location. In \cite{Hirano2005, football},
similarity of soccer player trajectories is considered based on
significant events such as passing or shooting.

Finally, as mentioned above, computation of similarity is often a requirement
for clustering trajectories. We refer to \cite{Lee2007, Elnekave2007,
  Piciarelli2008} for a few such
approaches. 

\mparagraph{Our contributions.}  Our first contribution is the
introduction of a new model for correspondences as defined using the
notion of assignments in Sec.~\ref{sec:global}. Together with this
model, we design an appropriate way to score assignments which can be
used to rank pairs of trajectories according to similarity. The model
and scoring function provides a unified framework which encompasses
all previous approaches for computing trajectory similarity such as
dynamic time warping, sequence alignment, edit-distance and others
while satisfying the requirements presented previously in the problem
statement. 
Our model may also be extended to spatio-temporal data by treating
time as an extra dimension. For conciseness, we ignore the temporal
component in the rest of the paper.

Next, in Sec.~\ref{sec:algos}, we describe a quadratic time dynamic
programming based algorithm for computing optimal assignments between
two trajectories $P$ and $Q$ under our scoring function. Our algorithm
combines ideas from sequence alignment and dynamic time warping and
matches the asymptotic running time complexity of these approaches. 

In Sec.~\ref{sec:local} we show that our model can also be used for
performing the so-called \emph{local assignment} for identifying most
similar sub-trajectories between two trajectories $P$ and $Q$. This is
done by adapting the model slightly; the algorithm remains the
same. Computing local assignments is similar to the notion of local
sequence alignment in computational biology.

If the points on $P$ and $Q$ are sampled non-uniformly and at
different rates, then mapping each point of $P$ to a point of $Q$ may
lead to unnatural alignments.  We therefore consider a
\emph{semi-continuous model} in which we interpolate each trajectory
between consecutive sampled points, say, by a linear function, and
allow a point of $P$ or $Q$ to be assigned to an interpolated point of
the other trajectory (cf. Sec.~\ref{sec:semi-cont}). This extension
leads to considerably better assignments in cases where sampling rate
differences are significant while measurement noise is less
significant.

Finally, we show the effectiveness of our model (in
Sec.~\ref{sec:experiments}) by evaluating it on a number of real
datasets and comparing it with sequence alignment, dynamic time
warping, and a modified dynamic time warping heuristic to distinguish
similar and deviating portions between pairs of trajectories. In
practice, our model captures the advantages of both dynamic time
warping and sequence-alignment based approaches with none of their
drawbacks.


\section{Model}\label{sec:global}

Let $P$ and $Q$ be two sequences of points as defined above. We
describe our model for measuring similarity between $P$ and $Q$ and
finding common portions of them. Our model builds on the strengths of
both dynamic time warping (DTW) and sequence alignment
. Recall
that
\begin{itemize}
\item DTW does not handle deviations well because it tries to match
  every point on $P$ and $Q$, but it can handle non-uniform sampling
  of points by allowing multiple points of $P$ to match with one point
  of $Q$, and vice versa; see Fig.~\ref{fig:toy}(b).
\item The sequence alignment model identifies deviations well while
  distinguishing them from noise but does not allow multiple points of
  $P$ to match with one point of $Q$, and hence, has issues with
  non-uniform sampling; see Fig.~\ref{fig:toy}(c).
\end{itemize}

The following notion of assignments models correspondences between $P$
and $Q$ which captures the advantages of the above approaches without
their drawbacks. Together with an appropriate scoring function defined
later, assignments accurately reflects the degree of similarity of one
trajectory to the other by satisfying the requirements posed in the
problem statement in Sec.~\ref{sec:intro}.

\begin{definition}\label{def:assignment}
  An \textbf{\emph{assignment}} for $P$
  and $Q$ is a pair of functions $\alpha:P\rightarrow Q\cup\{\gsymbol\}$
  and $\beta:Q\rightarrow P\cup\{\gsymbol\}$ for the points of $P$ and $Q$
  respectively. If $\alpha(p_i)=\gsymbol$ (or $\beta(q_j)=\gsymbol$), then
  $p_i$ (or $q_j$) is called a \emph{gap point}. A maximal contiguous 
  sequence of gap points in $P$ or $Q$ is called a \emph{gap}.

  An assignment is \textbf{\emph{monotone}} if it
  satisfies the following conditions: (i) if $\alpha(p_i)=q_j$ then
  for all $i' > i$, $\alpha(p_{i'})=\gsymbol$ or 
	$\alpha(p_{i'})=q_{j'}$ for some $j' > j$, 
  and (ii) if $\beta(q_j)=p_i$ then
  for all $j' > j$, $\beta(q_{j'})=\gsymbol$ or 
	$\beta(q_{j'})=p_{i'}$ for some $i' > i$.
\end{definition}

Intuitively, if a point $p_i\in P$ lies on a similar portion of the
two trajectories then $\alpha(p_i)$ defines the point on $Q$ to which
$p_i$ corresponds, and $p_i$ is a gap point otherwise. A similar
interpretation holds for $\beta(\cdot)$. Unlike traditional
alignment/matching models, our assignments are asymmetric which allows
it to better adapt to trajectories with different sampling rates. The
notion of gaps is introduced to identify deviations between the two
trajectories. Using gaps enables the identification of a ``good''
assignment even if there are only partial observations on any of the
trajectories; we can compute an assignment for the observed
portions. Distinguishing between noise and dissimilarity can be
accomplished by restricting to assignments where gaps are sufficiently
long (short deviations are more likely to be due to noise and thus,
the underlying portions of the trajectories are similar).

It will be easier to view an assignment of $P$ and $Q$
in terms of the complete directed
bipartite graph $G:=G(P,Q)=(P\cup Q, P\times Q\cup Q\times P)$, i.e.,
$G$ has a directed edge $(p_i,q_j)$ and another directed edge
$(q_j,p_i)$ for every pair $p_i\in P$ and $q_j\in Q$. We say that a
pair of edges $(p_i,q_j)$ (or $(q_j,p_i)$) and $(p_k,q_l)$ (or
$(q_l,p_k)$) \textbf{\emph{cross}} if $i<k$ and $j>l$ or vice
versa. Under our definition, the opposite edges $(p_i,q_j)$ and
$(q_j,p_i)$ do not cross each other.\footnote{Note that our definition
  of the crossing is topological, defined in terms of the graph
  $G$. the line segments $\overline{p_iq_j}$ and $\overline{p_kq_l}$
  corresponding to two non-crossing edges $(p_i,q_j)$ and $(p_k,q_l)$
  may cross each other (geometrically), and may be disjoint even if
  the graph edges are crossing.}

Hence, in this perspective, a \textbf{\emph{monotone}} assignment is a
set $E$ of pairwise non-crossing edges in $G$ so that $E$ has at most
one outgoing edge from every point in $P\cup Q$. Points with no
outgoing edges are gap points and, as in Def.~\ref{def:assignment}, a
maximal contiguous sequence of gap points in $P$ or $Q$ is called a
\textbf{\emph{gap}}, and the length of the sequence is called the
length of the gap. Let $\T(E)$ denote the set of gaps in $P$ and $Q$
for the assignment $E$. We define the \textbf{\emph{score}} of $E$,
denoted by $\score(P,Q;E)$, as
\begin{equation}
\score(P,Q;E) = \sum_{(u,v)\in E} \frac{1}{\c+\|u-v\|^2} + \sum_{g\in
  \T(E)} \a + \b|g|,\label{eqn:score}
\end{equation}
where $\a,\b$ and $\c$ are parameters which are chosen carefully, as
described in Sec.~\ref{sec:param-select}, $\|\cdot\|$ is the
$L_2$-norm and $|g|$ is the length of a gap $g$. For appropriate
comparison between different pairs of trajectories and their
corresponding assignments, the above score may be normalized in a
straightforward manner to provide a score between 0 and 1.
%
We can rewrite the score in terms of the functions $\alpha,\beta$ from
Def.~\ref{def:assignment} as well:
\begin{align}
\score(P,Q;\alpha,\beta) = &\sum_{\substack{p_i\in P\\\alpha(i)\neq
    \gsymbol}} \frac{1}{\c+\|p_i-\alpha(p_i)\|^2}  \nonumber\\
 &+ \sum_{\substack{q_j\in Q\\ \beta(j) \neq \gsymbol}}
 \frac{1}{\c+\|q_j-\beta(q_j)\|^2} 
  + \sum_{g\in \T(\alpha,\beta)} \Big(\a + \b\cdot |g|\Big),\label{eq:assignscore}
\end{align}
where $\T(\alpha,\beta)$ is the set of gaps in $P$ and $Q$ for the
assignment $\alpha,\beta$.

We define the similarity between $P$ and $Q$ as
\[
\score(P,Q) = \max_{\alpha,\beta} \score(P,Q;\alpha,\beta).
\]
The assignment $\alpha^*,\beta^*=\argmax_{\alpha,\beta}
\score(P,Q;\alpha,\beta)$ identifies the similar portions of $P$ and
$Q$.

\mparagraph{Why the directed graph?} We now explain why we chose a
directed graph model and an assignment. Both DTW and sequence
alignment can be formulated as computing a subset of non-crossing
edges in the complete \emph{undirected} bipartite graph $P\times
Q$. In particular, DTW finds a subset $D\subset P\times Q$ of
non-crossing edges such that each vertex of $P\cup Q$ is incident on
at least one edge of $D$ and the total ``length'' of edges in $D$ is
minimum. In some applications, an appropriate choice for the length of
edges would be the Euclidean distance whereas in other applications, a
different function is used. On the other hand, the sequence-alignment
model asks for computing a matching $M\subset P\times Q$ whose score
is maximum, where the score of a matching is similar to
(\ref{eqn:score}).

\begin{figure}[h]
\centering
\begin{minipage}{0.45\textwidth}
\centering
\begin{tabular*}{\textwidth}{@{\extracolsep{\fill}}cc}
\includegraphics[scale=0.8]{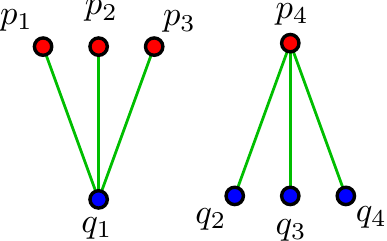} &
\includegraphics[scale=0.8]{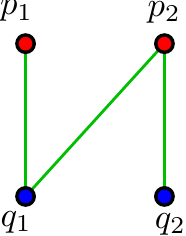} 
\\
(a) & (b) 
\end{tabular*}
\caption{Pros and cons of allowing multiple matches in an undirected
  graph:(a) A situation where allowing multiple matches is
  logical. (b) A situation where allowing multiple matches does not
  allow us to obtain a clear correspondence.}
\label{fig:undir_prob}
\end{minipage}
\hfil
\begin{minipage}{0.45\textwidth}
\centering
\begin{tabular*}{\textwidth}{@{\extracolsep{\fill}}cc}
\includegraphics[scale=0.8]{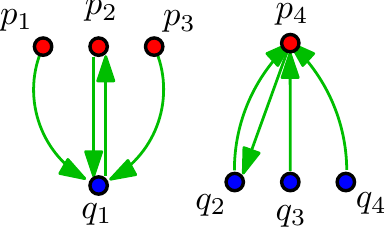} &
\includegraphics[scale=0.8]{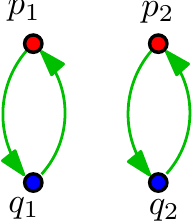}
\\
(a) & (b)
\end{tabular*}
\caption{Using a directed graph provides a logical set of
  correspondences always. Comparison with the examples of
  Fig.~\ref{fig:undir_prob}.}
\label{fig:dir_solves}
\end{minipage}
\end{figure}

It is tempting to describe an assignment as well with respect to the
undirected graph but this leads to difficulties. For example, we may
relax the condition of finding a ``matching'' from the
sequence-alignment model allowing multiple points of $P$ to match with
one point of $Q$, (see Fig.~\ref{fig:undir_prob}(a)) and simply ask
for finding a set of non-crossing edges whose score is maximum, but
this is not always meaningful and introduces additional edges that are
redundant. For example, in Fig.~\ref{fig:undir_prob}(b), this
approach will find three edges --- the diagonal edge $(p_2,q_1)$ is
spurious and is an artifact of the model because it is more meaningful
to match $p_1$ with $q_1$, $p_2$ with $q_2$ and vice versa.

The directed graph model avoids this problem by not requiring the
functions $\alpha$ and $\beta$ to be symmetric; see
Fig.~\ref{fig:dir_solves}.

There are a few other subtle advantages of using the directed graph
model, omitted from this abstract due to lack of space.
\swami{We can add it if no lack of space.}

\mparagraph{Remark.} (i) Our framework is not limited to the scoring
function (\ref{eqn:score}). For example, the sequence-alignment based
approach, DTW or other measures such as edit-distance are easily
incorporated into our model. (ii) Note that according to our
definition of assignments, vertices in $G$ which have incoming edges
but no outgoing edges are designated as gap points. This may be easily
modified so gap points have no incoming edges.

\begin{figure*}[ht]
\begin{minipage}{\textwidth}
\begin{align}
\score(i,j) = \max\Big\{& \sgapany(i,j), \sanygap(i,j), \snonany(i,j) +
\delta(i,j), \sanynon(i,j) + \delta(i,j)\Big\} \label{eqn:assrecur}\\
\sgapany(i,j) = \max\Big\{& \score(i-1,j) + \a + \b, \sgapany(i-1,j) +
\b, \sgapnon(i,j) + \delta(i,j), \nonumber\\
&\sgapgap(i,j)\Big\}\label{eqn:gapany} \\
\snonany(i,j) = \max\Big\{& \snongap(i-1,j),  \sanynon(i-1,j) +
\delta(i,j), \snonany(i,j-1) + \delta(i,j),\nonumber\\
 &\score(i-1,j) \Big\},\label{eqn:nonany}\\
\snongap(i,j) = \max\Big\{& \snonany(i,j-1) + \a + \b,  \snongap(i,j-1) + \b, \sanygap(i-1,j) \Big\},\label{eqn:nongap}\\
\sgapgap(i,j) = \max\Big\{& \sanygap(i-1,j) + \a + \b, \sgapgap(i-1,j)
+ \b, \sgapany(i,j-1) + \a + \b, \nonumber\\
&\sgapgap(i,j-1) + \b\Big\}.\label{eqn:gapgap}
\end{align}
\end{minipage}
\caption{Recurrence relations for $\score$ and each of the auxiliary
  functions. The relations for $\sanynon$, $\sgapnon$ are symmetric to
  those above for $\snonany$ and $\snongap$ respectively. Here, $\delta(i,j)=1/(\c+\|p_i-q_j\|^2)$.}
\label{fig:recur}
\end{figure*}
\section{Algorithms}\label{sec:algos}

We now describe an algorithm for computing the optimal score
$\score(P,Q)$ and the corresponding assignment. Our algorithm is
similar to that for sequence alignment~\cite{Needleman1970} and runs
in $O(mn)$ time. The main difference is that, because of the
asymmetric nature of the definition of assignments, the recurrence
relation is more complex and we need to compute a few auxiliary
assignments and score functions.

\mparagraph{Auxiliary functions.} For $1\leq i\leq m$, let
$P_i=\langle p_1,\dots,p_i\rangle$ denote the prefix of $P$ of length
$i$, and for $1\leq j\leq n$, let $Q_j=\langle q_1,\dots,q_j\rangle$
denote the prefix of $Q$ of length $j$. The algorithm computes the
similarity score $\score(P_i,Q_j)$ for all $1\leq i\leq m$ and $1\leq
j\leq n$. For brevity, we will use $\score(i,j)$ to denote
$\score(P_i,Q_j)$. Let $\Sigma(i,j)$ denote the (optimal) monotone
assignment corresponding to the score $\score(i,j)$. With a slight
abuse of notation, we will use both the graph representation --- a set
of non-crossing edges --- as well as the pair of functions
$\alpha,\beta$. It will be clear from the context which of the two
representations we are referring to.

For each pair $i,j$, to compute $\score(i,j)$ and $\Sigma(i,j)$
efficiently, we also compute a set of auxiliary functions described
below:
\begin{itemize}
\item $\sgapany(i,j)$ and $\sanygap(i,j)$: $\sgapany(i,j)$ denotes the
  score of the best monotone assignment, which is denoted by
  $\agapany(i,j)$, for $P_i$ and $Q_j$ with the restriction that $p_i$
  is a gap point. That is, there is no outgoing edge from $p_i$, but
  our model allows $\agapany(i,j)$ to have incoming edges to $p_i$. We
  similarly define $\sanygap(i,j)$ and $\aanygap(i,j)$.
\item $\sgapgap(i,j)$: the score of the best monotone assignment,
  denoted by $\agapgap(i,j)$, for $P_i$ and $Q_j$ with the restriction
  that both $p_i$ and $q_j$ are gap points.
\item $\snonany(i,j)$ and $\sanynon(i,j)$: $\snonany(i,j)$ is the
  score of the best monotone assignment for $P_i$ and $Q_j$, denoted
  by $\anonany(i,j)$, with the restriction that $p_i$ is not assigned
  to any point of $Q_j$ but points of $Q_j$ can be assigned to $p_i$,
  i.e., there is no outgoing edge from $p_i$ but there can be incoming
  edges to $p_i$. The difference between $\snonany(i,j)$ and
  $\sgapany(i,j)$ is that $p_i$ is considered as a gap point in the
  latter and $\sgapany(i,j)$ includes the gap score corresponding to
  $p_i$, namely $\a+\b$ if a new gap starts at $p_i$ in
  $\agapany(i,j)$ and $\b$ otherwise, while in the former no score is
  added corresponding to $p_i$. We define $\sanynon(i,j)$ and
  $\aanynon(i,j)$ analogously.
\item $\sgapnon(i,j)$ and $\snongap(i,j)$: $\sgapnon(i,j)$ is the
  score of the best monotone assignment for $P_i$ and $Q_j$, denoted
  by $\agapnon(i,j)$, with the restriction that $p_i$ is not assigned
  to any point of $Q_j$ as in the previous case, and $q_j$ is a gap
  point. Note that there are no outgoing edges from $p_i$ or $q_j$ but
  there may be incoming edges to one or both of them, and that
  $\sgapnon(i,j)$ does not include any score corresponding to $p_i$
  but does include a gap score for $q_j$. We define $\snongap(i,j)$
  and $\anongap(i,j)$ analogously.
\end{itemize}

\mparagraph{Recurrences.} We now describe recurrence relations for
each of the auxiliary score functions just described, so that each
value can be computed in $O(1)$ time using dynamic
programming. Fig.~\ref{fig:recur} describes the recurrence
relations. For brevity, we set
$\delta(i,j)=1/(\c+\|p_i-q_j\|^2)$. Because of lack of space, we derive
the recurrences only for $\score(i,j)$ and $\sgapany(i,j)$; others can
be derived in a similar manner. Also, the corresponding assignments
can be computed using the standard backtracking
method~\cite{Needleman1970}.

\noindent\textit{Recurrence for $\score(i,j)$.} Consider the optimal
assignment $\ass(i,j)$ for $P_i$ and $Q_j$. There are three
possibilities: (i) $p_i$ is a gap point, (ii) $q_j$ is a gap point, and
(iii) there are outgoing edges from both $p_i$ and $q_j$. In cases (i)
and (ii), $\ass(i,j)$ is $\agapany(i,j)$ and $\aanygap(i,j)$
respectively, by definition. So consider case (iii). Suppose
$\ass(i,j)$ has (directed) edges $(p_i,q_{j'})$ and
$(q_j,p_{i'})$. Since the edges in $\ass(i,j)$ are non-crossing,
$j'<j$ implies that $i'=i$ and similarly $i'<i$ implies that
$j'=j$. Hence, $\ass(i,j)$ contains at least one of $(p_i,q_j)$ and
$(q_j,p_i)$. In the former case, $\ass(i,j)\setminus \{(p_i,q_j)\}$ is
nothing but $\snonany(i,j)$, and in the latter case,
$\ass(i,j)\setminus \{(q_j,p_i)\}$ is $\aanynon(i,j)$. Thus,
$\score(i,j)$ satisfies recurrence (\ref{eqn:assrecur}).

\smallskip
\noindent\textit{Recurrence for $\sgapany(i,j)$.} There are three
cases for the assignment $\agapany(i,j)$: (i) $q_j$ is a gap point,
(ii) there is an incoming edge to $p_i$, (iii) there is no incoming
edge to $p_i$. In case (i), $\agapany(i,j)=\agapgap(i,j)$, since both
$p_i$ and $q_j$ are gap points. So assume $q_j$ has an outgoing
edge in $\agapany(i,j)$.

If there is an incoming edge to $p_i$, then, by the non-crossing
property, $\agapany(i,j)$ contains the edge $(q_j,p_i)$ and thus, we have
$\agapany(i,j)\setminus \{(q_j,p_i)\} = \agapnon(i,j)$. On the other
hand, if $\agapany(i,j)$ does not have an incoming edge to $p_i$, then
$\agapany(i,j)$ is the same as $\ass(i-1,j)$ or $\agapany(i-1,j)$
depending on whether a new gap starts at $p_i$ in $\agapany(i,j)$ or a
gap containing $p_i$ starts earlier. Putting everything together,
(\ref{eqn:gapany}) gives the recurrence for $\sgapany(i,j)$.

\smallskip We maintain a separate table for each auxiliary function
and compute the entries in increasing order of $i$ and $j$. For a
fixed pair $i,j$, we compute them in the following order:
$\sgapgap,\sgapnon, \snongap, \snonany, \sanynon, \sgapany, \sanygap ,
\score$. It can be verified from
(\ref{eqn:assrecur})--(\ref{eqn:gapgap}) that each of them can be
computed in $O(1)$ time. Hence, the total time spent in computing the
final $\score(P,Q)$ and $\ass(P,Q)$ is $O(mn)$. If we maintain the
entire tables, the space used is also $O(mn)$ but it can be reduced to
$O(m+n)$~\cite{Hirschberg:1975:LSA:360825.360861}. We thus obtain the
following:
\begin{theorem}\label{thm:algo}
Given two sequences of points $P$ and $Q$ in $\reals^d$ of lengths $m$
and $n$ respectively, the similarity score $\score(P,Q)$ and the
corresponding assignment can be computed in $O(mn)$ time using
$O(m+n)$ space.
\end{theorem}

\section{Local Assignment}\label{sec:local}

We now describe how we modify our model and algorithm, described in
Sections~\ref{sec:global} and \ref{sec:algos}, for finding the most
similar sub-trajectories between two trajectories, or the so-called
local assignments. Intuitively, during the course of the execution of
the algorithm from Sec.~\ref{sec:algos}, when we find that the score
of aligning initial portions of the trajectories is too small, we
should discard them from further consideration and start
afresh. However, the score of assigning a point $p_i$ to $q_j$, or
vice versa, is always positive. So, under this model, there is no
incentive for starting afresh. We therefore, modify the score of an
assignment $E$, represented as a set of non-crossing edges, as
follows:
\begin{align}
\score(P,Q;E) = &\sum_{(u,v)\in E} \left[\frac{1}{\c+\|u-v\|^2} -
  \tau\right] + \sum_{g\in \T(E)} \left(\a + (\b-\tau)|g|\right).\label{eqn:localscore}
\end{align}
Here, $\tau$ is a threshold parameter that we subtract from each term
in (\ref{eqn:score}). The value of $\tau$ indicates how low we would
like the score to go before we decide that it is too low for local
similarity and choose to start afresh. As earlier, our goal is to
compute the maximum score and the corresponding assignment:
\begin{align*}
\score(P,Q) = \max_{E} \score(P,Q;E);\quad \ass(P,Q) = \argmax_{E} \score(P,Q;E),
\end{align*}
where the maximum is taken over all monotone assignments.

Following the same ideas as in the algorithm for local sequence
alignment~\cite{Smith1981}, we can compute $\score(P,Q)$ in $O(mn)$
time. More precisely, we modify the recurrences in
(\ref{eqn:assrecur})--(\ref{eqn:gapgap}). For example,
\begin{align*}
\score(i,j)=\max \big\{&\sgapany(i,j), \sanygap(i,j), 
    \snonany(i,j)+\delta(i,j), 
    \sanynon(i,j)+\delta(i,j), 0\big\}.
\end{align*}
The others are modified accordingly, by adding the 0 term. Finally,
instead of returning $\score(m,n)$, we return the score
$\max_{i,j} \score(i,j)$. Omitting details, we conclude the
following:

\begin{theorem}\label{thm:localalgo}
Given two point sequences $P$ and $Q$ of lengths $m$ and $n$
respectively, an optimal local assignment for them can be computed in
$O(mn)$ time.
\end{theorem}

\section{Semi-Continuous Model}\label{sec:semi-cont}

\begin{figure}[h]
\centering
\begin{tabular}{cc}
\includegraphics[width=0.4\textwidth]{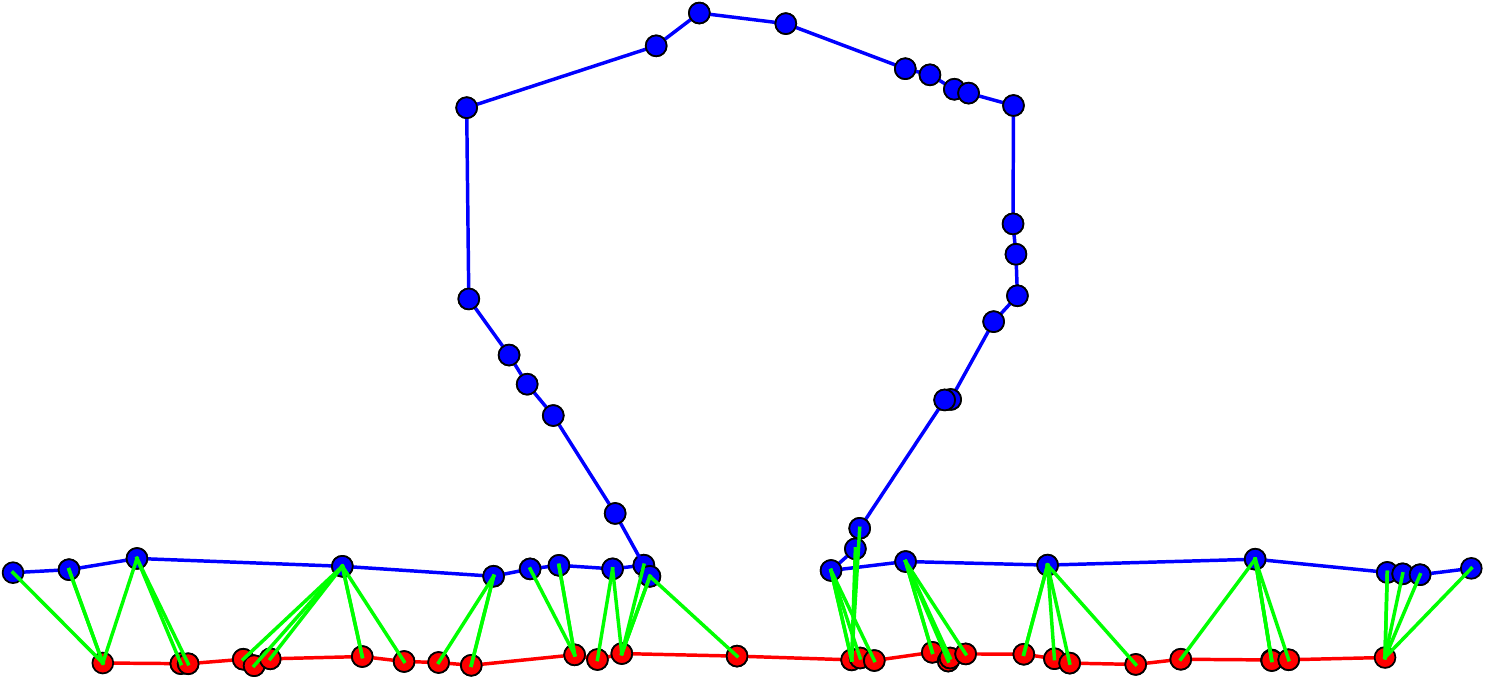} &
\includegraphics[width=0.4\textwidth]{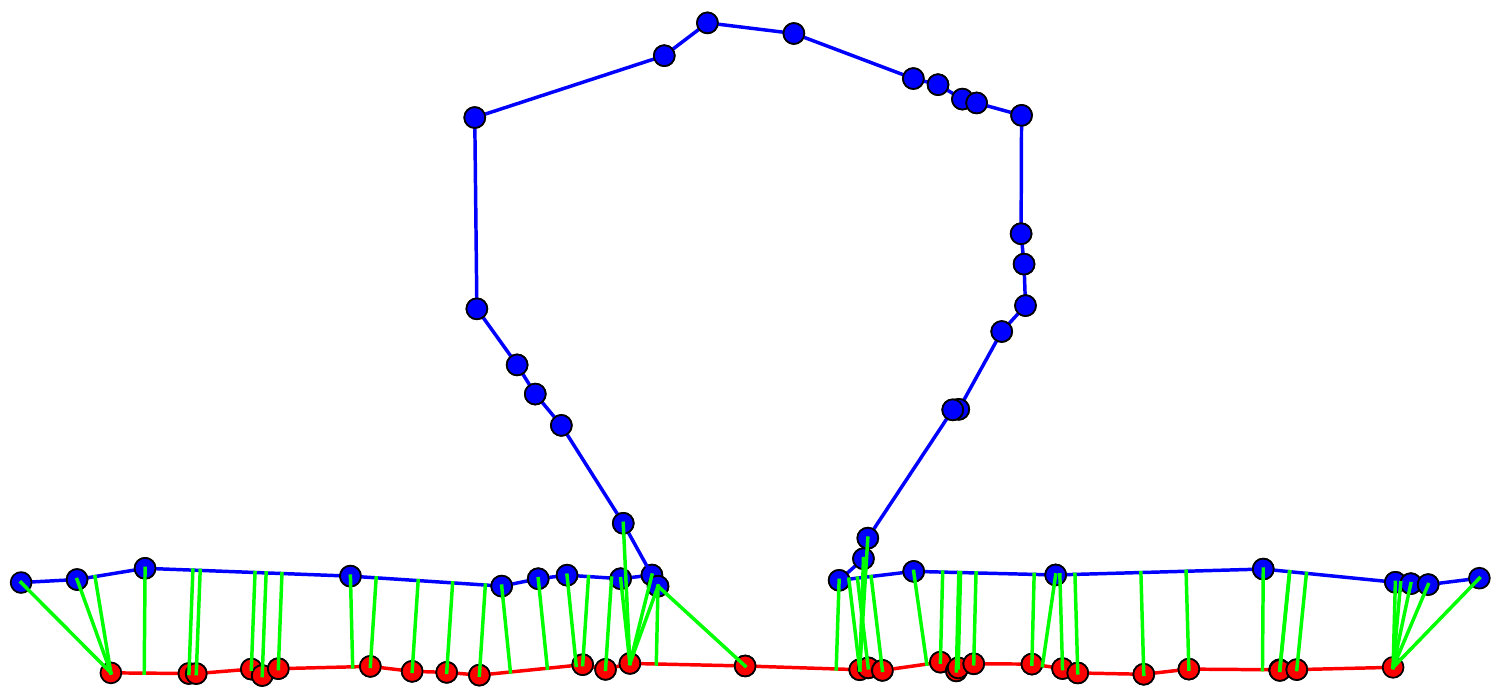} \\
(a) & (b) 
\end{tabular}
\caption{Discrete (a) vs semi-continuous (b).}
\label{fig:semicont_toy}
\end{figure}

As discussed above, the (asymmetric) assignment model is robust
to sampling rates since it tries to assign each point in one
trajectory to a similar point in the other trajectory if one
exists. In many applications, where data is being acquired by sensors
on a network and energy or communication is expensive, the sampling
may be very sparse at some portions. If we assume that $P$ and $Q$ are
points sampled from a continuous process, then we may wish to model
this continuous process and assign each point of $P$ (or $Q$) to a
point on this continuous trajectory which is not necessarily one of
the input points. For example,
consider Fig.~\ref{fig:semicont_toy}. Due to the different sampling
rates of the trajectories, it is better to assign endpoints in one
trajectories to points in between endpoints of the other trajectory.

A commonly used model for trajectories is to connect two consecutive
sampled points by a line segment --- resulting in a polygonal
curve. Let $\overline{P}$ and $\overline{Q}$ denote these curves for
$P$ and $Q$ respectively. Ideally, we would like to define a monotone
assignment as a pair of functions $\alpha:P\rightarrow
\overline{Q}\cup \{\gsymbol\}$ and $\beta:Q\rightarrow
\overline{P}\cup\{\gsymbol\}$ and the goal will be to compute a monotone
assignment with the maximum score, as defined in
(\ref{eq:assignscore}). This is however, very hard to compute because
of the algebraic complexity of the assignment; the description of the
points on $\overline{P}$ and $\overline{Q}$ to which $Q$ and $P$ are
mapped can be quite large. Because of lack of space, we omit a
lower-bound construction from this paper.

We circumvent this problem by sampling points on each edge of
$\overline{P}$ and $\overline{Q}$ --- for each point $p_i\in P$ and
every edge $\overline{q_jq_{j+1}}$ of $Q$, we sample the closest point
of $p_i$, $\xi_{i,j} =  \arg$ $\min_{x\in \overline{q_jq_{j+1}}}
\|p_i-x\|$ on $\overline{q_jq_{j+1}}$. We do the same for every point
$q_k\in Q$ and every edge $\overline{p_ip_{i+1}}$ of
$\overline{P}$. Let $P^*$ and $Q^*$ be the resulting sequence of
points. We can then run the previous algorithm on $P^*$ and
$Q^*$. Since $|P^*|,|Q^*|\leq mn$, the total running time is
$O(m^2n^2)$. Besides the high running time, the sampling step is
somewhat ad-hoc.

To improve efficiency and avoid this large sampling step, we
propose a method of ``cheating'' to assign points in one trajectory to
points in between endpoints in the other trajectory if no endpoint is
a good match. At each step of the dynamic programming algorithm, we
examine prefixes $P_i$ and $Q_j$. At this step, we may choose to
assign $\alpha(p_i)=q_j$ or $\beta(q_j)=p_i$ depending on their
contribution to the overall score which is based on the distance
$\|p_i-q_j\|$. We modify this slightly by assigning
$\alpha(p_i)=\hat{q}_{j-1,j}$ where $\hat{q}_{j-1,j}$ is the closest
point on the segment preceding $q_j$. Similarly, we assign
$\beta(q_j)=\hat{p}_{i-1,i}$.

Although the assignment is now no longer restricted to be monotone, in
practice, we expect that in the similar portions of the trajectories,
such an assignment would be monotone and further, it would capture the
correspondences better than the discrete setting. Moreover, the
running time is now $O(mn)$ which matches the efficiency of the
discrete algorithms. Fig.~\ref{fig:semicont_toy} shows a comparison
of the two settings for a toy example. This example is very similar to
that of Fig.~\ref{fig:toy} but with different sample points. Note
that the semi-continuous model obtains a much more regular set of
correspondences of the positions along the trajectories than the
discrete setting.


\section{Parameter Selection}\label{sec:param-select}

Parameter selection is an important issue when choosing the scoring
function. During the course of the algorithm, when examining a pair of
points $p_i\in P$ and $q_j\in Q$, the difference in values
$1/(\c+\|p_i-q_j\|^2$ versus $\b$ dictate the choice of whether to
assign $\alpha(p_i)=q_j$ or $\beta(q_j)=p_i$ versus assigning one or
both as gap points.

We work with the hypothesis that all ``matched'' points have roughly
the same distance.  Let the threshold on the distance beyond which
points are dissimilar be $r$. We suggest the following choices of the
parameters:
\[
\c=\frac{r}{2},\quad \b=\frac{1}{c+r^2},\quad \a=-\l\b,
\]
where $l$ indicates a minimum gap length. If $l=0$, the algorithm
simply chooses points which are farther apart than $r$ to be gap
points. Otherwise, it only chooses to start gaps when at least $l$
points are farther than $r$ apart since, if not, $\a+\b\cdot|g|$ will
be negative.

The choice of $r$ is clear if we have semantic information about the
trajectories such as what type of entities generated them. For
example, if we have GPS trajectories from road networks and wish to
classify similar portions as those following the same road, then the
width of the road and the sampling rate or GPS accuracy would
determine the choice of $r$.

On the other hand, in many situations, the choice of $r$ is not
clear. For example, if we are given road trajectories but the roads
may consist of both highways as well as side streets or alleyways, the
widths may vary widely. In such cases, we wish to infer the value of
$r$ from the dataset. Unfortunately, this is a hard inference
problem. One could use a Bayesian method for this problem, but this is
beyond the scope of this paper.



Instead, we use a simple iterative algorithm which proceeds according
to the following steps:
\begin{itemize}
\item[(1)] Start with a rough guess of the upper bound $\hat{r}$ and
  compute an assignment with $r=\hat{r}$. 
\item[(2)] Discard a percentage of the larger distances in this assignment
  and compute the $\rms$ of the remaining distances.
\item[(3)] Choose a new threshold which is a small factor of the $\rms$,
  say $r=c_1\cdot \rms$.
\item[(4)] Repeat steps (2) and (3) until the assignments converge.
\end{itemize}

When we identify all similar portions and dissimilar portions, we do
not expect the $\rms$ to change significantly when we discard some of
the larger distances. Hence, the assignments should converge and the
algorithm should terminate.

We may set different criteria for termination, particularly if we have
additional information. For example, if we have a lower bound on the
threshold, we can terminate when we reach this value, since the
algorithm need not necessarily terminate at this point. If we know the
characteristics of the noise on the measurements, for example, if it
is Gaussian, we can fit a Gaussian to the remaining distances and
converge when it is a good enough fit.


\def\dtw{\textsc{DTW}\xspace}
\def\dtwp{\textsc{DTW-Pruned}\xspace}
\def\seqalign{\textsc{Seq-Align}\xspace}
\def\ass{\textsc{Assignment}\xspace}

\section{Experiments}\label{sec:experiments}

We have conducted an experimental study on real datasets to evaluate
the effectiveness of our model and to compare with previous
approaches. We used the DTW and sequence-alignment based approaches to
provide a basis for comparison of our approach. We refer to these
algorithms as \dtw and \seqalign and to our algorithm from
Sec.~\ref{sec:algos} as \ass throughout this section.

Since \dtw tries to find correspondences for every point on a
trajectory and consequently, does not aim to distinguish between
similar and deviating portions, we applied a simple heuristic to make
the comparison with \ass and \seqalign fairer: after computing the
\dtw correspondences, we pruned all correspondences where the distance
between the two constituent points was greater than a
threshold. Recall that the parameter selection for our scoring
function (cf. Sec.~\ref{sec:param-select}) uses a similar distance
threshold $r$. We call this modified \dtw heuristic \dtwp.

We present two types of results in this section: (i) an analysis of
the nuances of \ass, \dtw and \seqalign by examining individual pairs
of trajectories and their resulting correspondences, and (ii)
characteristics of the data based on the correspondences computed over
all pairs of trajectories in the datasets. The goal of our experiments
was not only to present a qualitative study but also to observe the
characteristics of the data and how they impact the model parameters.

\subsection{Description of Datasets.}  
\begin{figure}[h]
  \centering
  \begin{tabular}{ccc}
    \includegraphics[width=0.3\textwidth]{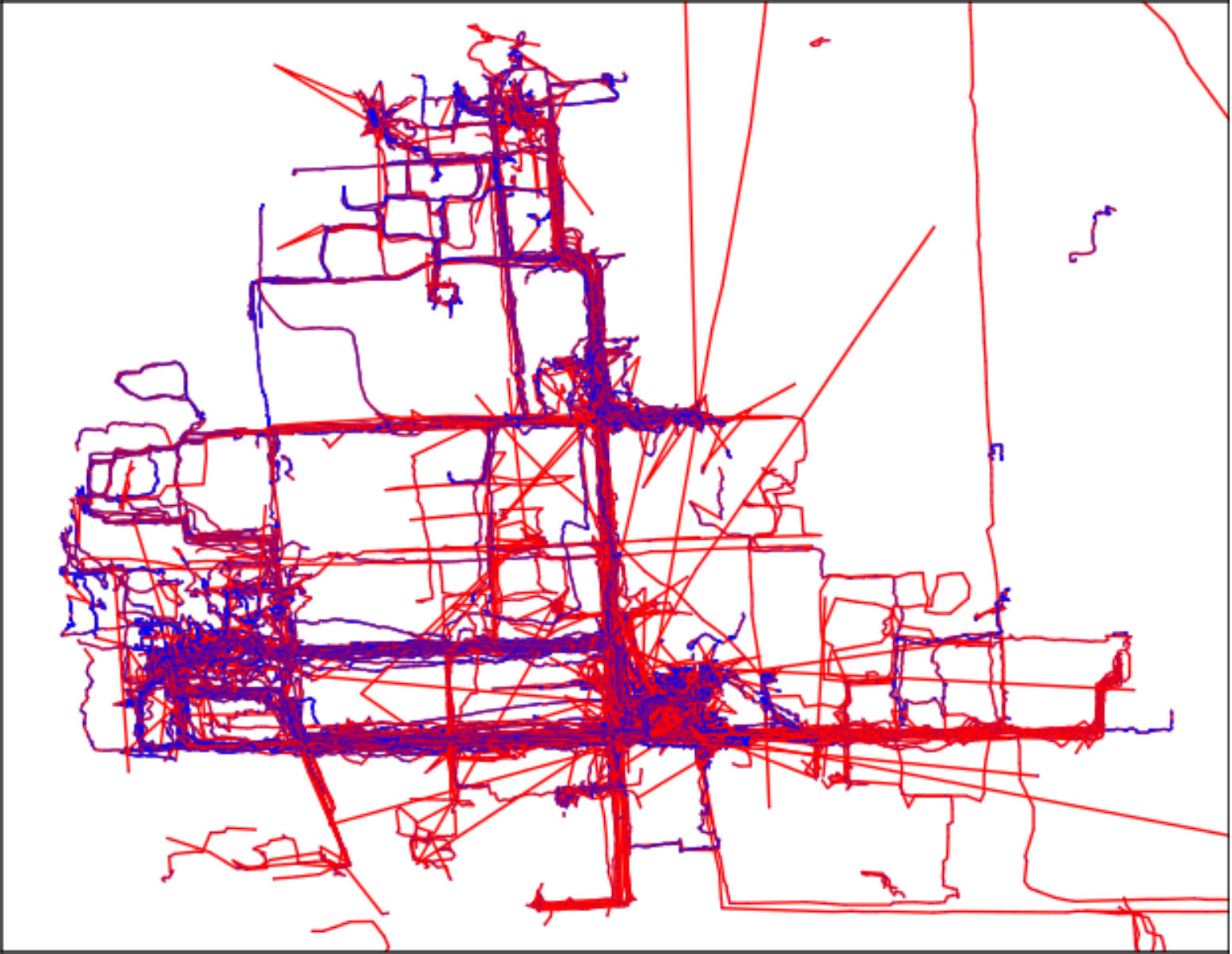}
    &
    \includegraphics[width=0.3\textwidth]{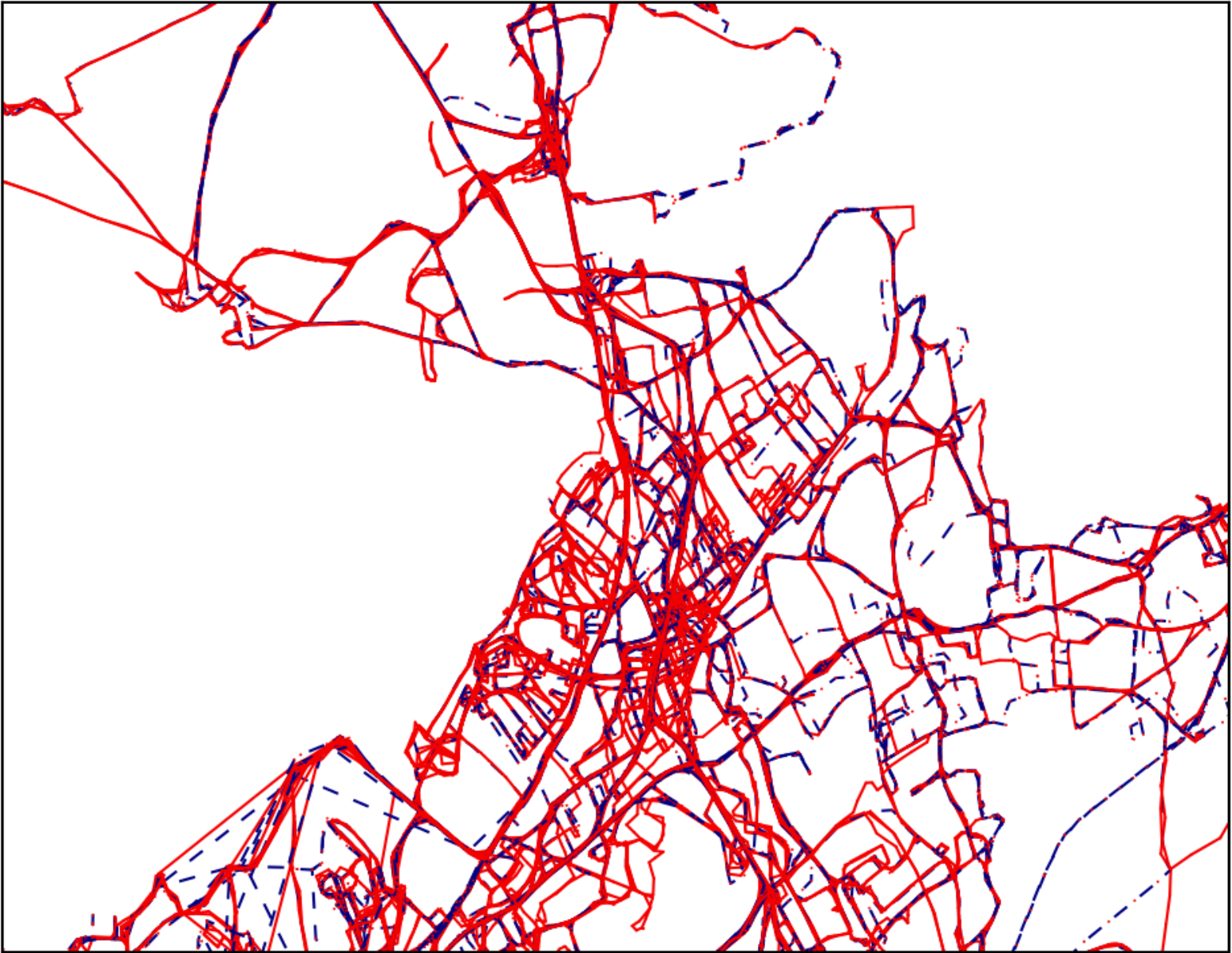}
    &
    \includegraphics[width=0.3\textwidth]{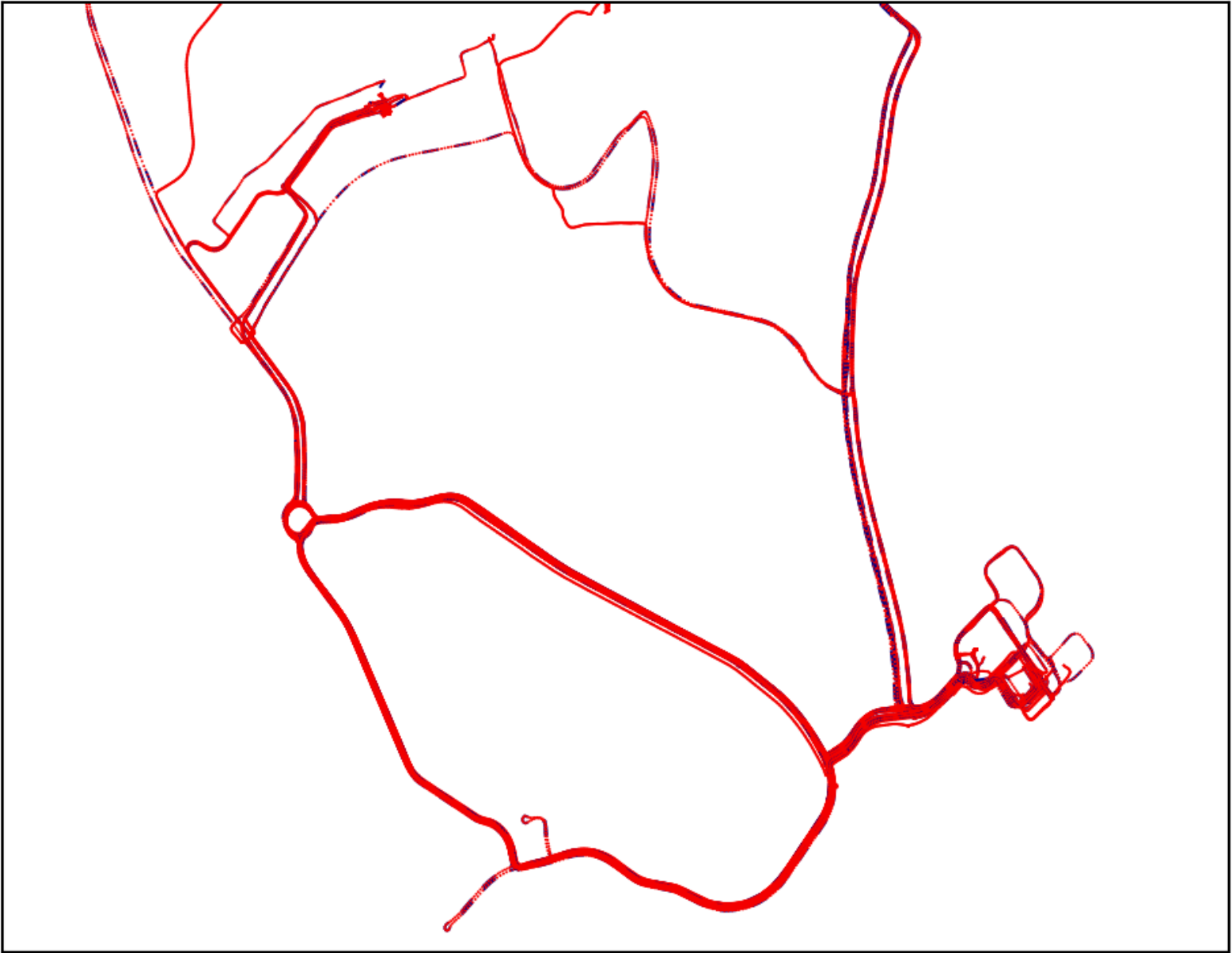}
    \\
    (a) & (b) & (c)
  \end{tabular}
  \caption{Datasets: (a) GeoLife datasets of trajectories with
    transportation modes in \{biking, walking, running\}
    respectively. (b) Buses dataset, (c) trajectories from road
    cycling exercises recorded with a fitness GPS device. Sample
    points are blue while the edges between two points are red.}
  \label{fig:datasets}
\end{figure}

We have used three real
datasets in our experiments: (i) the Geolife project by Microsoft
Research
Asia~\cite{DBLP:journals/debu/ZhengXM10,Zheng:2009:MIL:1526709.1526816,Zheng:2008:UMB:1409635.1409677}
consisting of 17,621 trajectories of 182 users in China, (ii) 145
trajectories of school buses in Athens,
Greece~\cite{springerlink:10.1007/11535331_19}, and (iii) 38
trajectories from road cycling exercises captured by a fitness GPS
device at a constant one second sample rate. Many of the trajectories
in the GeoLife dataset are labeled with the mode of transportation
used from the set \{biking, walking, running, bus, car, taxi, train,
subway, airplane\}. We extracted the trajectories with labels in
\{biking, walking, running\} and extracted a sample of trajectories
for our experiments. Our subset consists of a total of 883
trajectories. Fig.~\ref{fig:datasets}(a) shows the portion of this
dataset where trajectories have the highest concentration which is,
not surprisingly, in Beijing. Fig.~\ref{fig:datasets}(b) and
\ref{fig:datasets}(c) shows the other two datasets.

Although all three datasets are GPS trajectories, they differ
significantly from each other due to the different modes of
transportation and sensor equipment as well as due to road network
characteristics
In the buses dataset, the sampling rate is fairly infrequent and the
data also contains significant measurement noise. On the other hand,
although the sampling rate is more frequent for the biking, walking
and running trajectories in the GeoLife dataset, here too, it contains
significant measurement noise. In addition, there seem to be a large
number of partially observed trajectories (long stretches with no
samples). The dataset of cycling exercise routes collected is highly
accurate and extremely densely sampled making it a good representation
of ideal conditions.

\subsection{Results on Pairs of Trajectories}\label{sec:pairresults}

To show the effectiveness of our algorithms, we present results on a
single pair of trajectories from each of the datasets. We chose the
pairs in such a manner that they exhibit significant similar portions
as well as dissimilar portions and have sufficiently dissimilar
sampling rates in the case of the pairs from the Bus and GeoLife
datasets (the cycling exercise trajectories all have a uniform
sampling rate). Consequently, the results exhibited on these pairs are
representative of the results on other trajectories as well. In all
cases, we perturbed one of the trajectories slightly in the figures to
present the results more clearly.

\begin{figure}[hp]

\begin{minipage}{\textwidth}
  \centering
  \begin{tabular}{cc}
    \includegraphics[width=0.45\textwidth]{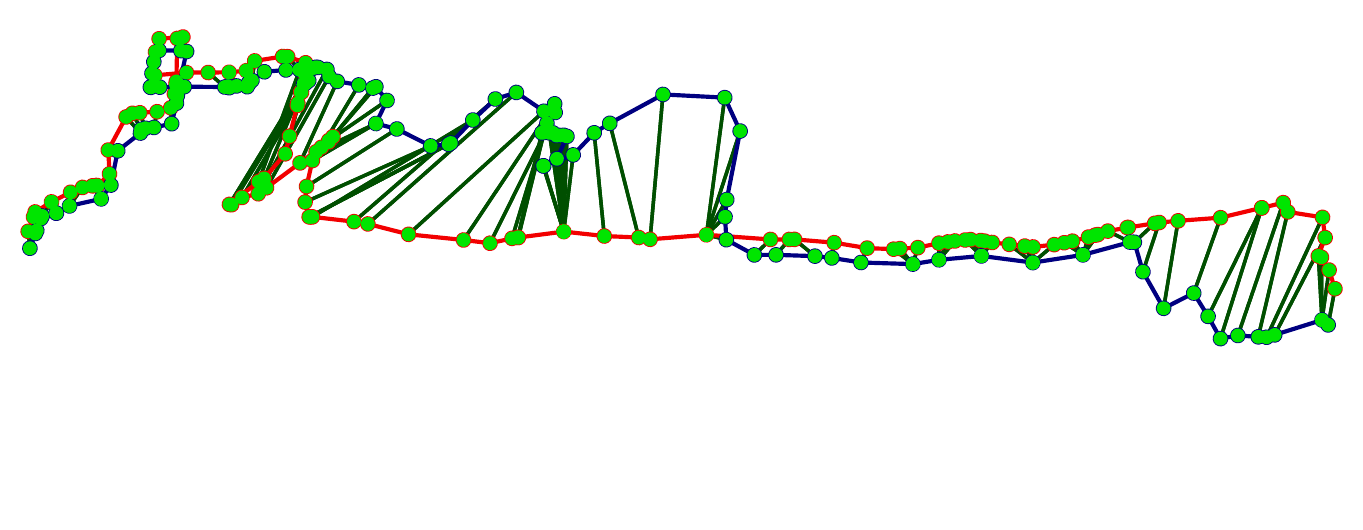} &
    \includegraphics[width=0.45\textwidth]{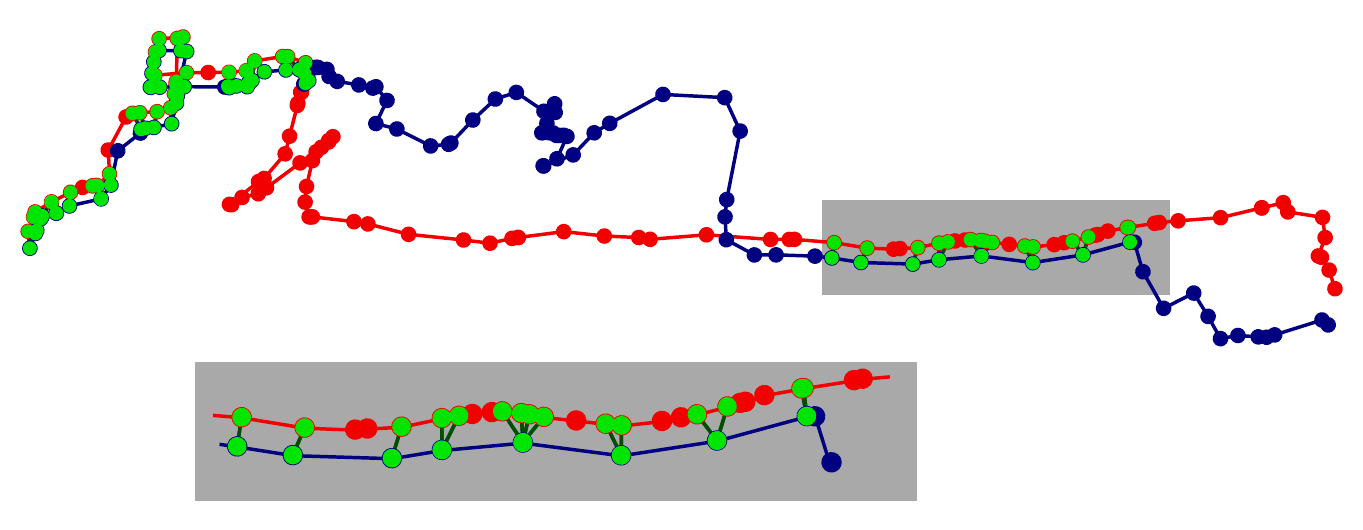} \\
    \scriptsize{(a) \dtw} & \scriptsize{(b) \dtwp} \\
    \includegraphics[width=0.45\textwidth]{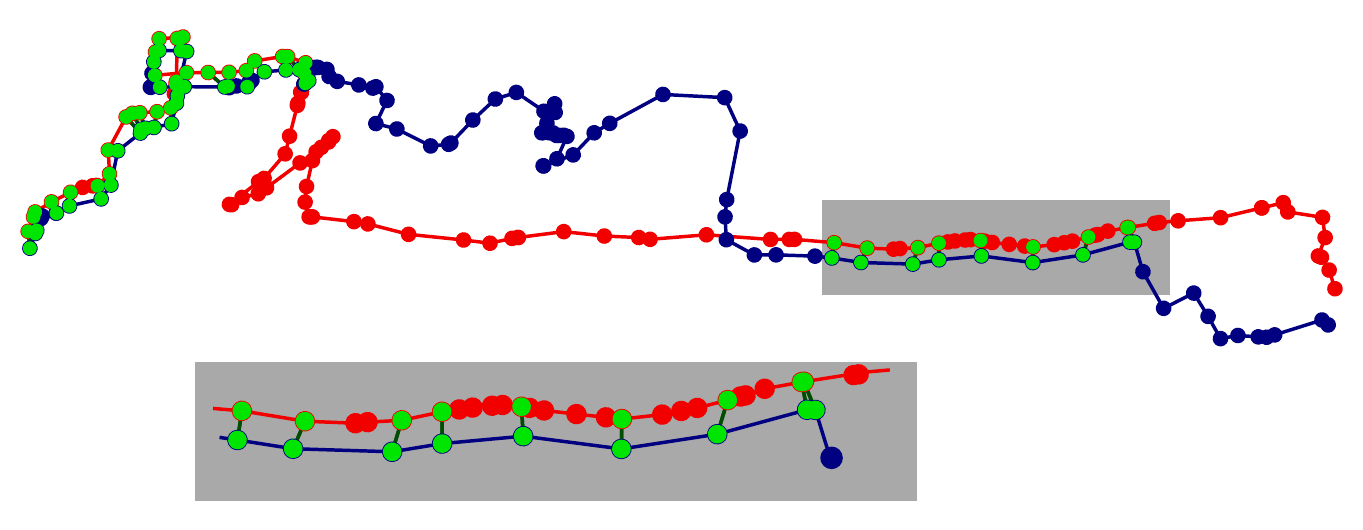} &
    \includegraphics[width=0.45\textwidth]{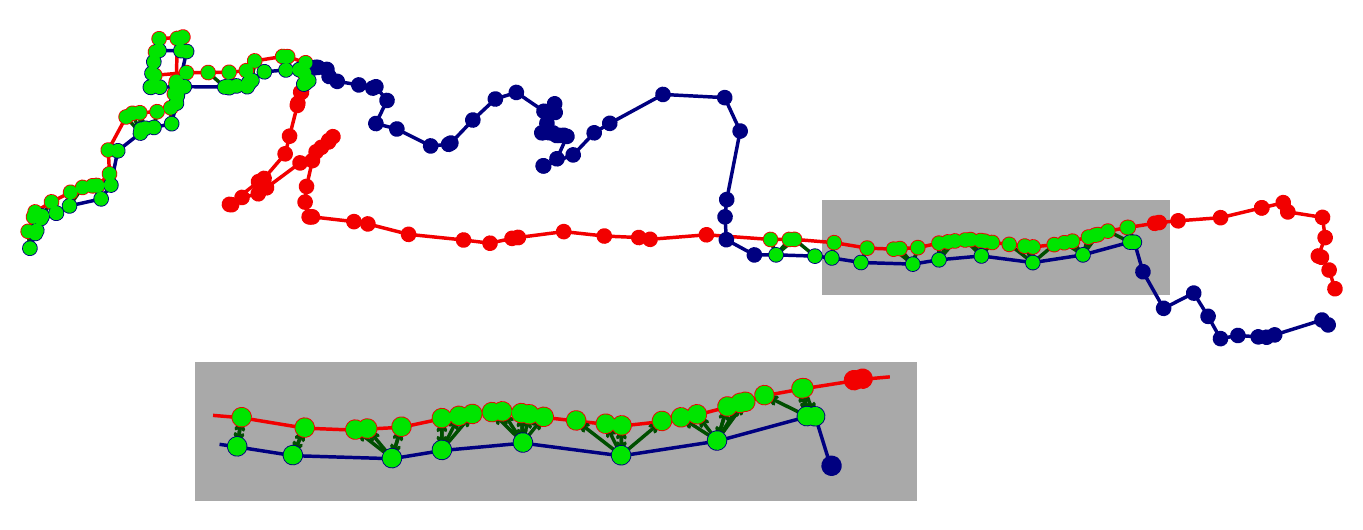} \\
    \scriptsize{(c) \seqalign} & \scriptsize{(d) \ass}
  \end{tabular}
  \caption{Results on a trajectory pair from the buses dataset.}
  \label{fig:buses_alignments}
\end{minipage}
\vspace{0.1in}

\begin{minipage}{\textwidth}
  \centering
  \begin{tabular}{cc}
    \includegraphics[width=0.45\textwidth]{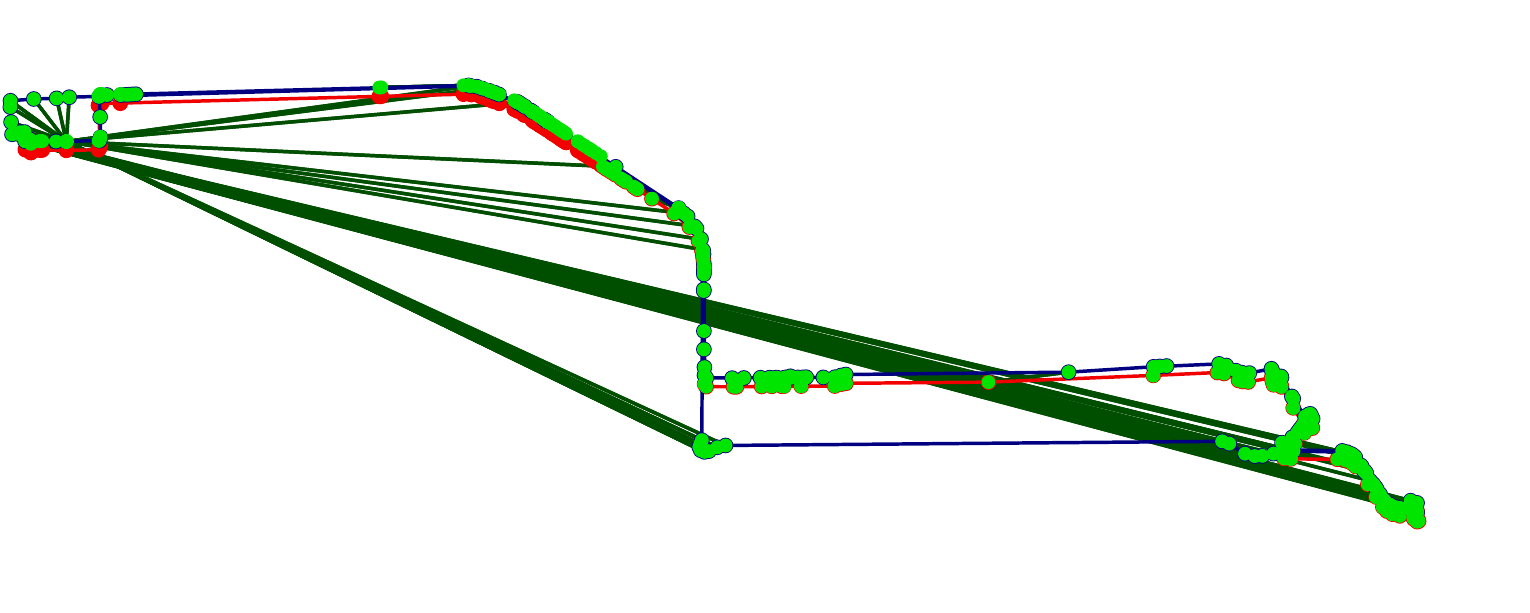} &
    \includegraphics[width=0.45\textwidth]{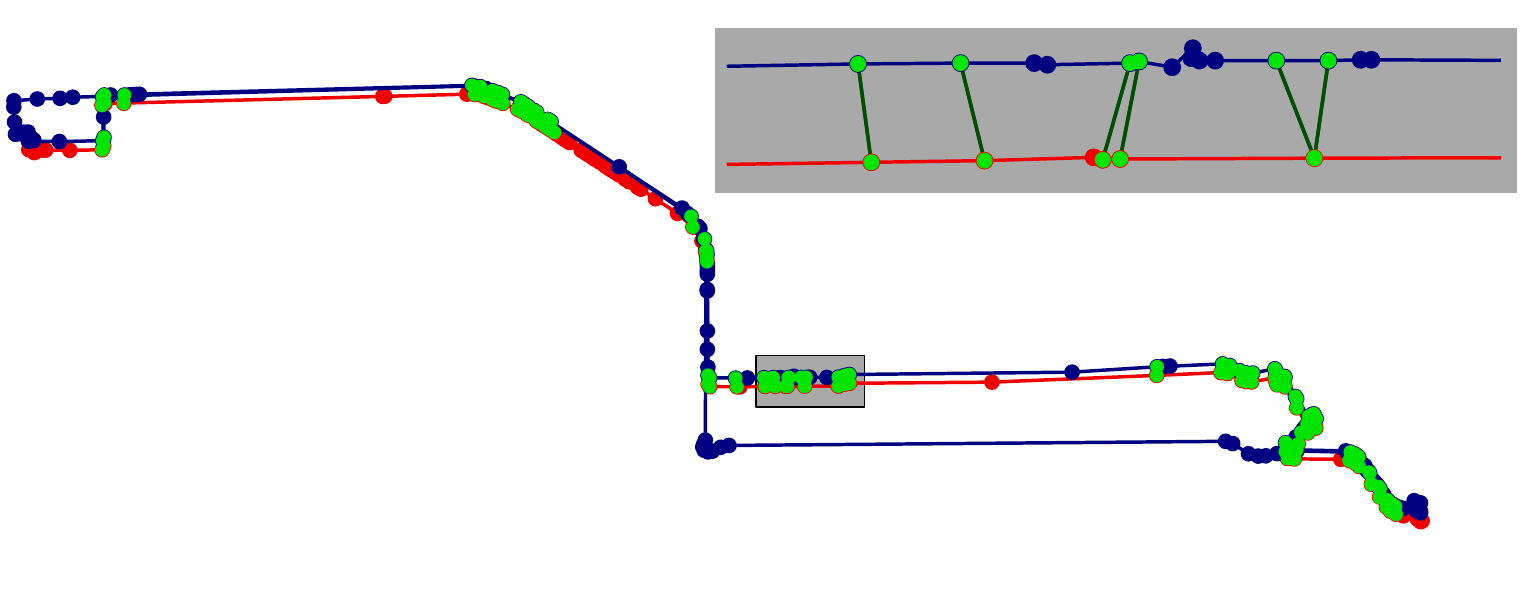} \\
    \scriptsize{(a) \dtw} & \scriptsize{(b) \dtwp} \\
    \includegraphics[width=0.45\textwidth]{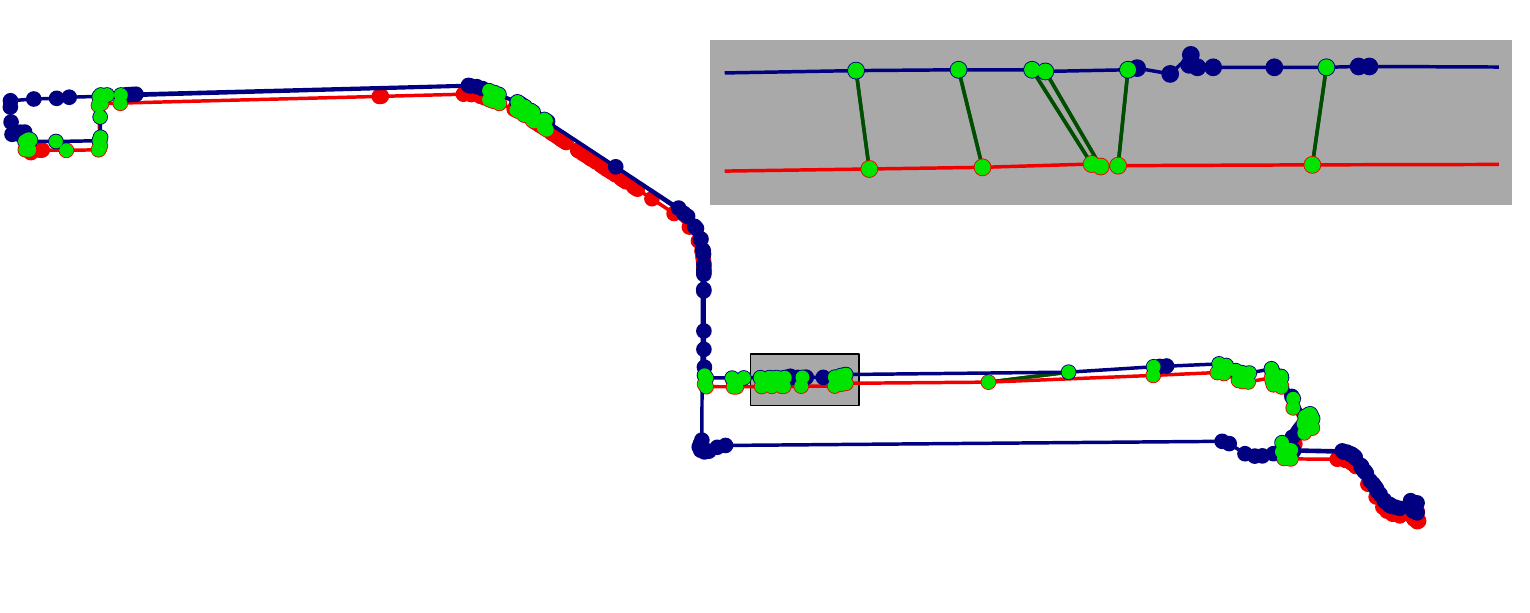} &
    \includegraphics[width=0.45\textwidth]{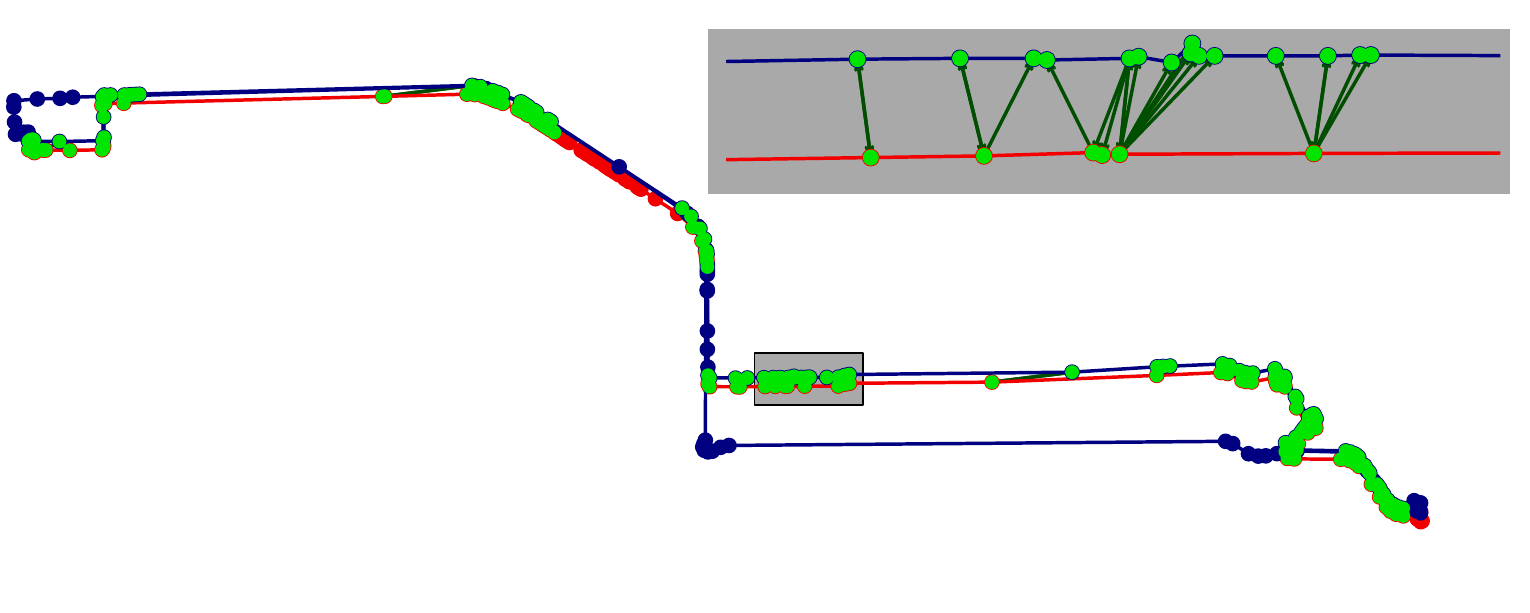} \\
    \scriptsize{(c) \seqalign} & \scriptsize{(d) \ass}
  \end{tabular}
  \caption{Results on a trajectory pair from the GeoLife dataset.}
  \label{fig:geo_alignments}
\end{minipage}    
\vspace{0.1in}

\begin{minipage}{\textwidth}
  \centering
  \begin{tabular}{cc}
    \includegraphics[width=0.3\textwidth]{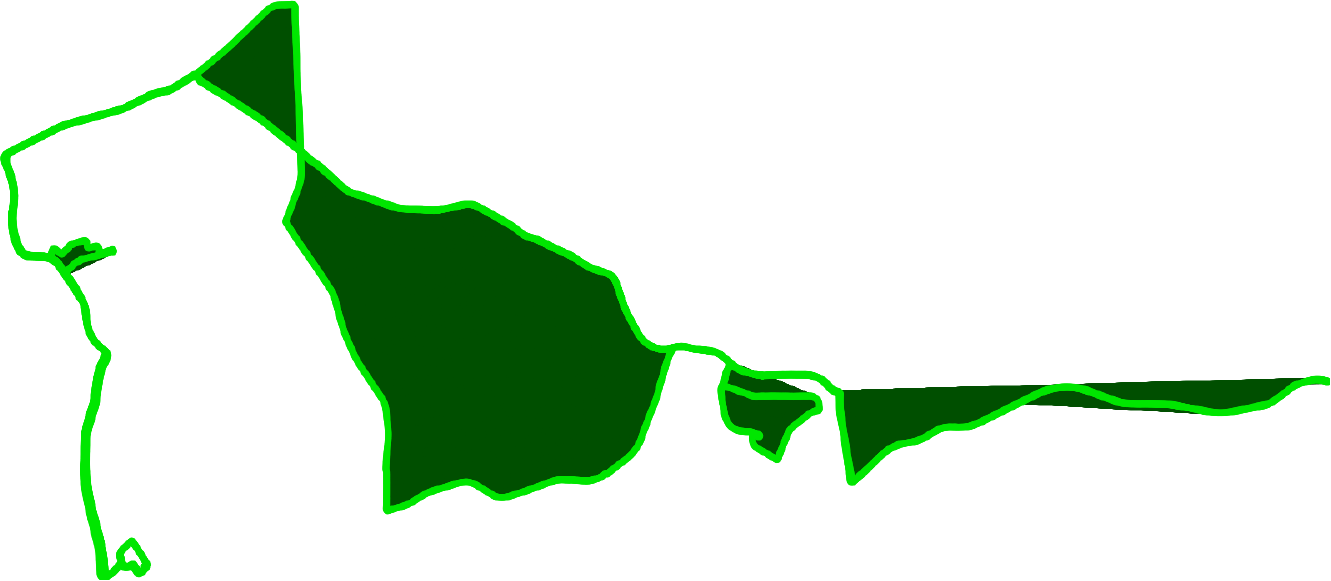} &
    \includegraphics[width=0.3\textwidth]{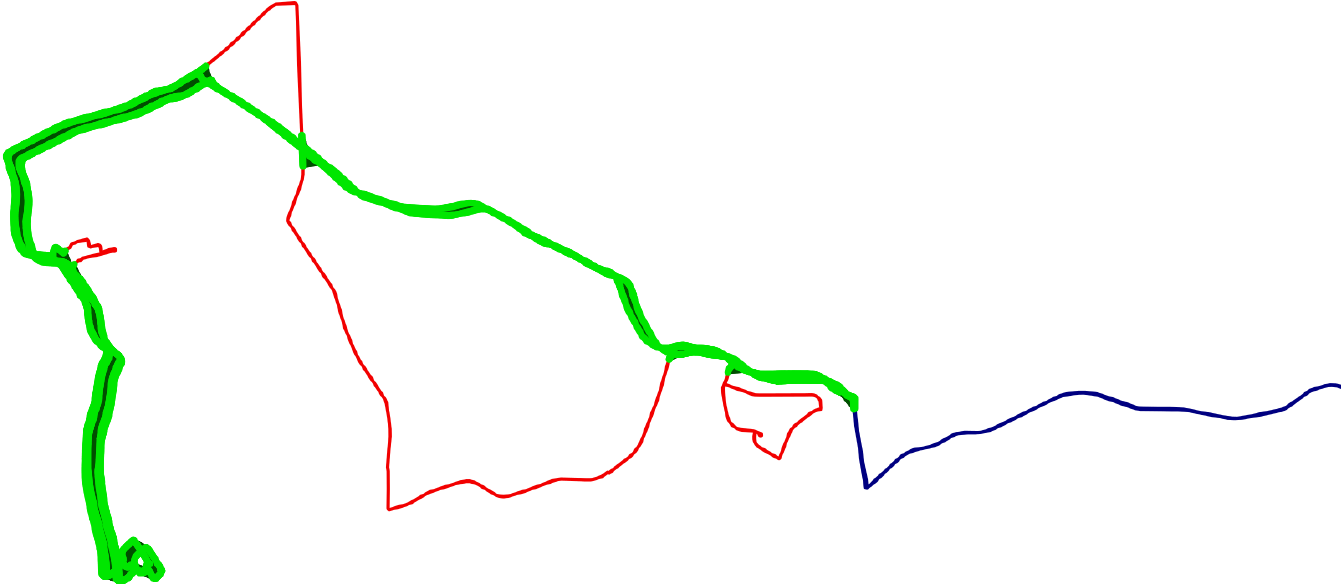} \\
    \scriptsize{(a) \dtw} & \scriptsize{(b) \dtwp} \\
    \includegraphics[width=0.3\textwidth]{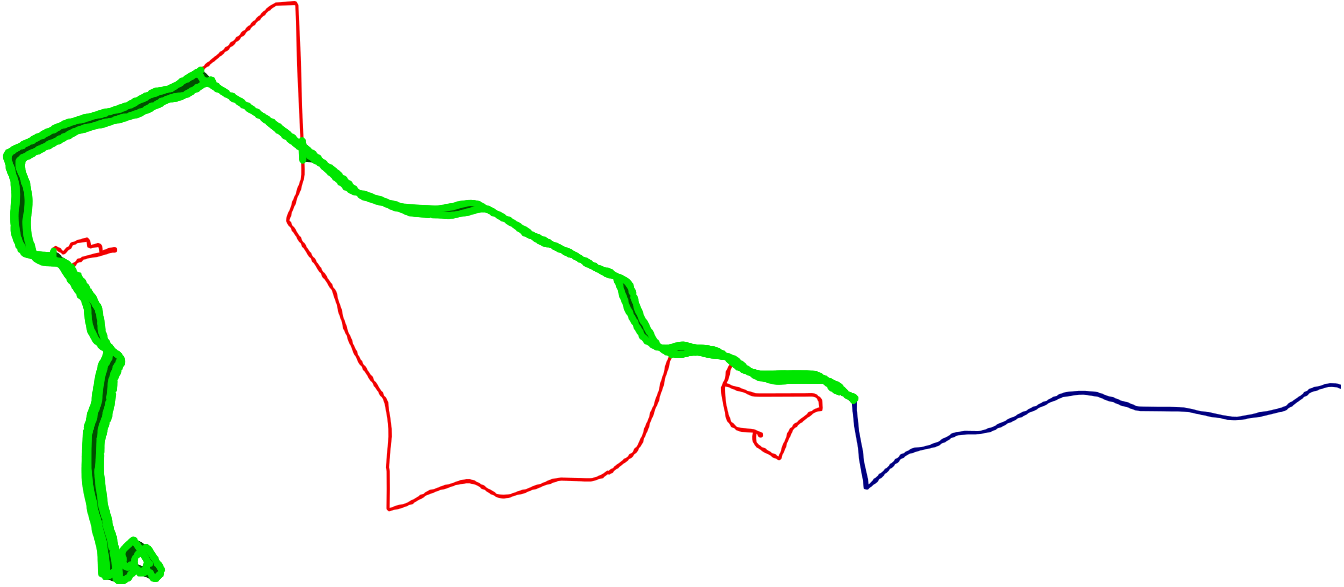} &
    \includegraphics[width=0.3\textwidth]{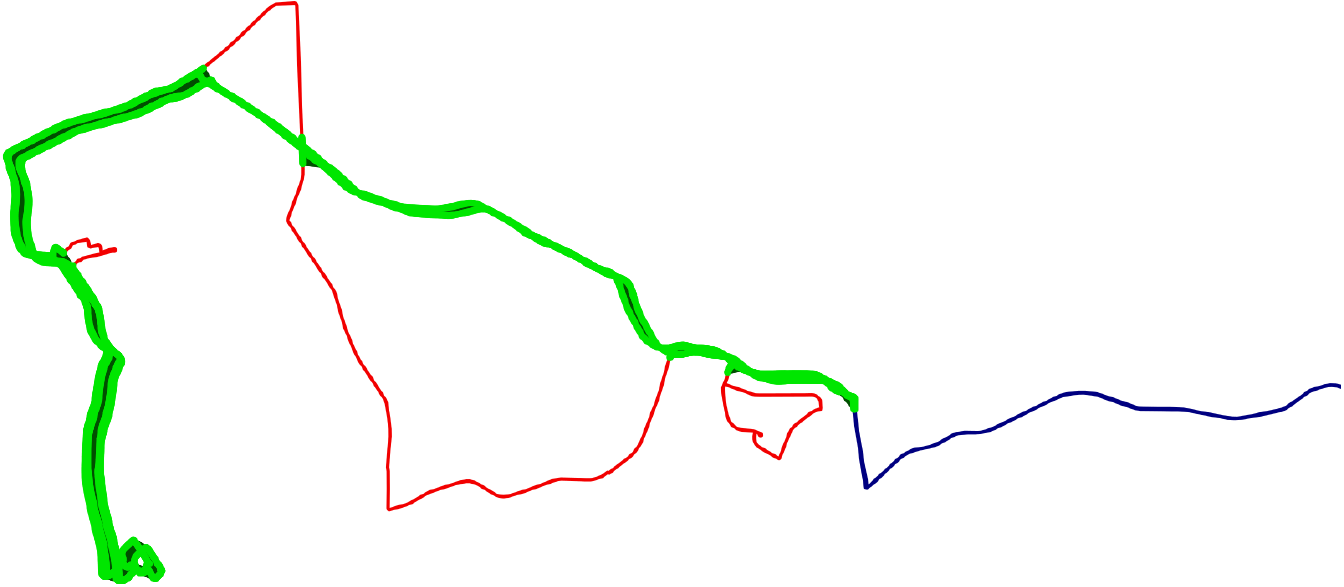} \\
    \scriptsize{(c) \seqalign} & \scriptsize{(d) \ass}
  \end{tabular}
  \caption{Results on a trajectory pair from the cycling dataset.}
  \label{fig:bike_alignments}
\end{minipage}    
\end{figure}

\mparagraph{Global Assignment.}  In each case, we compare the results
of \ass to the results of \dtw, \dtwp as well as \seqalign. We chose a
distance threshold $r=100$\,m for our parameter selection process for
\ass (see Sec.~\ref{sec:param-select}) as well as for \dtwp. The
minimum gap length was set at $4$ for both \seqalign and \ass.
Figures~\ref{fig:buses_alignments}, \ref{fig:geo_alignments} and
\ref{fig:bike_alignments} show the results of computing assignments on
the pair of trajectories chosen from each dataset. In the former two,
we have also shown in an inset in some cases: a zoomed in portion of
the trajectories where the issues with the different approaches are
clearer.

The correspondences computed by \dtw are clearly not meaningful in
differentiating deviating and similar portions (see
Figures~\ref{fig:buses_alignments}(a), \ref{fig:geo_alignments}(a) and
\ref{fig:bike_alignments}(a)). On the other hand, for the buses and
GeoLife data, \dtwp clearly does much better in identifying deviations
(see Figures~\ref{fig:buses_alignments}(b) and
\ref{fig:geo_alignments}(b)) but generates a number of smaller gaps
due to some correspondences which may be desired having larger
distance than the threshold possibly due to measurement
noise. \seqalign which does impose a minimum gap length to get around
this issue also has a similar behavior. Here, however, this is caused
by the restriction to one-to-one correspondences. Finally, \ass
captures the advantages of the other approaches by generating
correspondences for almost all points on the similar portions (see
Figure~\ref{fig:buses_alignments}(d) and
\ref{fig:geo_alignments}(d)). In this case, the imposition of a
minimum gap penalty generates correspondences which may be desired in
spite of having distance larger than the threshold due to the
surrounding correspondences. Due to the allowance of multiple incoming
edges for a point in \ass, the sampling rate issues are also handled
well. 
In the exercise cycling dataset (see Fig.~\ref{fig:bike_alignments}),
all approaches perform well due to the uniform sampling rate, velocity
and accuracy of sampling.

\begin{figure}[t]
  \centering
  \begin{minipage}{\textwidth}
    \centering
    \begin{tabular}{cc}
    \includegraphics[scale=0.25]{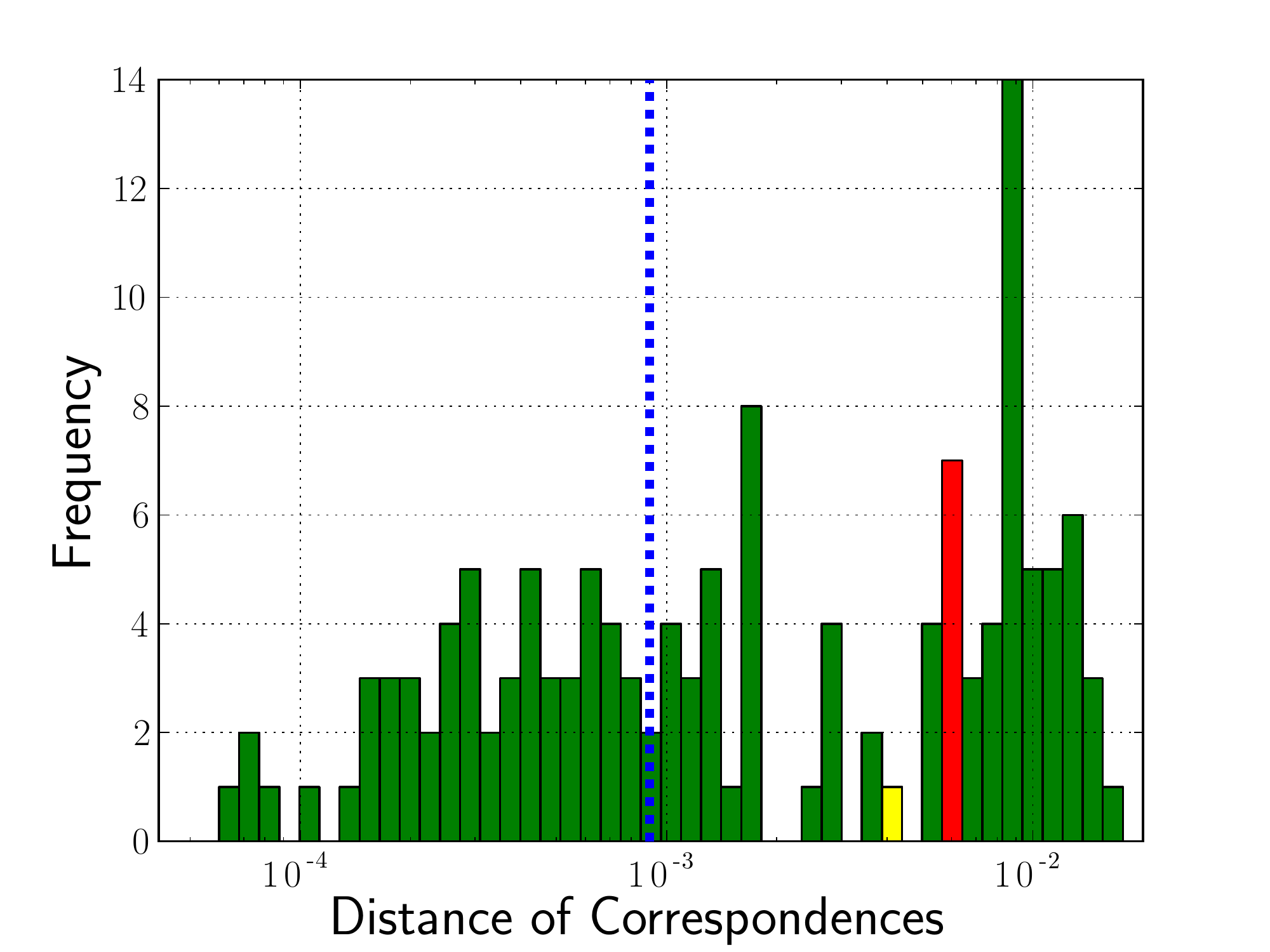} &
    \includegraphics[scale=0.25]{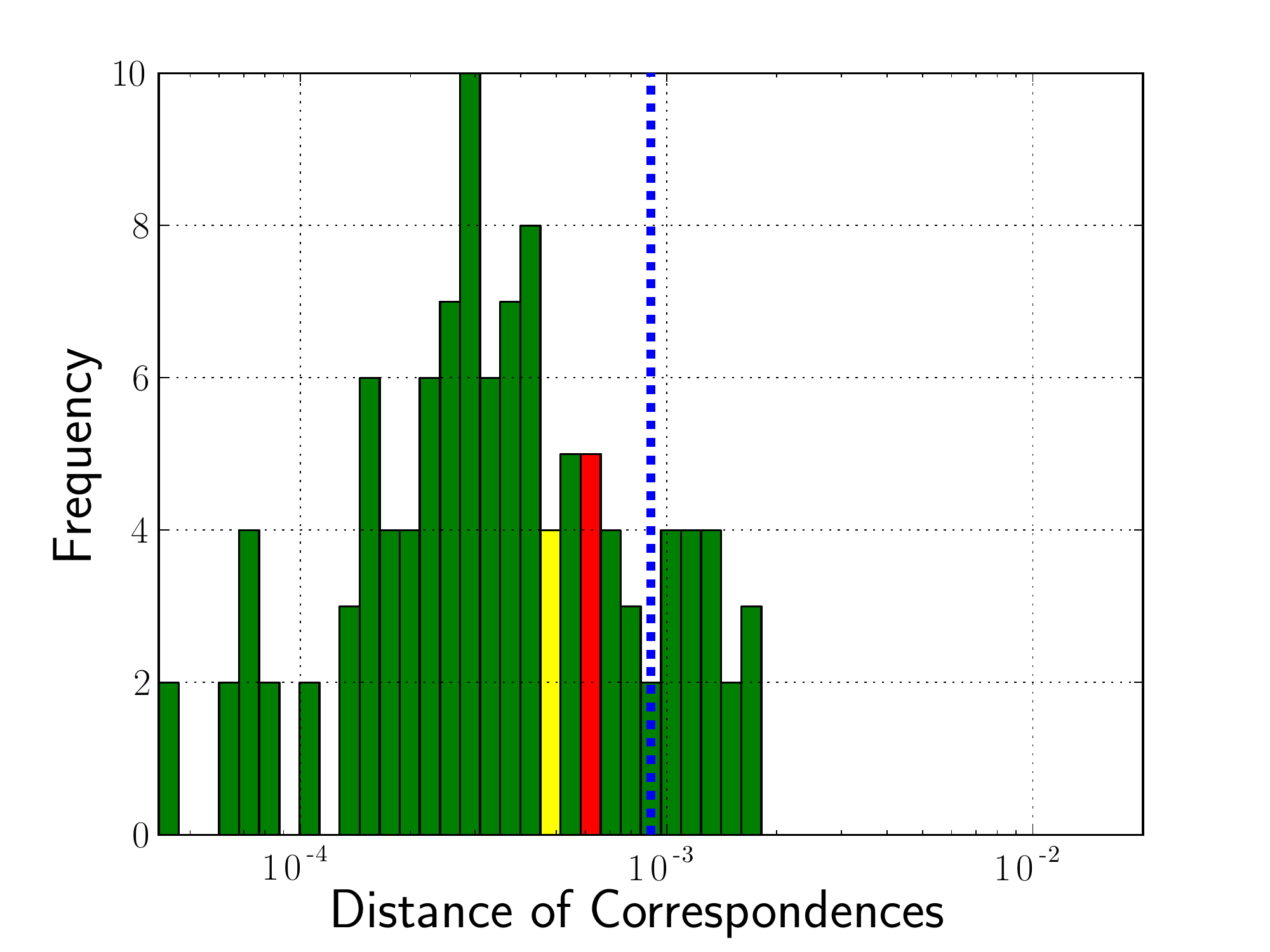} \\
    \scriptsize{(a) Buses dataset, \dtw} & \scriptsize{(b) Buses dataset, \ass}
  \end{tabular}
  \end{minipage}
  \smallskip
  \begin{minipage}{\textwidth}
    \centering
    \begin{tabular}{ccc}
    \includegraphics[scale=0.25]{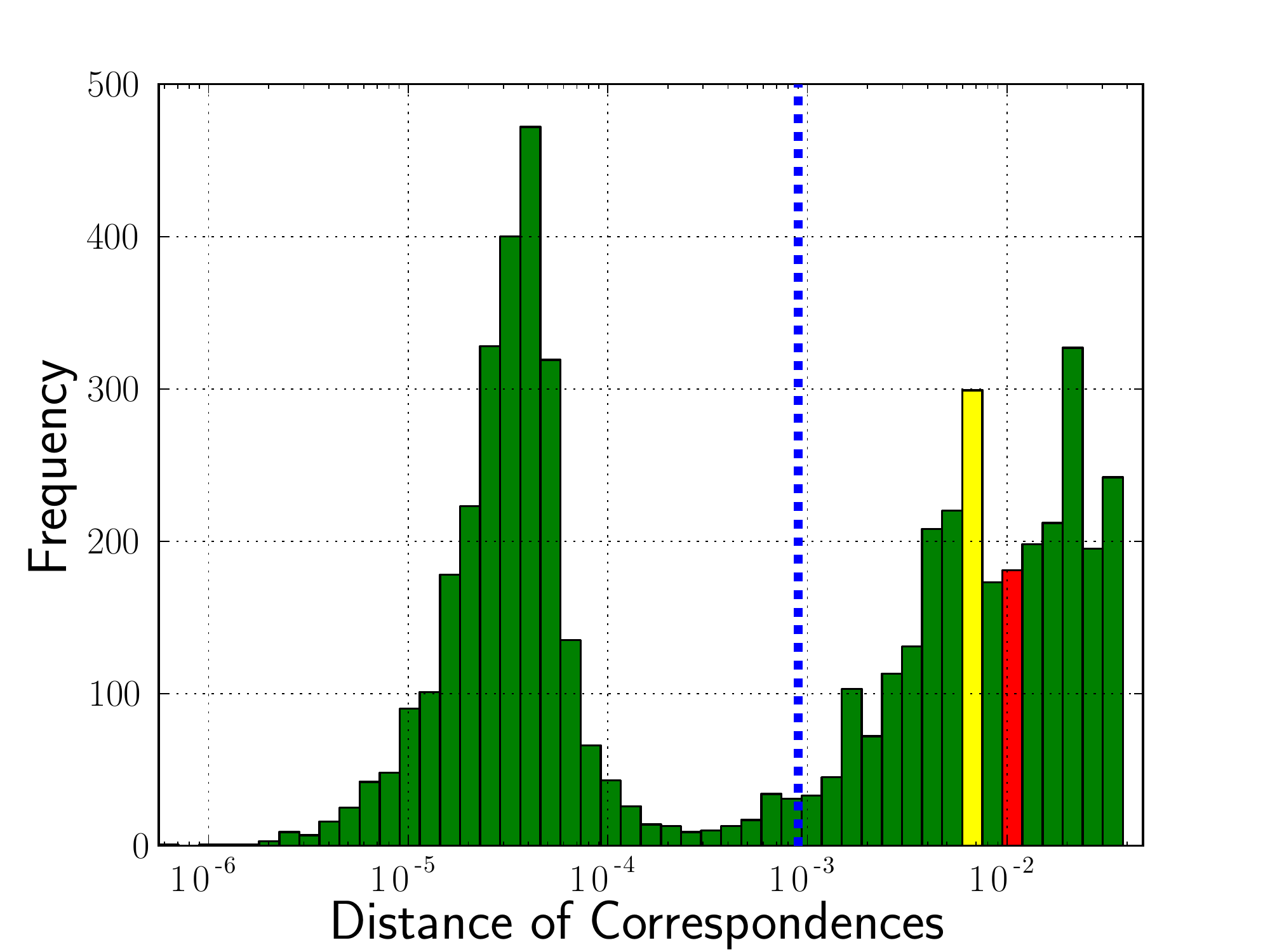} &
    \includegraphics[scale=0.25]{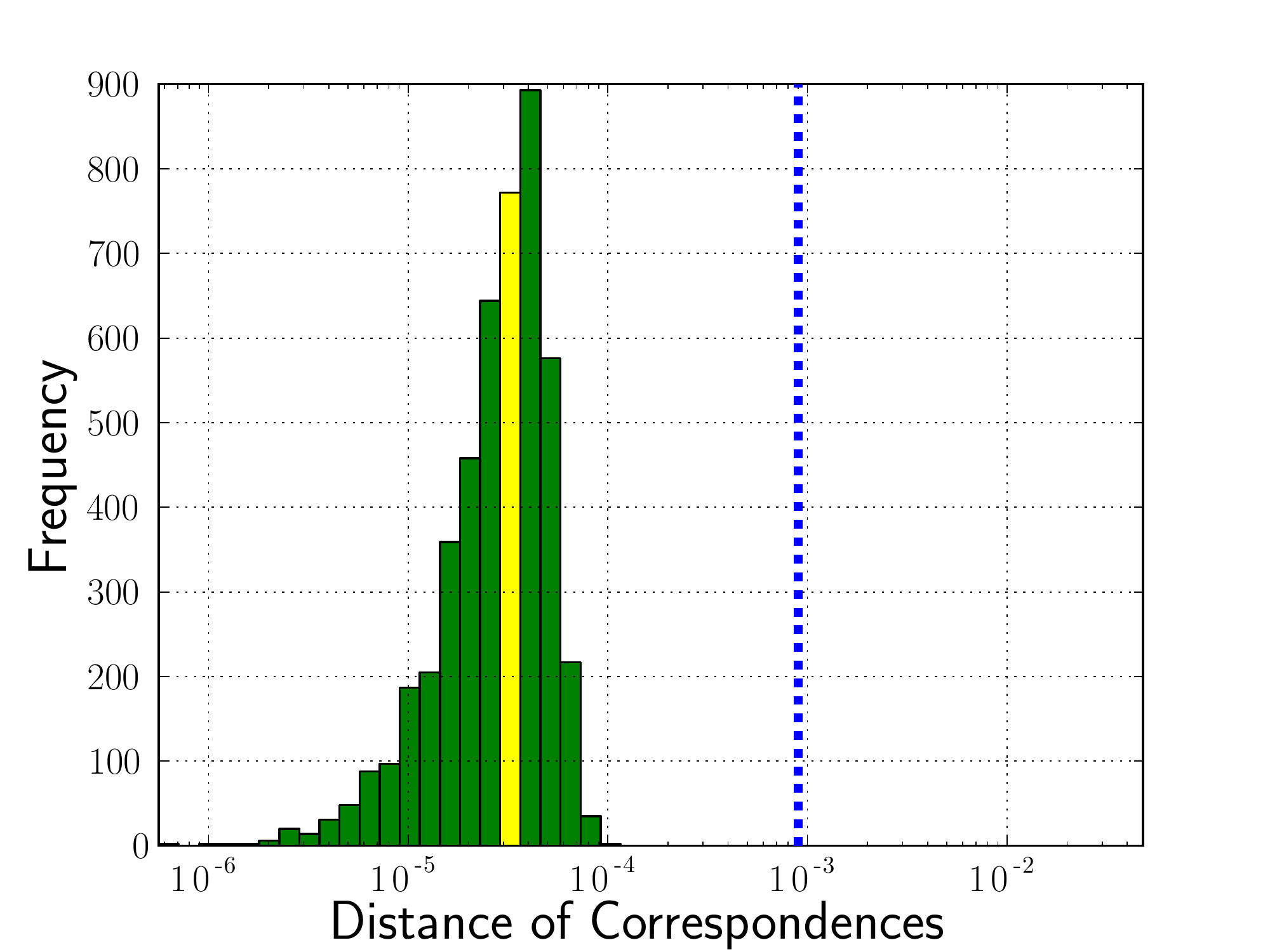} \\
    \\
    \scriptsize{(c) Cycling dataset}, \dtw & \scriptsize{(d) Cycling dataset, \ass}
  \end{tabular}
  \end{minipage}
  \caption{These histograms summarize the distance between
    corresponding pairs in the results from
    Figures~\ref{fig:buses_alignments} and
    \ref{fig:bike_alignments}. (a) and (b) correspond to
    Figures~\ref{fig:buses_alignments}(a) and
    \ref{fig:buses_alignments}(d) respectively while (c) and (d)
    correspond to Figures~\ref{fig:bike_alignments}(a) and
    \ref{fig:bike_alignments}(d) respectively. The bins containing the
    mean and $\rms$ values are highlighted in yellow and red
    respectively with no red bin indicating that the $\rms$ and mean
    lie in the same bin. The vertical blue line denotes the threshold
    distance.}
\label{fig:comparehisto}
\end{figure}

\thomas{This could be tightened by replacing ``corresponding pairs''
  by a concrete assignment/alignment $E$, but may not be worth it}

We also present histograms of the distances of all corresponding pairs
for \dtw and \ass for the buses and cycling datasets in
Fig.~\ref{fig:comparehisto}. The $x$-axis is set at a log-scale to
better visualize the histograms. The bins containing the mean and
$\rms$ values are highlighted in yellow and red respectively with no
red bin indicating that the $\rms$ and mean lie in the same
bin. Comparing Figures~\ref{fig:comparehisto}(a) and
\ref{fig:comparehisto}(b), we notice that the assignment retains some
edges whose distance is larger than the threshold $r$ (shown as a blue
dashed line). This is due to the minimum gap length constraint which
is necessary due to noise in the data. As mentioned above, including
correspondences nearby affect the chance of including a correspondence
in the result which is more meaningful than simply pruning based on
distance as in \dtwp.

\begin{figure}[ht]
\centering
\begin{minipage}{0.48\textwidth}
\centering
    \begin{tabular}{cc}
    \includegraphics[width=0.4\textwidth]{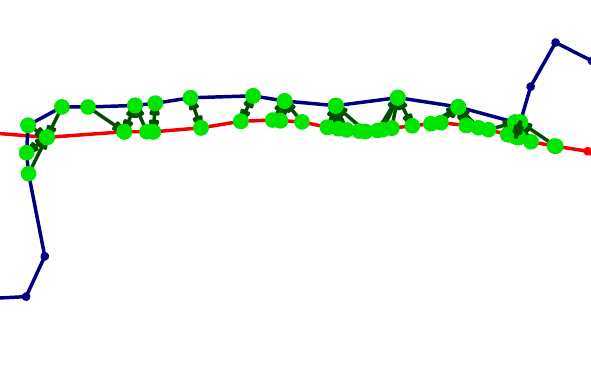} &
    \includegraphics[width=0.4\textwidth]{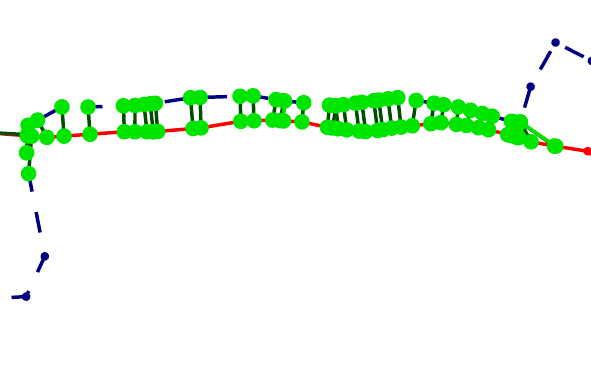} \\
    (a) Discrete & (b) Semi-continuous
  \end{tabular}
\caption{Comparison of discrete and semi-continuous assignments.}
\label{fig:sc}
\end{minipage}
\hspace{0.1in}
\begin{minipage}{0.48\textwidth}
\centering
\begin{tabular}{cc}
\includegraphics[width=0.5\textwidth]{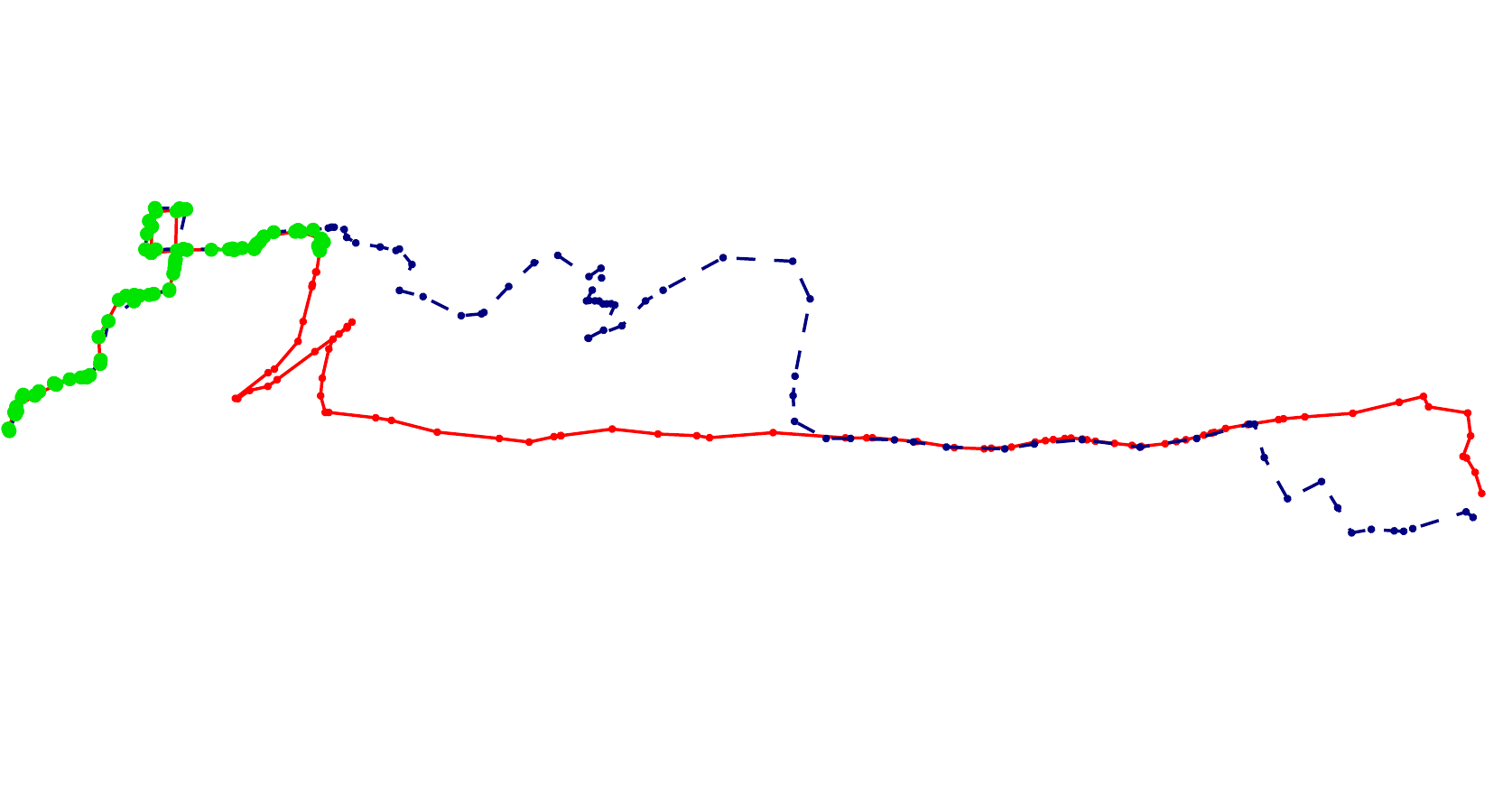} &
\includegraphics[width=0.5\textwidth]{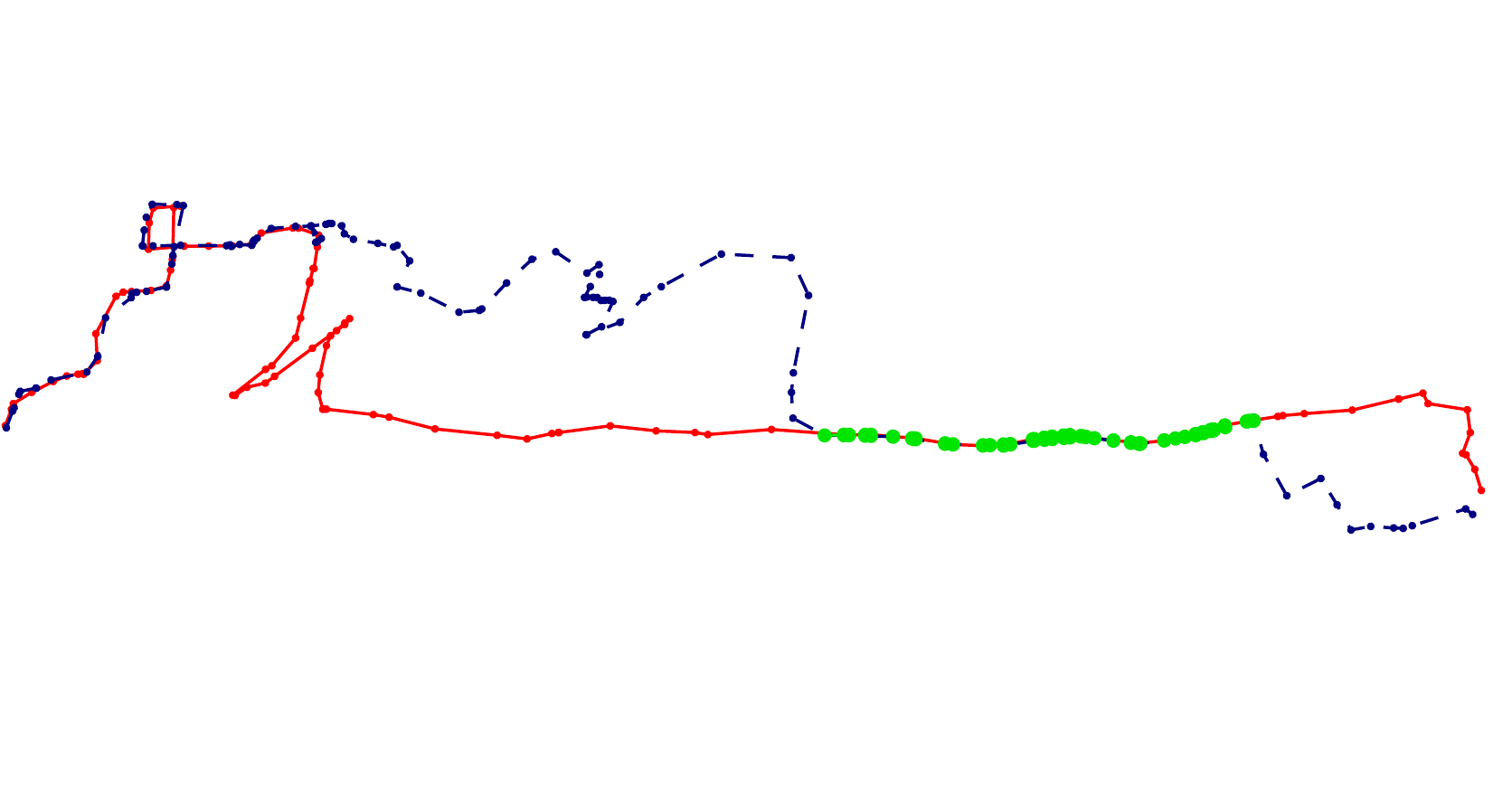}
\\
(a) Discrete & (b) Semi-continuous
\end{tabular}
\caption{Local assignments.}
\label{fig:local}
\end{minipage}
\end{figure}

\mparagraph{Local and Semi-Continuous Assignment.}\label{sec:extension}
We present qualitative results for local and semi-continuous
assignments. Fig. \ref{fig:sc} compares the discrete and
semi-continuous assignments for the trajectories chosen from the bus
dataset. For the semi-continuous case, we have used the method of
choosing the closest point during the course of our algorithm
described in Section~\ref{sec:semi-cont} and not the up-sampling
approach. Note the regularity of the segments indicating the
correspondences (shown in green) as compared to the discrete
setting. This is also a good indication that the variance of assigned
distances is much smaller and reflects the characteristics of the
routes taken by the trajectories.

Fig.~\ref{fig:local} shows the best of the individual local
assignments computed for the pair of trajectories chosen from the bus
dataset under both the discrete and semi-continuous setting. We chose
the parameter $\tau$ used to decide a lower bound on the scores of the
local assignment (see Sec.~\ref{sec:local}) to be $1.5\b$ and $2\b$
respectively for the discrete and semi-continuous cases. On the other
hand, the local assignment computed in the discrete setting captures a
shorter portion where the sampling is more uniform.

\begin{figure}[h]
\centering
\includegraphics[width=0.4\textwidth]{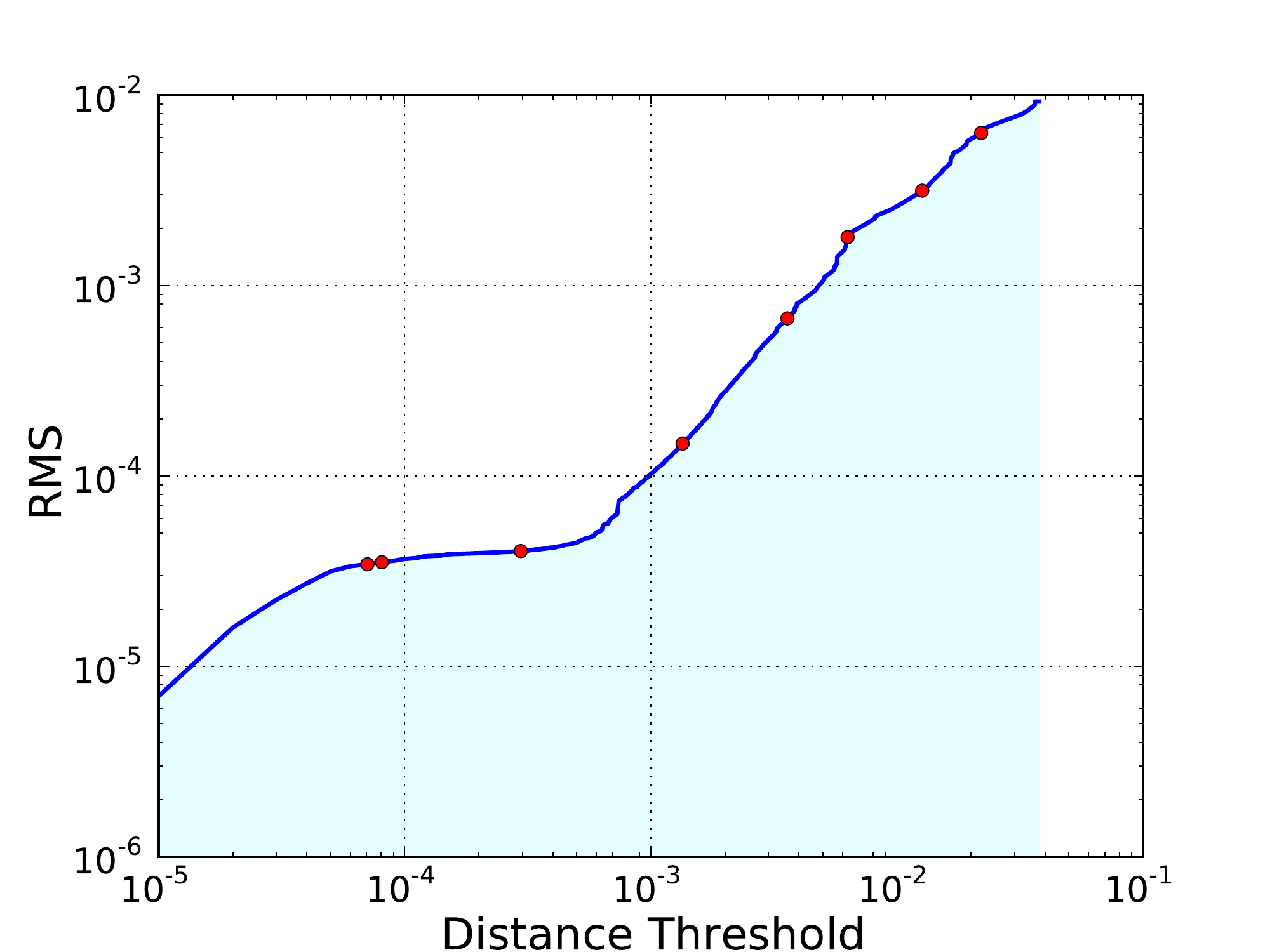} \\
\caption{Illustration of the root-mean-square of distances between the
  pairs in the optimal assignment as a function of the distance
  threshold used to define the parameters of $\score$
  (cf. Section~\ref{sec:param-select}). The red dots on the curve
  corresponds to the $\rms$ values produced during the iterative
  parameter selection algorithm.}
\label{fig:dvsr}
\end{figure}

\begin{figure}[ht]
    \centering
    \begin{tabular}{ccc}
    \includegraphics[width=0.3\textwidth]{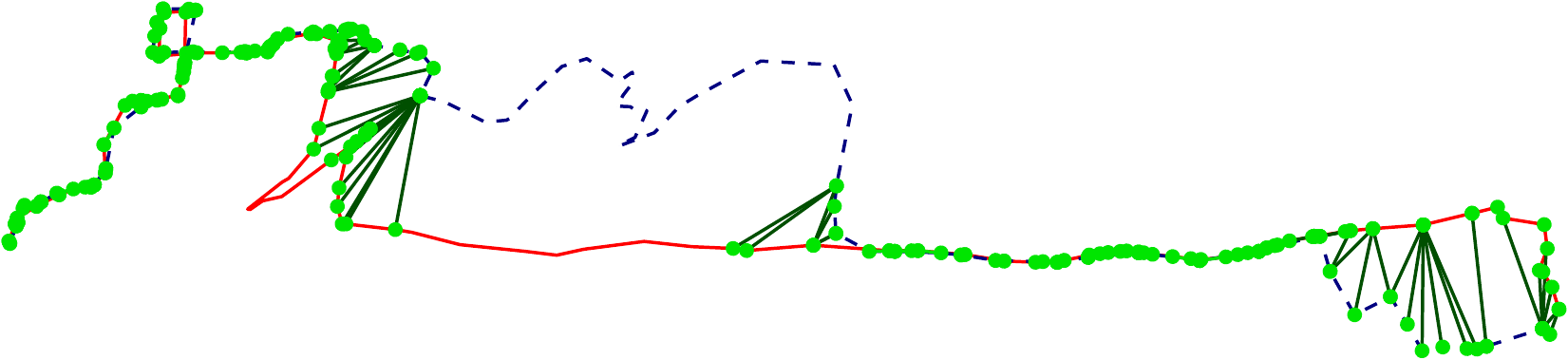} &
    \includegraphics[width=0.3\textwidth]{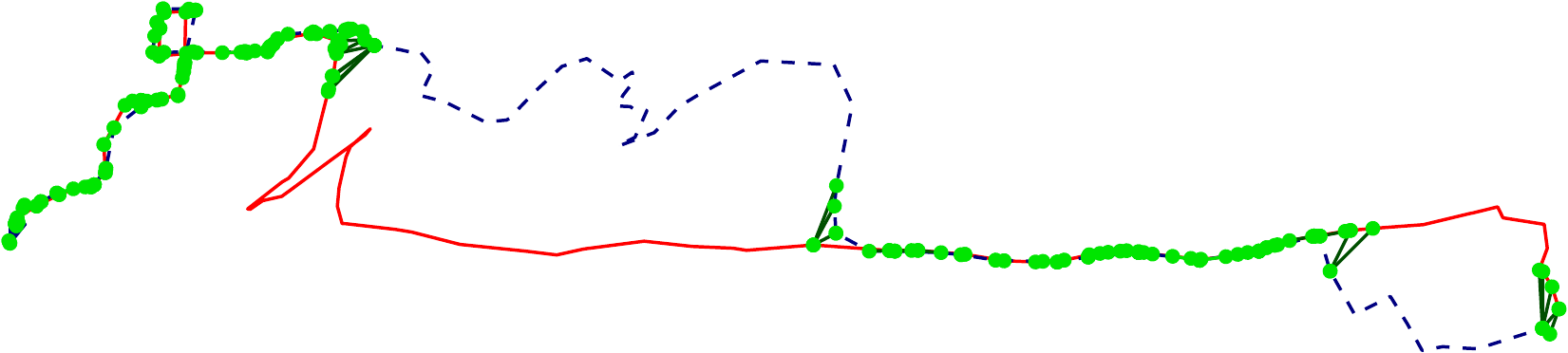} &
    \includegraphics[width=0.3\textwidth]{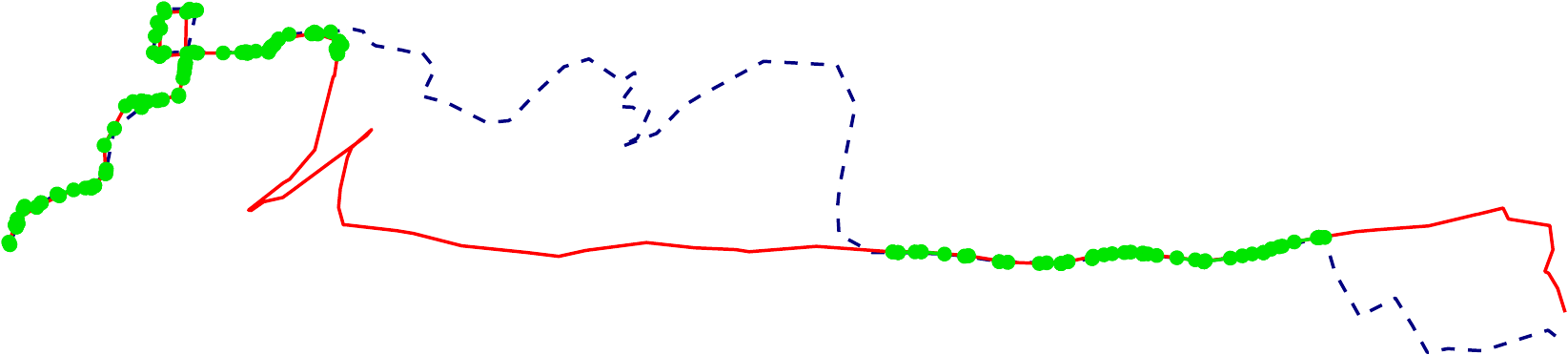} \\
    (a) & (b) & (c) \\
    \includegraphics[width=0.3\textwidth]{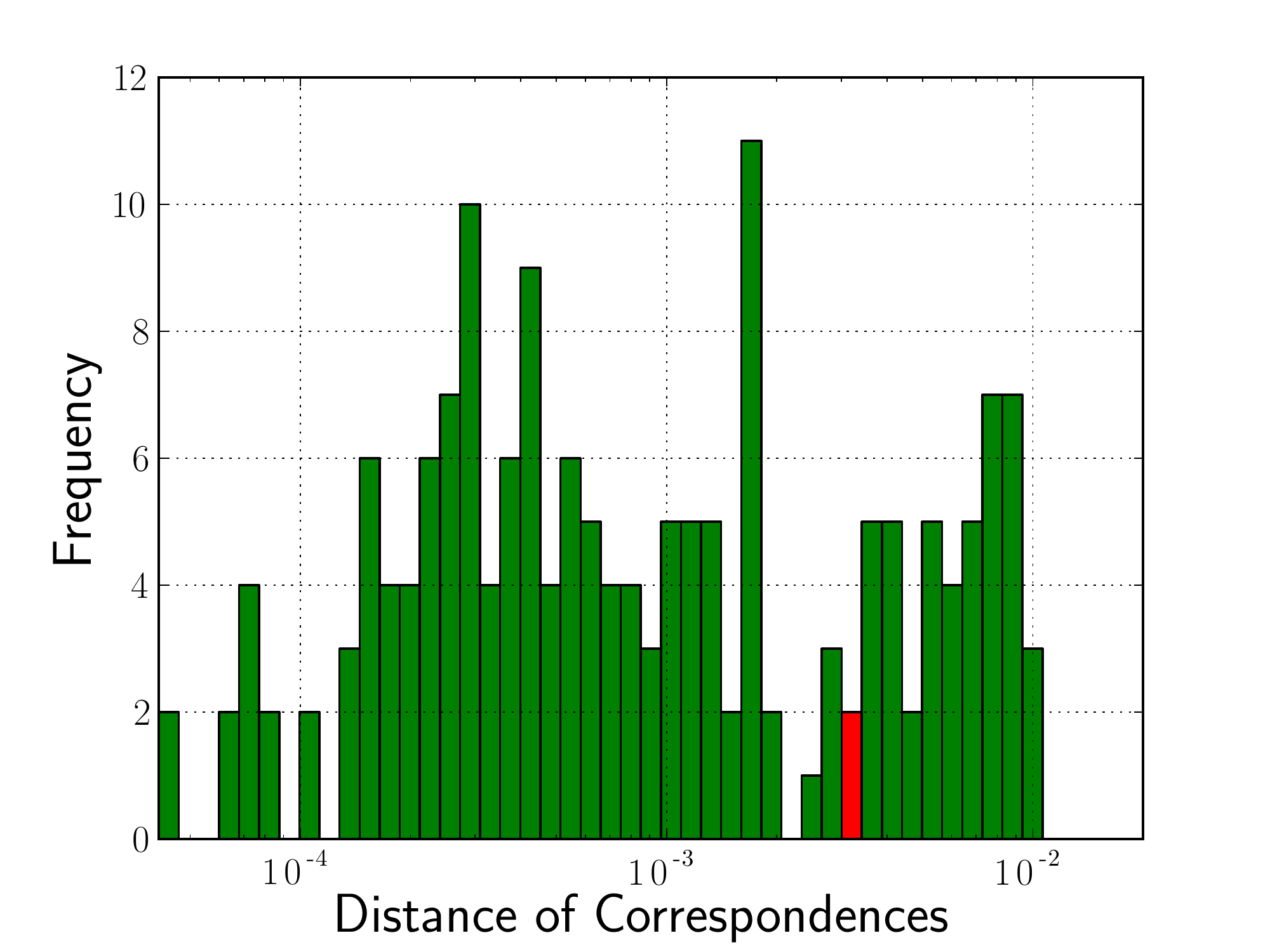} &
    \includegraphics[width=0.3\textwidth]{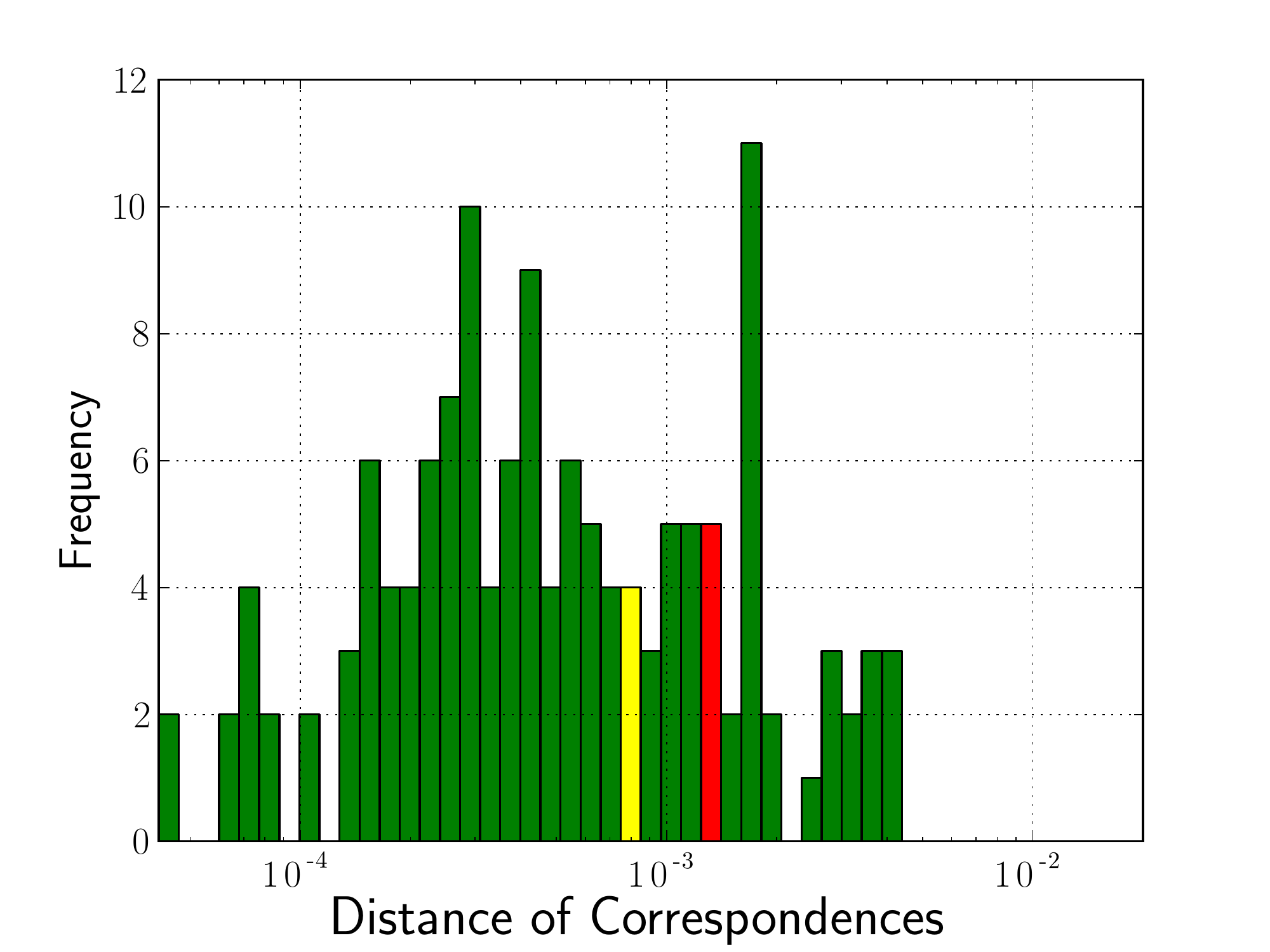} &
    \includegraphics[width=0.3\textwidth]{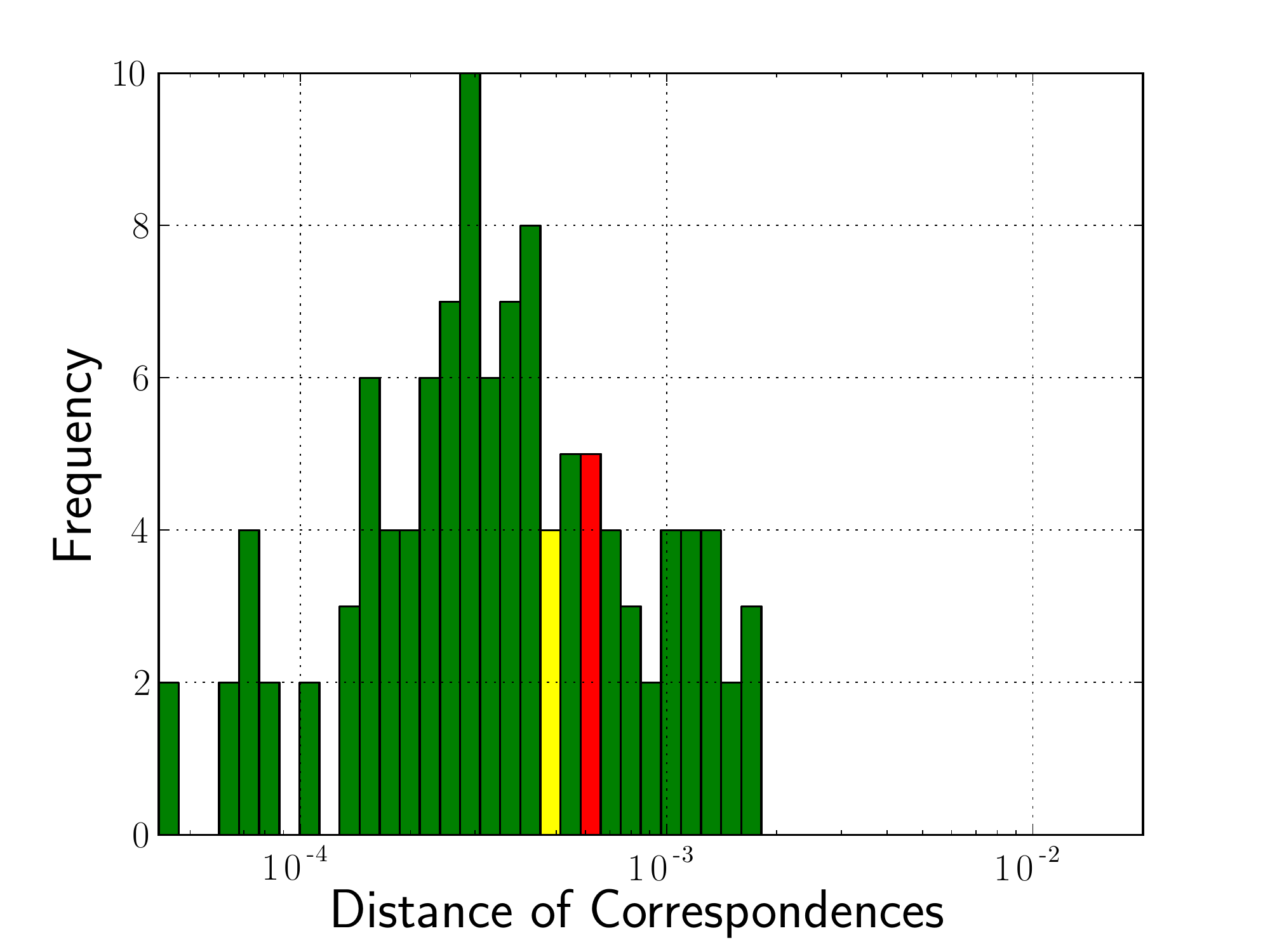} \\
    (d) & (e) & (f) \\
    \includegraphics[width=0.3\textwidth]{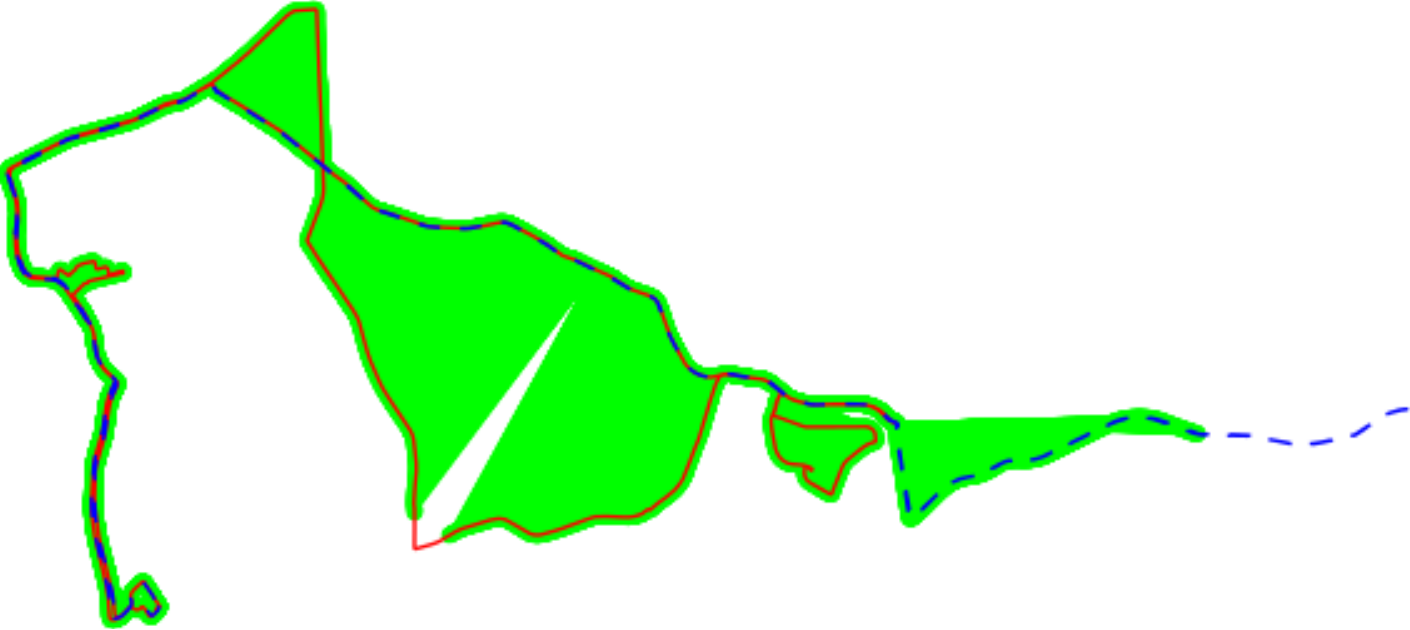} &
    \includegraphics[width=0.3\textwidth]{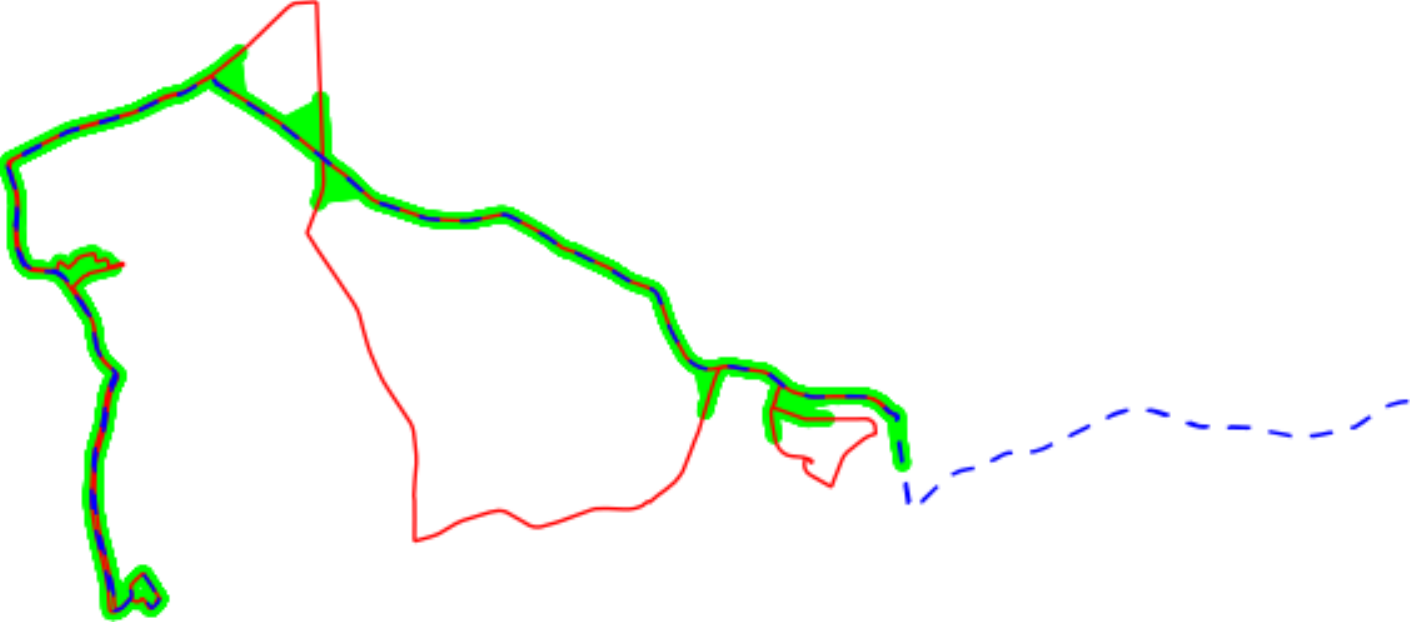} &
    \includegraphics[width=0.3\textwidth]{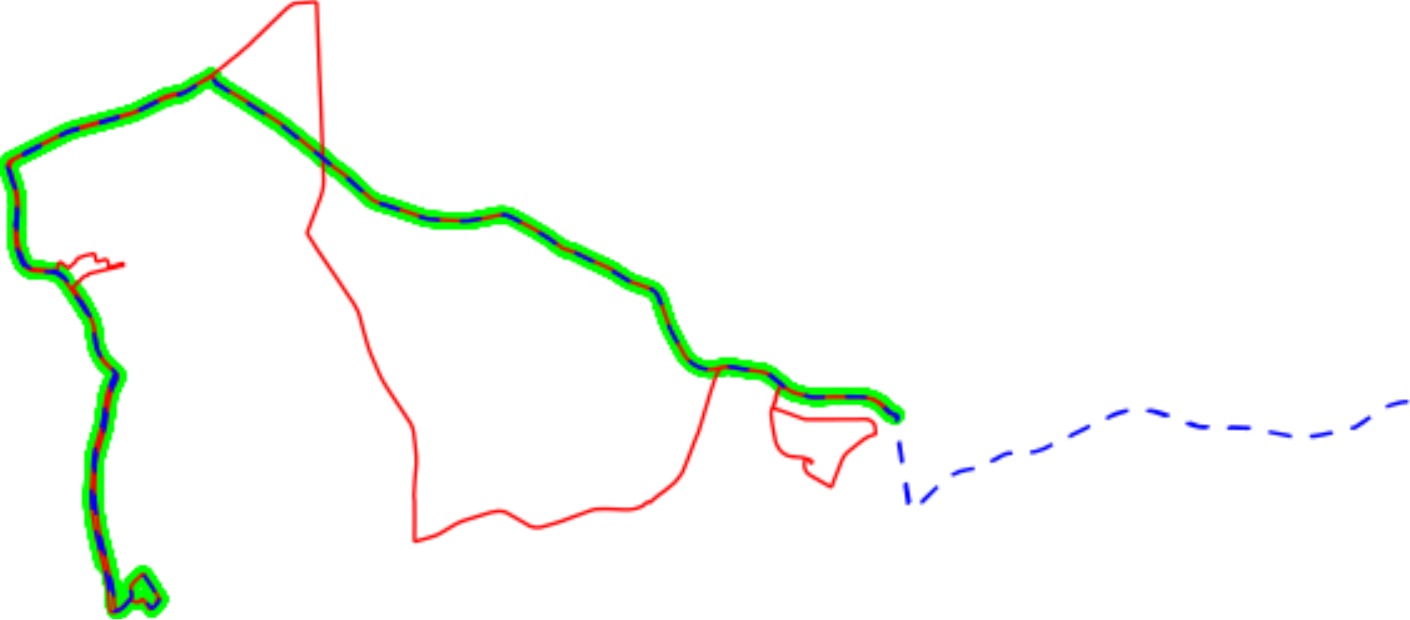} \\
    (g) & (h) & (i) \\
    \includegraphics[width=0.3\textwidth]{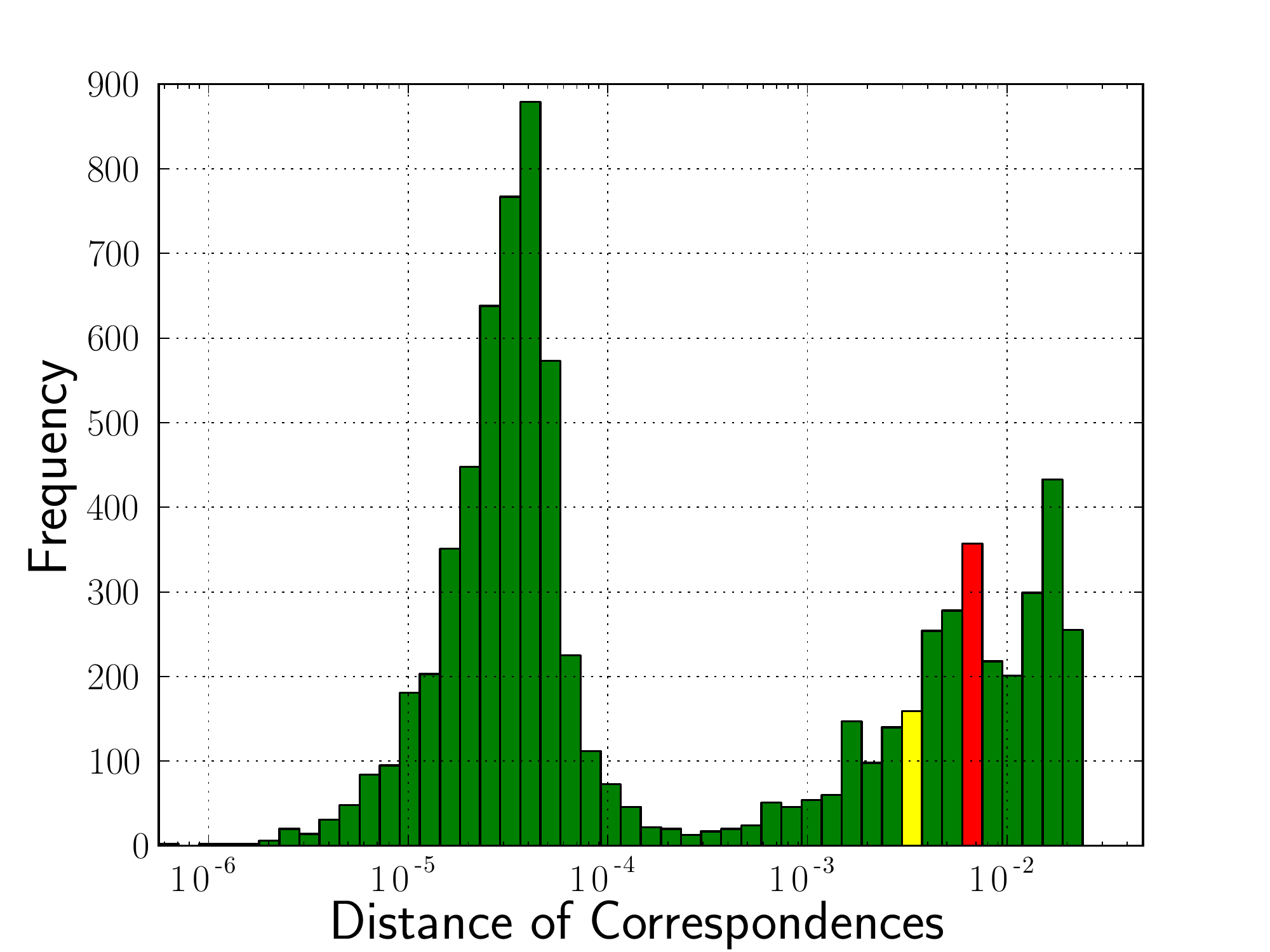} &
    \includegraphics[width=0.3\textwidth]{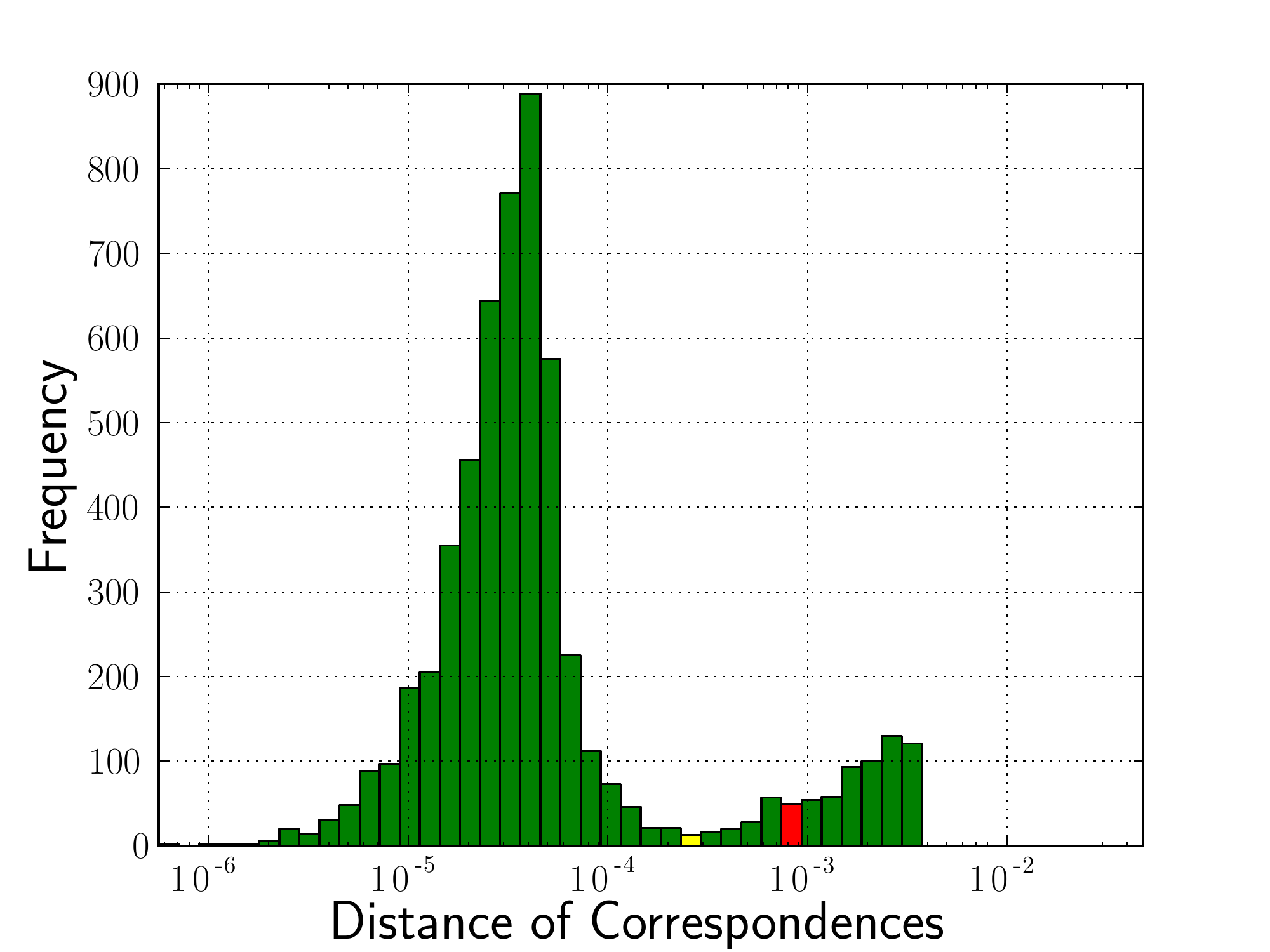} &
    \includegraphics[width=0.3\textwidth]{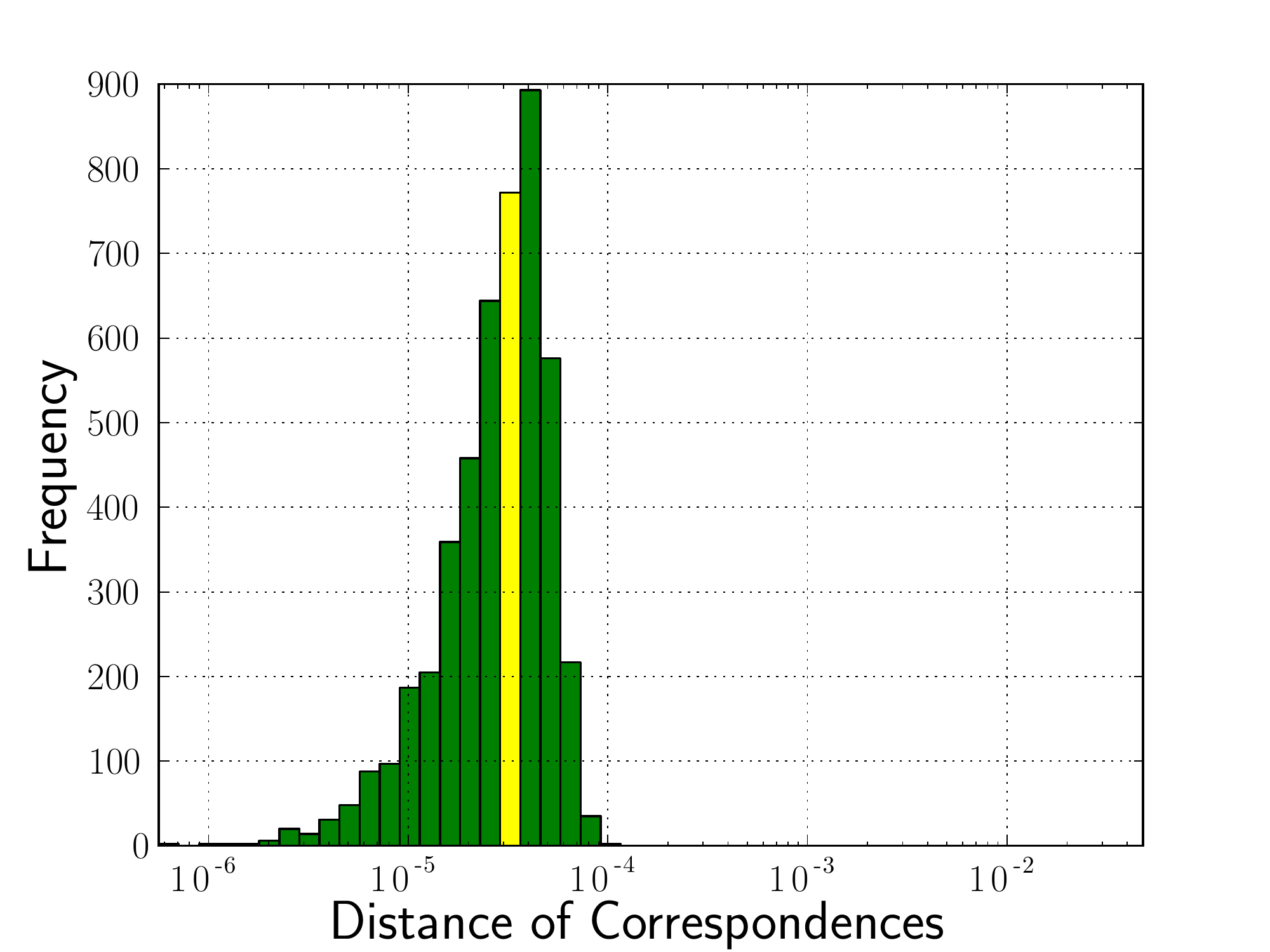} \\
    (j) & (k) & (l)
  \end{tabular}
  \caption{The assignments and corresponding distance histograms
    computed at various stages of the iterative algorithm. (a), (b),
    (c), (d), (e) and (f) correspond to the buses dataset pair while
    (g), (h), (i), (j), (k) and (l) correspond to the cycling dataset
    pair. (a), (d), (g) and (j) are at the first iteration, (b), (e),
    (h) and (k) are at an intermediate stage and (c), (f), (i) and (l)
    are at the point of convergence.}
  \label{fig:iter}
\end{figure}

\mparagraph{Effect of Parameters.}\label{sec:paramselectexp}
Recall that the parameters in the scoring function are all set based
on a threshold distance for gaps (see
Sec.~\ref{sec:param-select}). Fig.~\ref{fig:dvsr} shows how the $\rms$
of the optimal assignment varies as a function of the distance
threshold $r$ used to define the parameters of $\score$. For each
distance $r$ on the $x$-axis we defined the parameters $a,\Delta,c$
and computed the $\rms$ of the optimal assignment based on those
parameters. The axes in Fig.~\ref{fig:dvsr} are set at a log-scale due
to the much larger variation in distances and extremely low distance
threshold in the converged assignment. When we choose the ``correct''
$r$ for the algorithm, we expect variation in distances of assignment
edges to be only due to noise. Hence, the $\rms$ of the assignment
edges will be comparable to the chosen $r$ (if the actually deviating
portions are at sufficiently large distances relative to the
noise). This is reflected in the flat portion of the graph in
Fig~\ref{fig:dvsr}. The red points in Fig.~\ref{fig:dvsr} are the
chosen thresholds of the iterative procedure and as can be seen, it
converges (leftmost red point) when the chosen threshold is comparable
to the $\rms$.

Finally, Fig.~\ref{fig:iter} 
shows the assignment and histograms for the pair from the buses and
cycling datasets at the beginning, end and an intermediate step of the
iterative approach. As we can see, larger distances are slowly pruned
away until we reach the optimal assignment. In the bus dataset,
however, there is still significant variance in the data at the time
of convergence showing the low sampling rate and noise inherent in the
data.

\begin{figure}[ht]
\begin{minipage}{\textwidth}
  \centering
  \begin{tabular}{ccc}
    \includegraphics[width=0.3\textwidth]{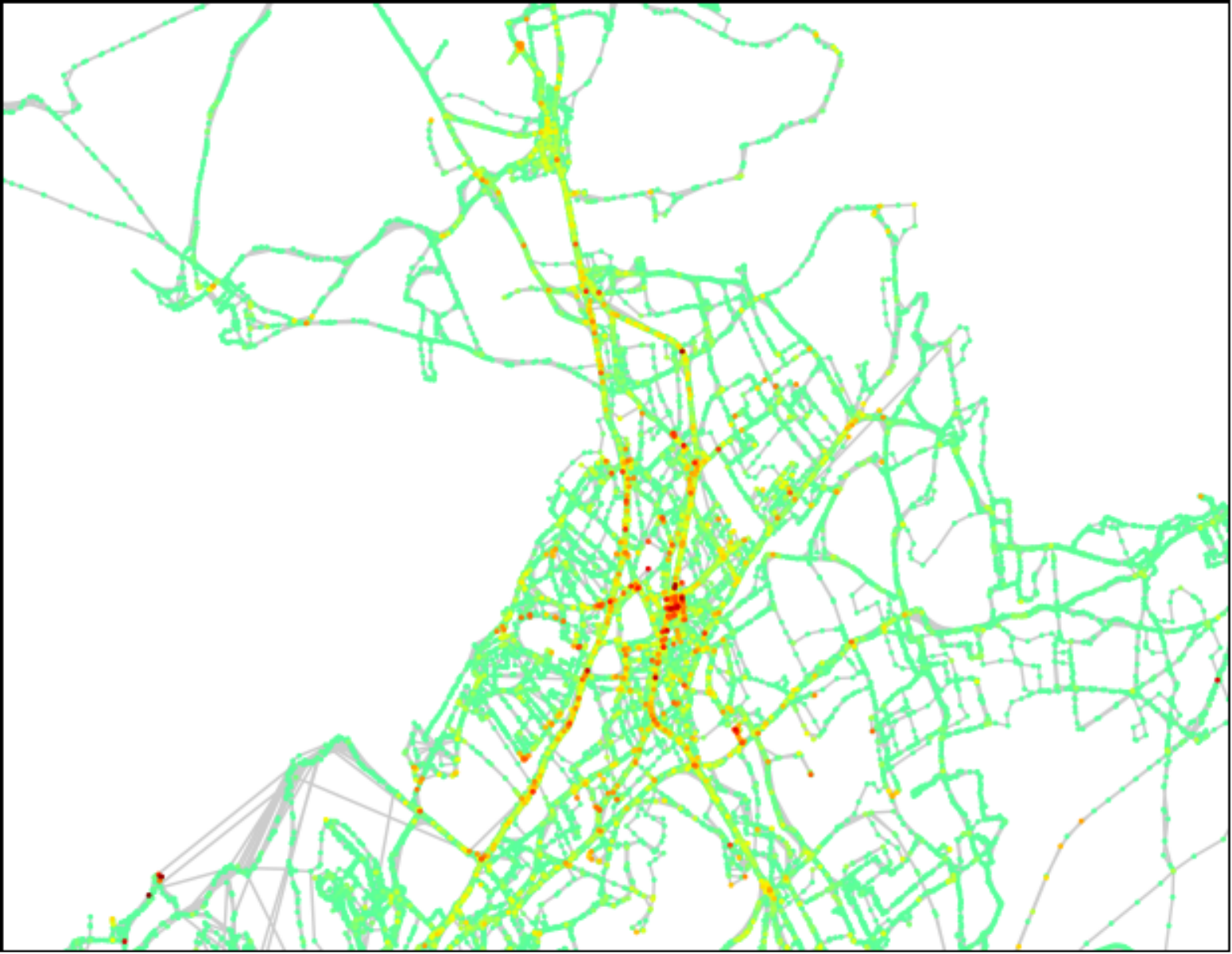}
    &
    \includegraphics[width=0.3\textwidth]{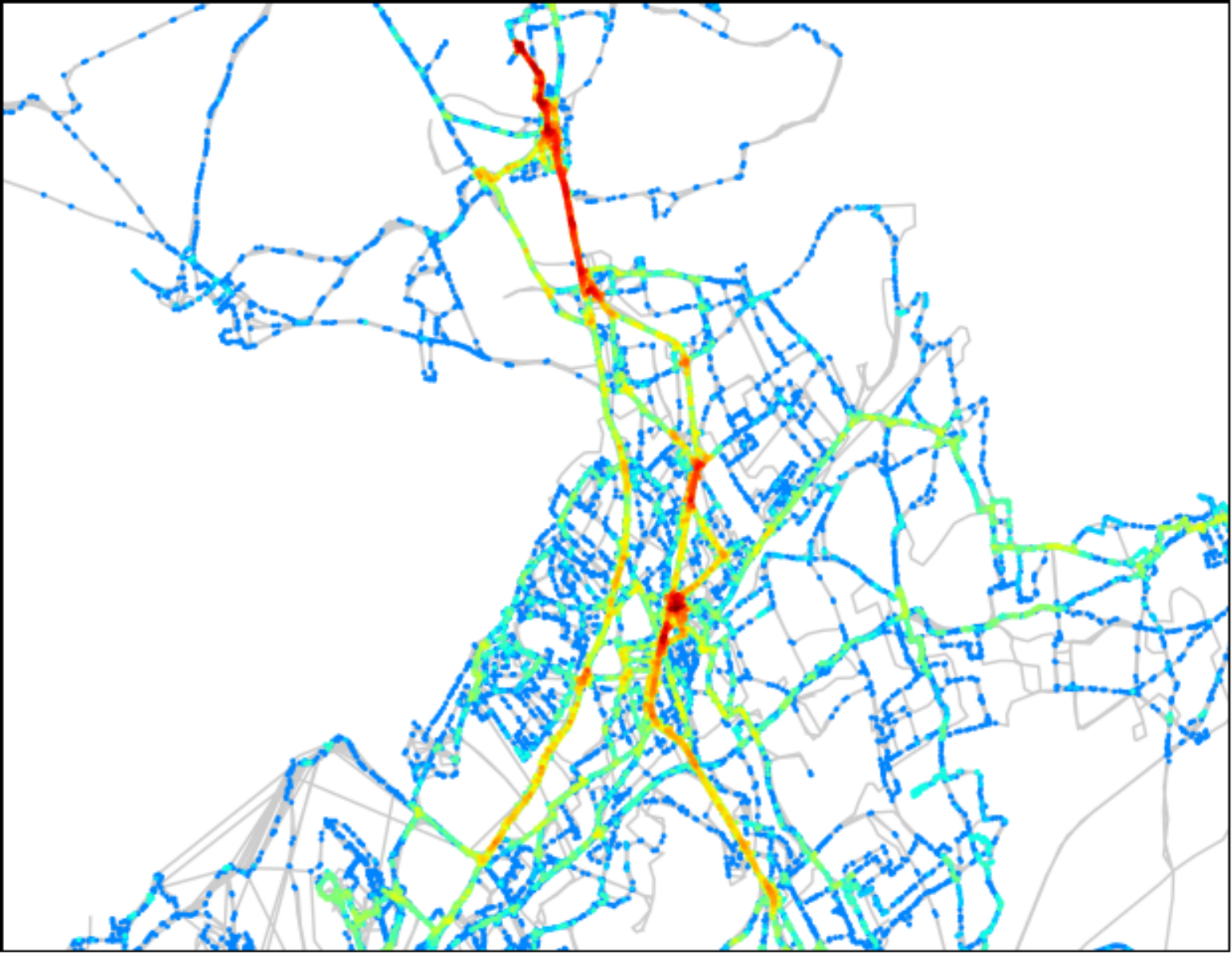}
    &
    \includegraphics[width=0.3\textwidth]{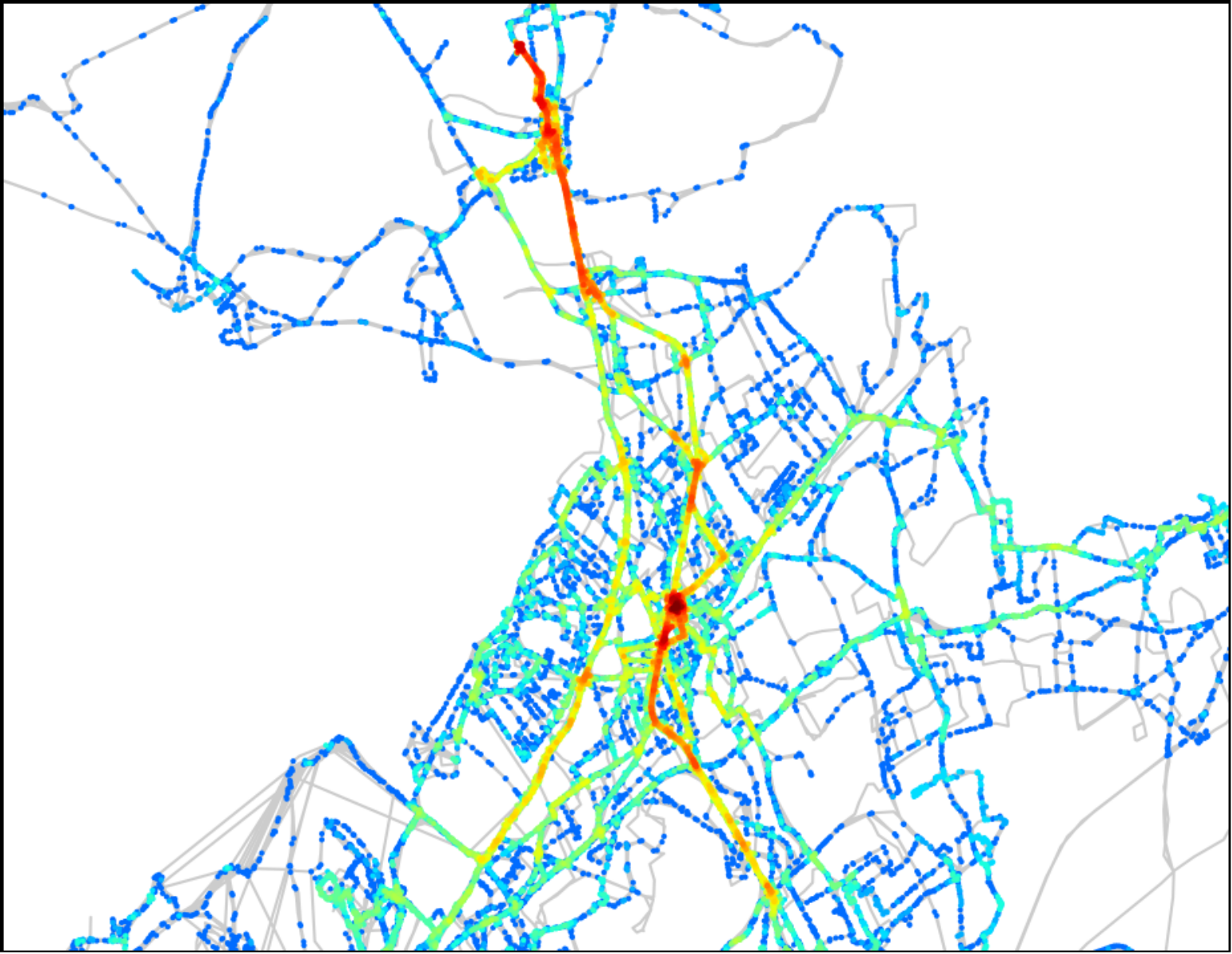}
    \\
    \scriptsize{(a) \dtw} & \scriptsize{(b) \dtwp} & \scriptsize{(c) \ass} \\
    \includegraphics[width=0.3\textwidth]{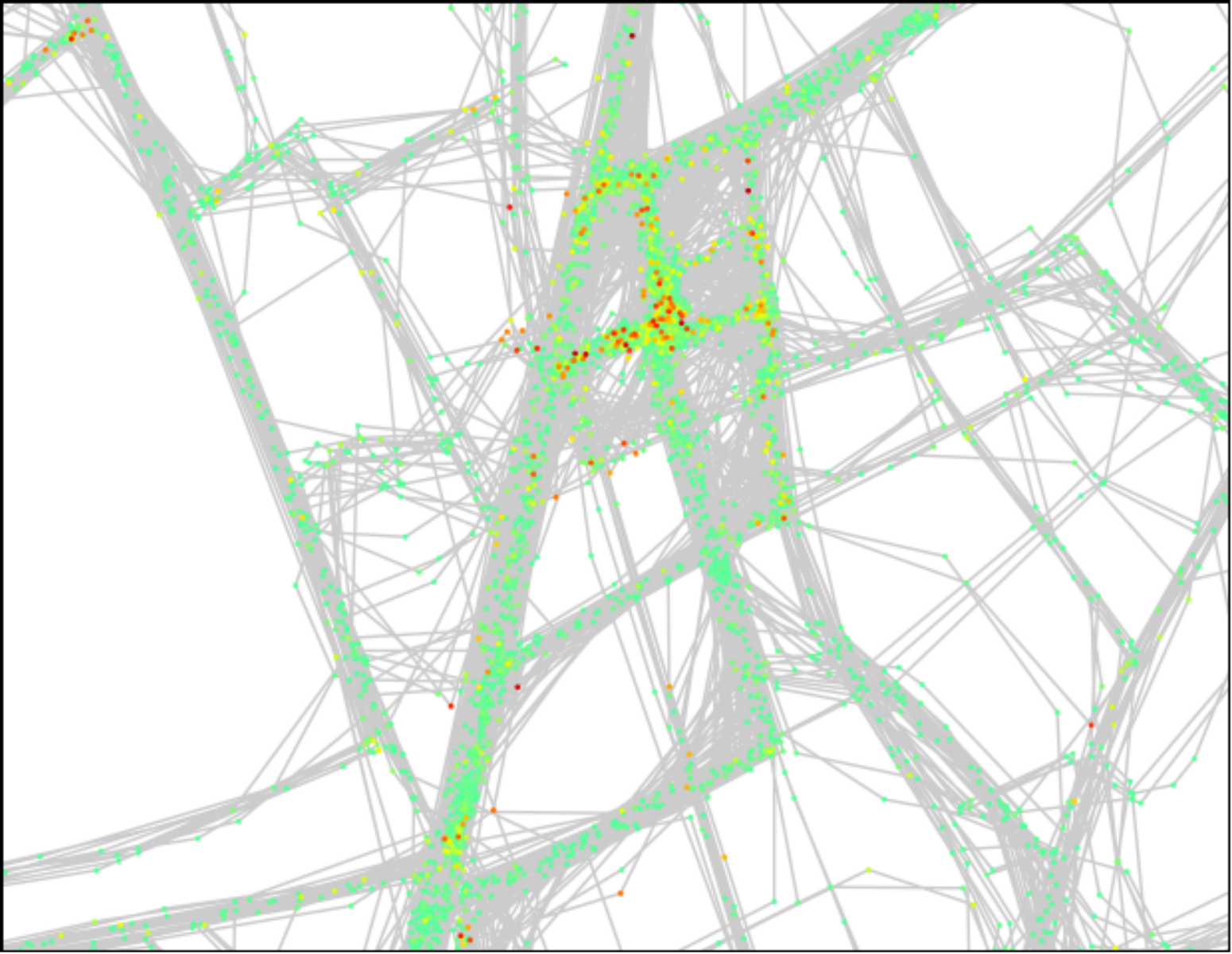}
    &
    \includegraphics[width=0.3\textwidth]{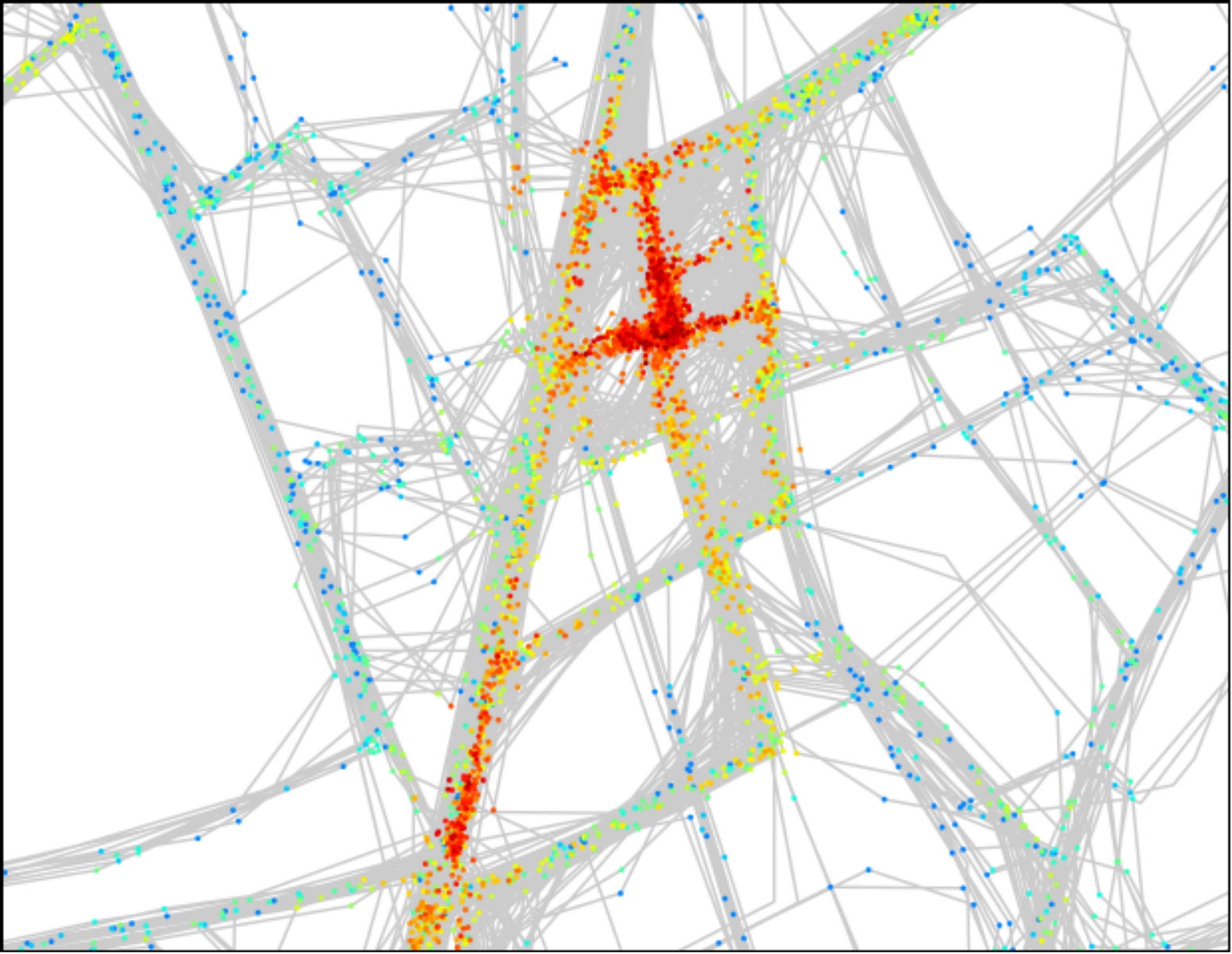}
     &
     \includegraphics[width=0.3\textwidth]{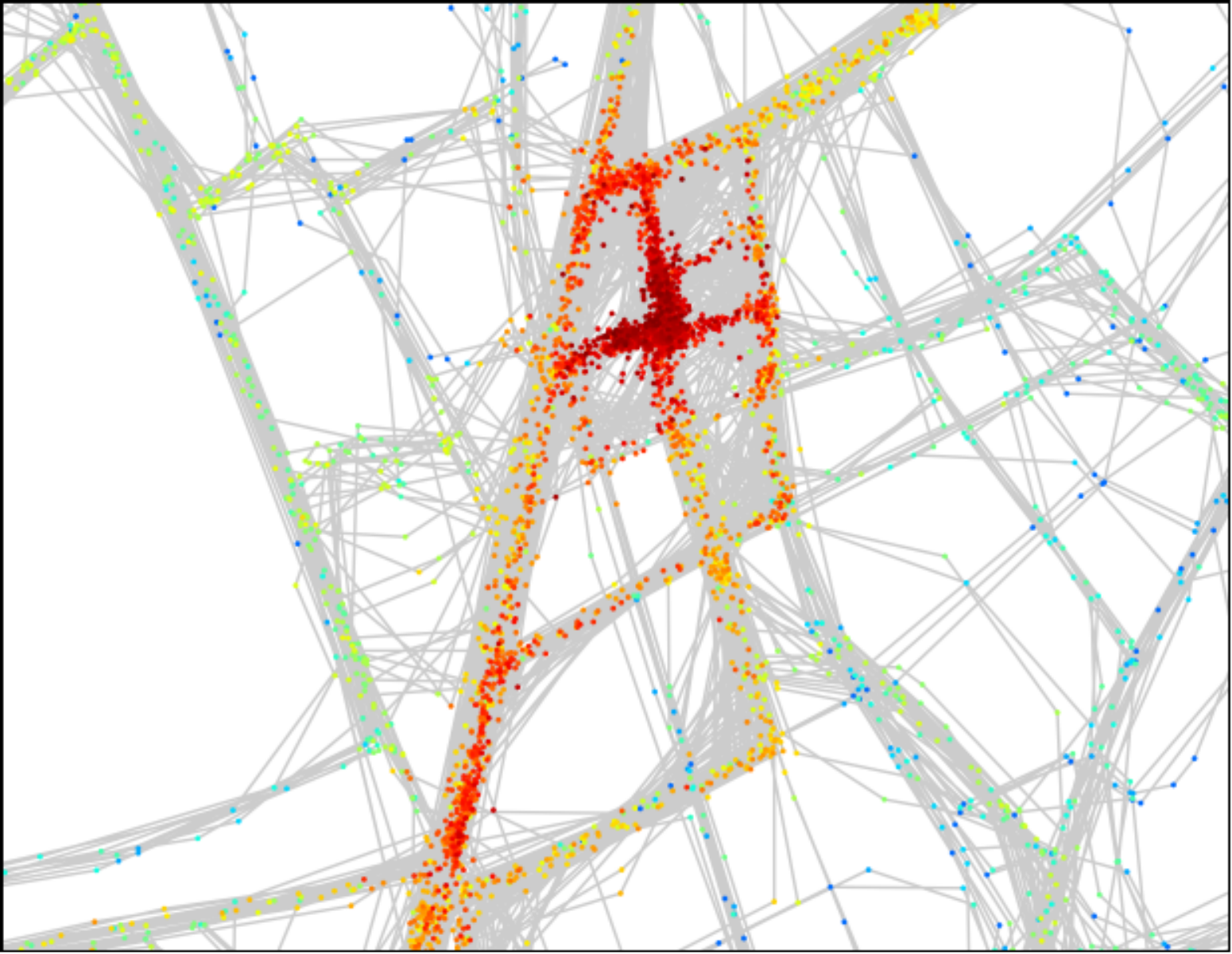}
    \\
    \scriptsize{(d) \dtw} & \scriptsize{(e) \dtwp} & \scriptsize{(f) \ass} \\
  \end{tabular}
  \caption{Heat maps of the importance of sample points for the buses
    dataset. (d), (e) and (f) show a zoomed in portion of (a), (b) and
    (c) respectively.}
  \label{fig:buses_hm}
\end{minipage}
\vspace{0.1in}

\begin{minipage}{\textwidth}
  \centering
  \begin{tabular}{ccc}
    \includegraphics[width=0.3\textwidth]{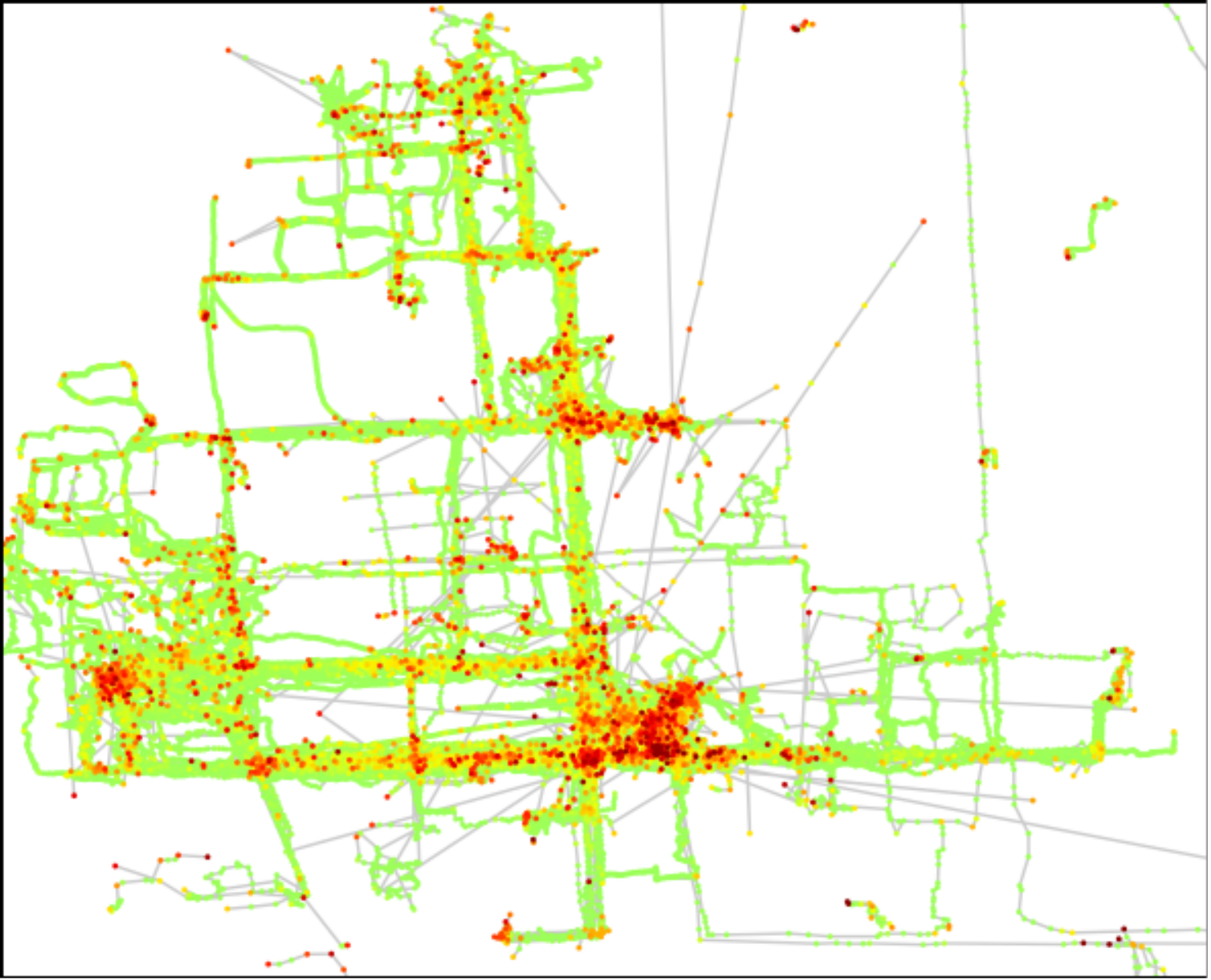}
    &
    \includegraphics[width=0.3\textwidth]{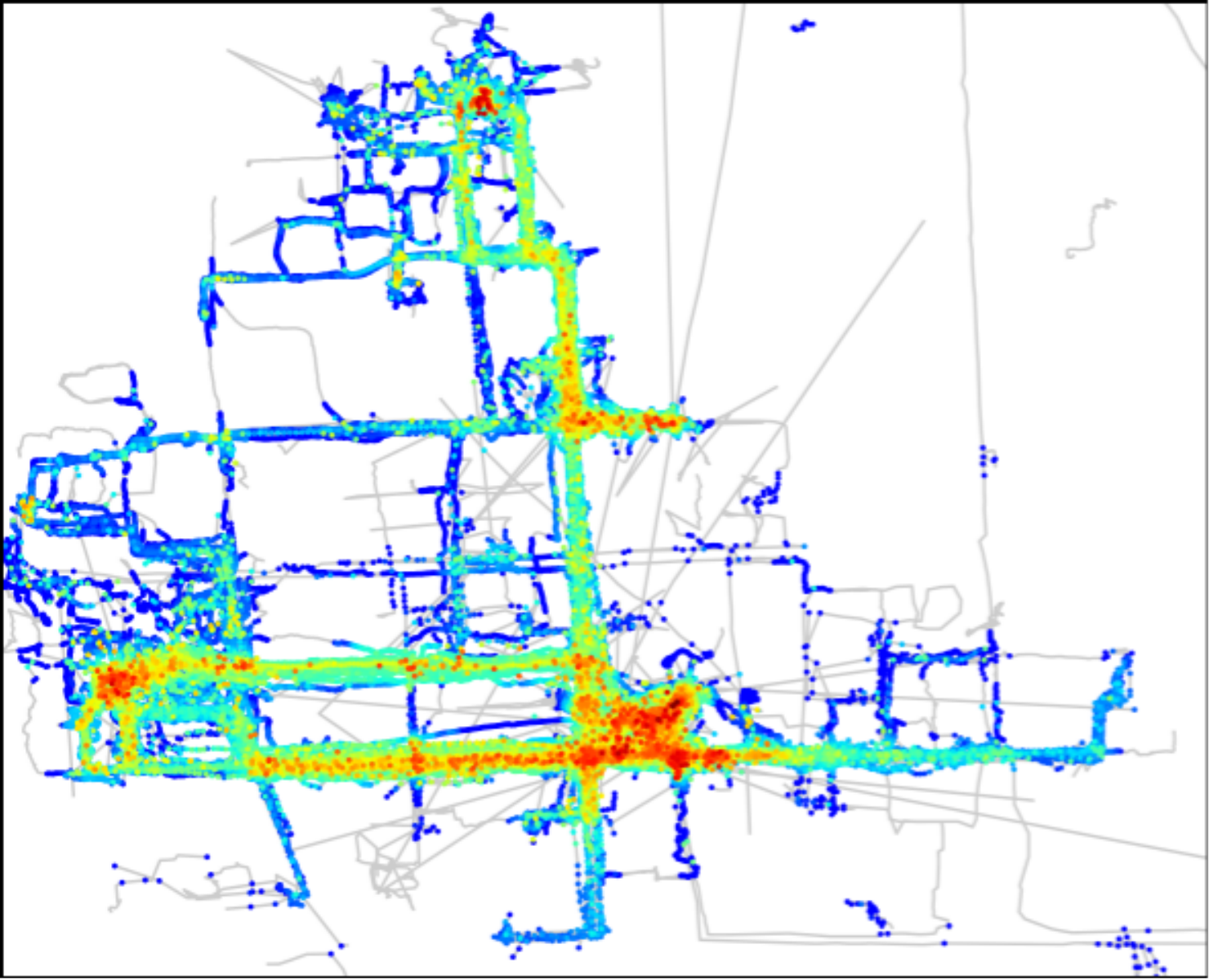}
    &
    \includegraphics[width=0.3\textwidth]{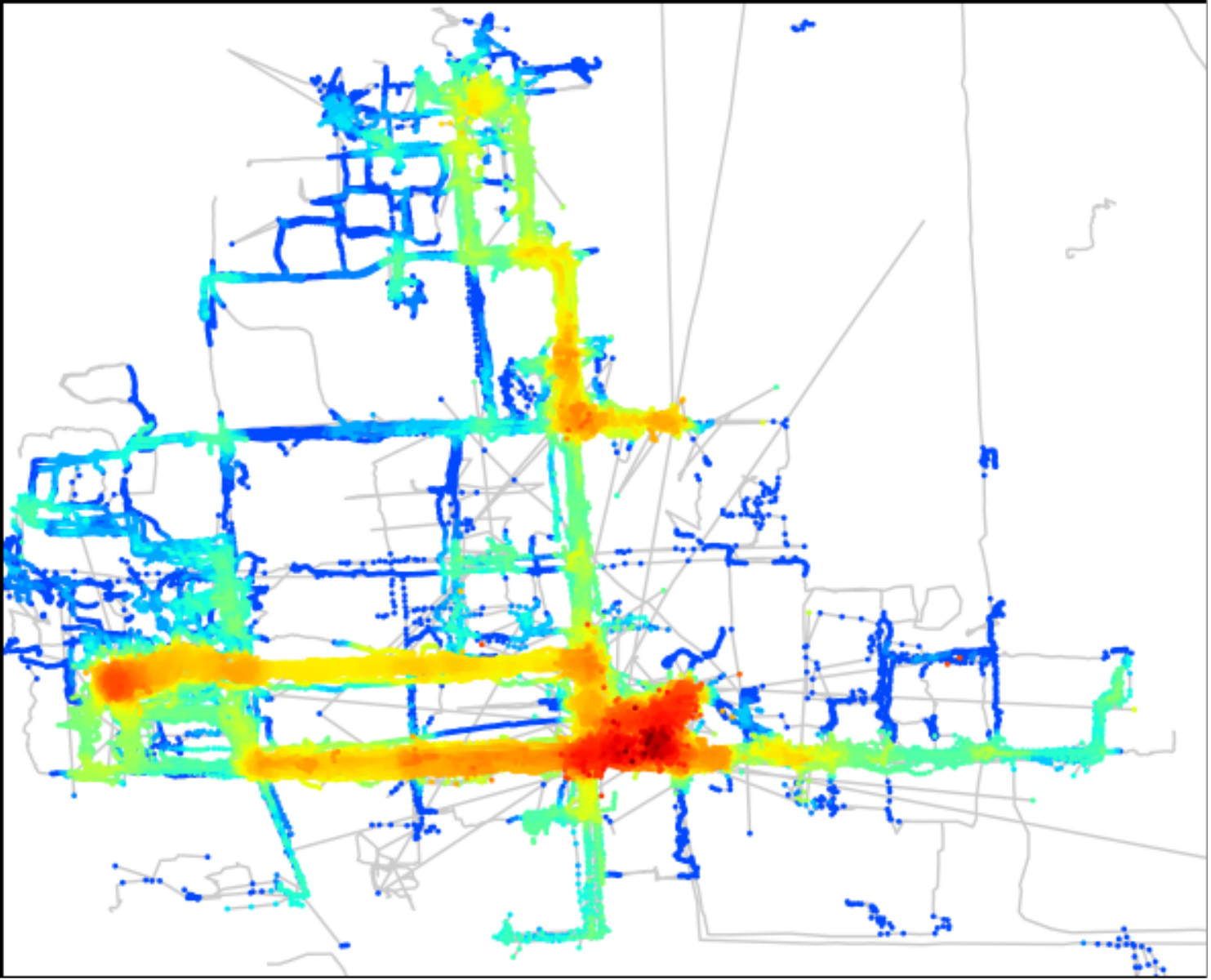}
    \\
    \scriptsize{(a) \dtw} & \scriptsize{(b) \dtwp} & \scriptsize{(c) \ass} \\
  \end{tabular}
  \caption{Heat maps of the importance of sample points for the
    GeoLife dataset.}
  \label{fig:geo_hm}
\end{minipage}
\vspace{0.1in}

\begin{minipage}{\textwidth}
  \centering
  \begin{tabular}{ccc}
    \includegraphics[width=0.3\textwidth]{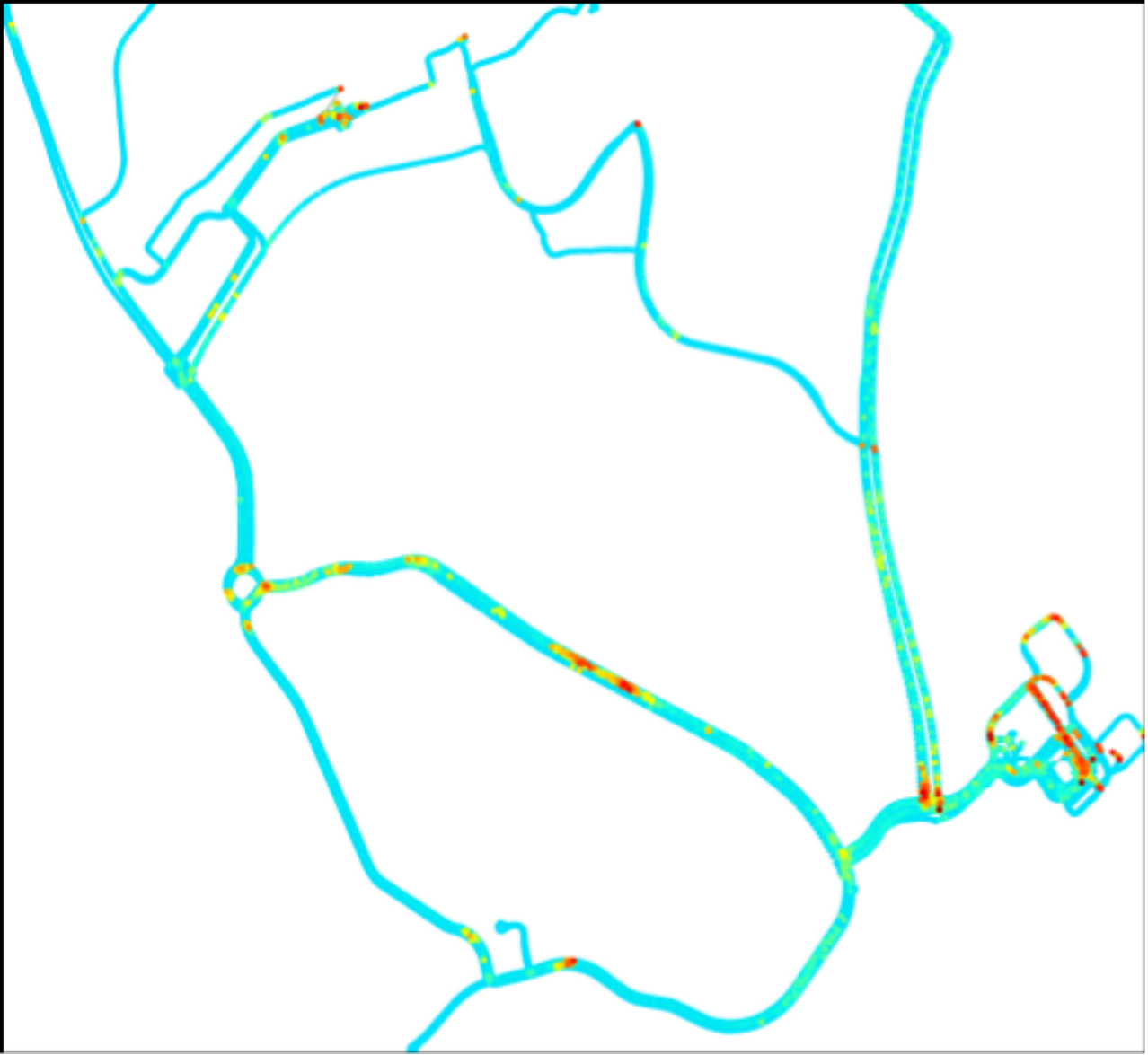}
    &
    \includegraphics[width=0.3\textwidth]{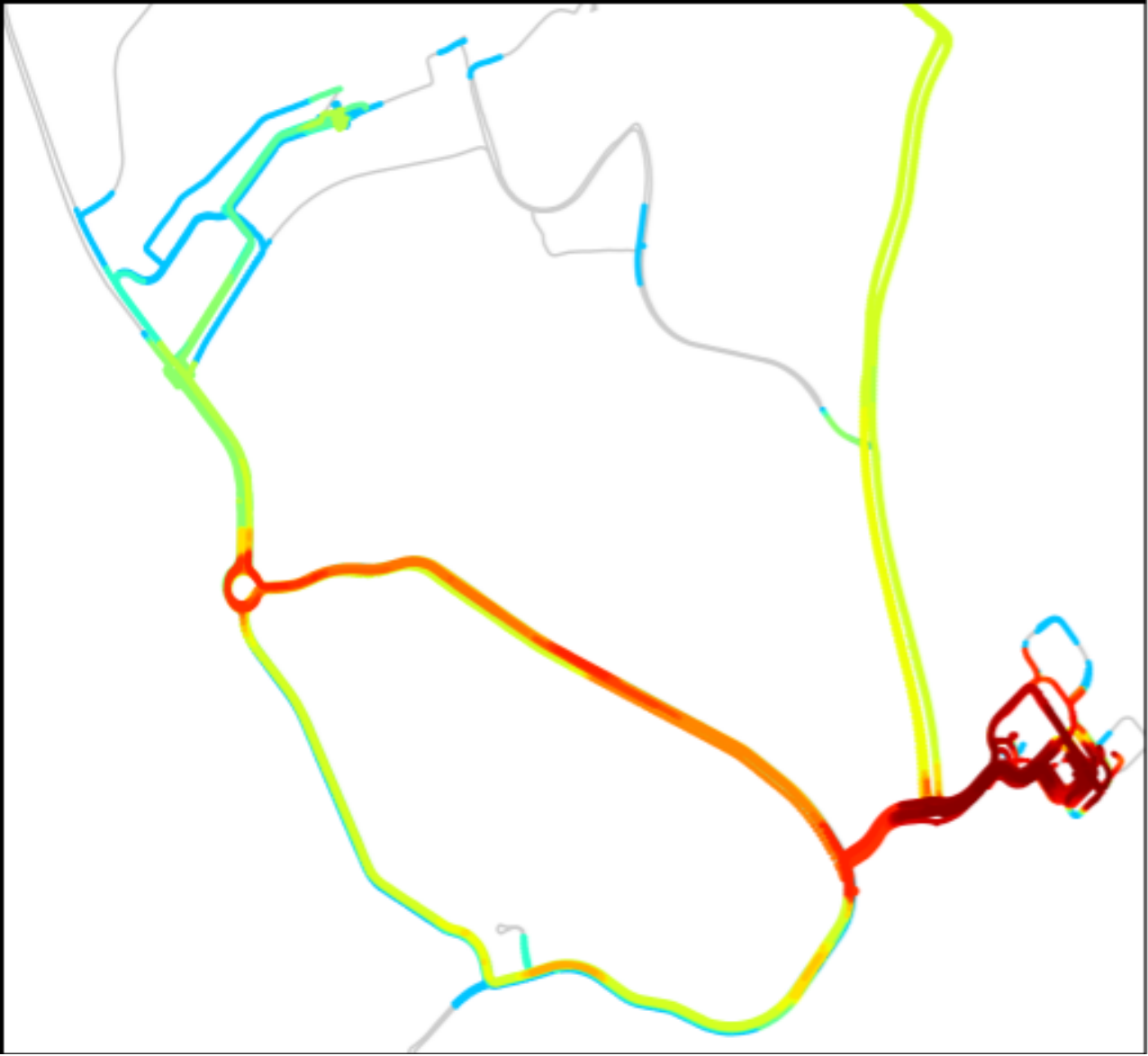}
    &
    \includegraphics[width=0.3\textwidth]{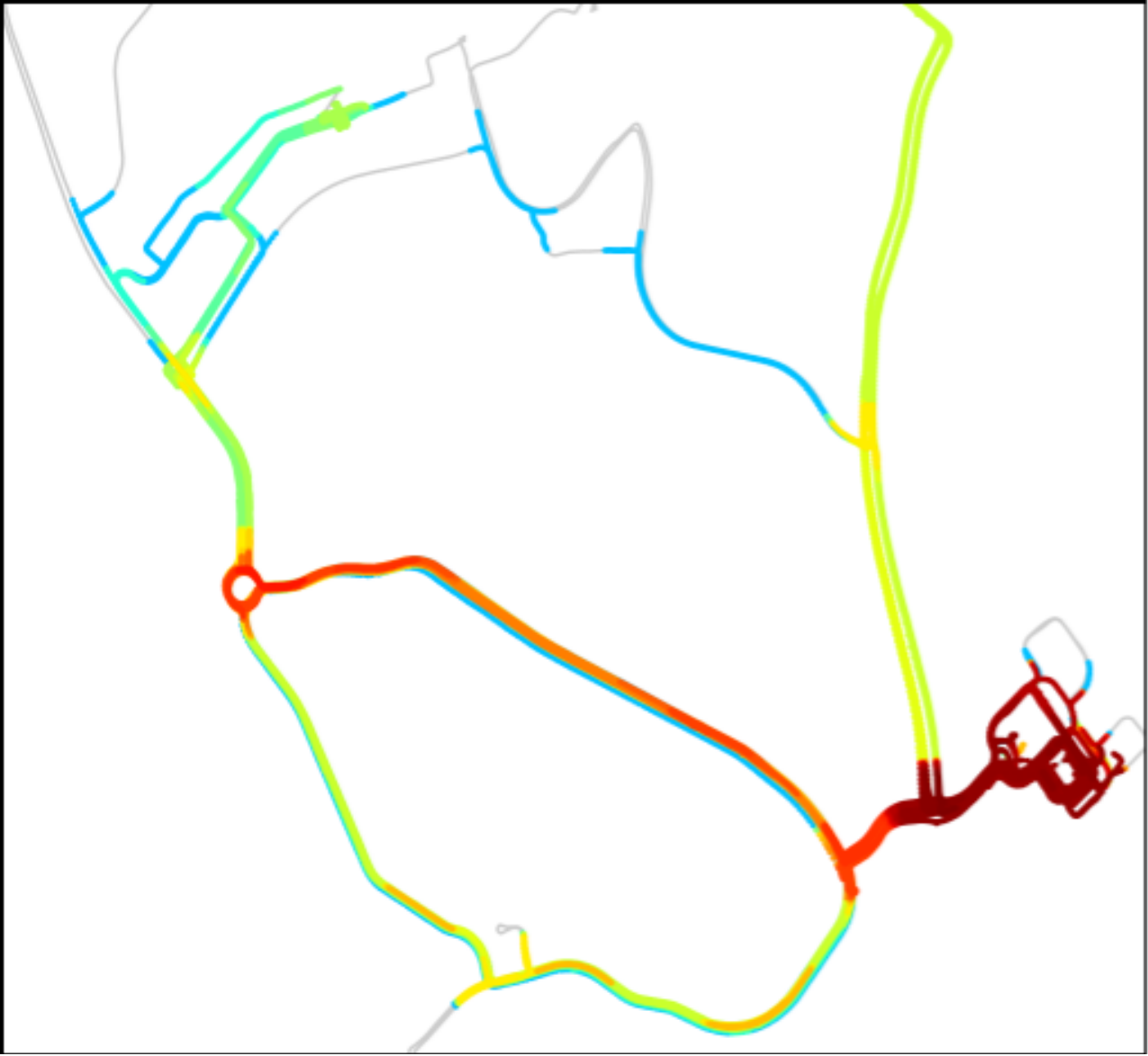}
    \\
    \scriptsize{(a) \dtw} & \scriptsize{(b) \dtwp} & \scriptsize{(c) \ass} \\
  \end{tabular}
  \caption{Heat maps of the importance of sample points for the
    cycling dataset.}
  \label{fig:cycle_hm}
\end{minipage}
\end{figure}

\subsection{Results on Complete Datasets}\label{sec:multipleresults}

To further demonstrate the effectiveness of our algorithms, we present
here results showing the behavior on the datasets as a whole. Based on
pairwise correspondences computed by the \dtw, \dtwp and \ass
algorithms, we analyze the {\em importance} of a sample point on a
trajectory based on its correspondences with all other
trajectories. Specifically, we wish to measure the importance of a
portion of a trajectory with respect to the dataset using the
importance of individual points on it. In all cases, we compute
pairwise correspondences for all pairs of trajectories in the
dataset. Based on these, we define the importance of a sample point
$p$ on some trajectory $P$ in three different ways for the three
algorithms such that the comparison is as fair as possible: (i) for
\ass, we define it as the number of outgoing edges for a sample point,
i.e., the number of trajectories $Q$ such that for some $q\in Q$,
$\alpha(p)=q$ in the assignment., (ii) for \dtwp, we define it in a
similar manner, the number of trajectories $Q$ such that $(p,q)$ for
some $q\in Q$ is part of the set of correspondences between $P$ and
$Q$, (iii) for \dtw, it is not possible to define similar to \dtwp
since then, the importance of all points is the same; hence, we define
it as the number of correspondence pairs that $p$ is part of over all
pairwise assignments. For \ass, the outdegree of a point is an
intuitive way to define the importance since it provides a natural way
of defining the number of trajectories which share a common portion
with $P$ at $p$. We chose the definition of importance of \dtwp to
reflect the same intuition.

Again, we chose 100\,m as the distance threshold for \dtwp and \ass
and 4 as the minimum gap length for \ass. Figures~\ref{fig:buses_hm},
\ref{fig:geo_hm} and \ref{fig:cycle_hm} show the heat maps of the
importance of the points belonging to all trajectories in the buses,
GeoLife and cycling datasets respectively. As is clear, the \dtw
approach does not provide any meaningful results, so we present it for
completeness. Comparing \dtwp and \ass, the results are quite
similar. Both of these seem to do a good job of identifying commonly
traveled routes and landmark points in the dataset. \ass seems to do a
slightly better job of identifying more central points along the
routes or points on a so-called ``mean'' trajectory although this is
apparent only upon close examination. Further analyzing the behavior
of trajectories based on the complete dataset is outside the scope of
this paper and we present these results here as a flavor of what our
framework can provide for this purpose.

\subsection{Summary}\label{sec:expsummary}
We have shown that our framework captures the advantages of both DTW
and sequence alignment based approaches for identifying trajectory
similarity, and that it is able to exceed their accuracy. Experiments
show that the approach is highly accurate in identifying similar
portions of trajectories from real datasets. Further, even without an
accurate prior knowledge of distances between points based on which to
compute similarity, our iterative procedure is able to converge at the
point where similar portions are identified and distinguished
accurately from dissimilar portions. We also show the effectiveness of
the semi-continuous setting and that the shift of the score for
computing local assignment is highly dependent on the variance of
distances between similar points. 

\thomas{I expanded this slightly based on Pankaj's comments and added
  the summary header, if we end up writing a discussion section we
  could probably move some of this stuff down there}

Finally, our results on importance of sample points based on pairwise
correspondences computed over the entire dataset shows that our
framework does as well or better than other approaches while provided
a principled way to compute correspondences and measure similarity
between two trajectories. We feel that, for the purposes of analyzing
complete datasets for highly conserved portions of trajectories and
performing clustering of trajectories on this basis, our framework of
assignments provides a solid foundation to work with. This direction
of research certainly seems like a rich one to undertake.


\sloppy

\arxiv{
\bibliographystyle{plain}
}
\sstd{
\bibliographystyle{splncs}
}
\bibliography{refs}

\end{document}